\documentclass[12pt]{article}
\pdfoutput=1
\usepackage{setspace}

\usepackage{graphicx}
\usepackage{epstopdf}
\usepackage[body={17.5cm, 21.5cm},right=2.05cm]{geometry}
\usepackage{amssymb}
\usepackage{amsmath}
\usepackage{dcolumn}% Align table columns on decimal point

\usepackage{graphicx}
\usepackage{epstopdf}
\usepackage{amsmath}
\usepackage{amsfonts}
\usepackage{amssymb}
\usepackage[usenames]{color}
\usepackage{mathrsfs}
\usepackage{array}
\usepackage{nonfloat}

\usepackage{floatrow}
\newfloatcommand{capbtabbox}{table}[][\FBwidth]

\usepackage{caption}

\usepackage{multirow}
\usepackage{psfrag}
\pdfoutput=1

\usepackage[colorlinks,bookmarks]{hyperref}
\definecolor{linkblue}{rgb}{0,0,0.8}
\definecolor{linkgreen}{rgb}{0,0.5,0}

\hypersetup{pdfpagemode=UseNone, pdfstartview=FitH, linkcolor=linkblue, %
            citecolor=linkgreen, urlcolor=linkblue}

\newcommand\nn{\nonumber}
\newcommand\eea{\end{eqnarray}}
\newcommand\bea{\begin{eqnarray}}

\def\la{\langle}
\def\ra{\rangle}
\def\beq{\begin{equation}}
\def\eeq{\end{equation}}
\def\d{\partial}

\newcommand{\be}{\begin{equation}}
\newcommand{\ee}{\end{equation}}
\newcommand{\ba}{\begin{align}}
\newcommand{\ea}{\end{align}}
\newcommand{\bg}{\begin{gather}}
\newcommand{\eg}{\end{gather}}
\newcommand{\bseq}{\begin{subequations}}
\newcommand{\eseq}{\end{subequations}}

\newcommand{\vk}{\boldsymbol{k}}
\newcommand{\vkp}{\boldsymbol{q}}
\newcommand{\vkkp}{\boldsymbol{k}-\boldsymbol{q}}
\newcommand{\vq}{\boldsymbol{q}}

\newcommand{\vx}{\boldsymbol{x}}

\newcommand{\vv}{\boldsymbol{v}}
\newcommand{\hn}{\hat{\boldsymbol{n}}}

\def\lcdm{$\Lambda$CDM}
\newcommand{\knl}{k_{\rm NL}}

\def\H{{\cal H}}
\newcommand{\hinvMpc}{h\,$Mpc$^{-1}}

\newcommand{\invMpc}{\,h\, {\rm Mpc}^{-1}\,}

\def\Ommnow{\Omega_{{\rm m},0}}
\def\Omm{\Omega_{\rm m}}

\newcommand{\ta}{\tilde a}

\newcommand{\lp}{\left(}
\newcommand{\rp}{\right)}
\newcommand{\lb}{\left[}
\newcommand{\rb}{\right]}
\def\co{c_{s  (1)}^2}
\def\ct{c_{s  (2)}^2}

\definecolor{purple}{rgb}{0.78,0.18,0.77}

\newcommand{\kren}{k_\text{ren}}
\newcommand{\gammai}{(\d\tau)_{\rho_l}}

\def\orderone{\mathcal{O}(1)}

\def\poneloop{P_\text{1-loop}}
\def\ptwoloop{P_\text{2-loop}}
\def\pthreeloop{P_\text{3-loop}}
\def\ptreecs{P_\text{tree}^{(c_{\rm s})}}

\def\poneloopcs{P_\text{1-loop}^{(c_{\rm s})}}

\def\kfail[#1]{k_{\rm fail}^{(#1)}}

\def\epsdl{\epsilon_{\delta<}}
\def\epssg{\epsilon_{s>}}

\makeatletter
\newlength{\apb@width}
\newcommand{\autoparbox}[2][c]{\settowidth{\apb@width}{#2}\parbox[#1]{\apb@width}{#2}}

\makeatother

\begin{document}

\vspace{5mm}
\vspace{0.5cm}
\begin{center}

\def\thefootnote{\fnsymbol{footnote}}
{\Large \bf  The EFT of Large Scale Structures at All Redshifts:\\[0.5cm] 
Analytical Predictions for Lensing
}
\\[0.8cm]

{\large Simon Foreman and Leonardo Senatore}
\\[0.5cm]

{\normalsize { \sl Stanford Institute for Theoretical Physics and Department of Physics, \\Stanford University, Stanford, CA 94306}}\\
\vspace{.3cm}

{\normalsize { \sl Kavli Institute for Particle Astrophysics and Cosmology, \\ Stanford University and SLAC, Menlo Park, CA 94025}}\\
\vspace{.3cm}

\end{center}

\vspace{.8cm}

\hrule \vspace{0.3cm}
{\small  \noindent \textbf{Abstract} \\[0.3cm]
\noindent 
We study the prediction  of the Effective Field Theory of Large Scale Structures (EFTofLSS) for the matter power spectrum at different redshifts. In previous work, we found that the two-loop prediction can match the nonlinear power spectrum measured from $N$-body simulations at redshift zero within approximately~2\% up to $k\sim 0.6\invMpc$  after fixing a single free parameter, the so-called ``speed of sound". We determine the time evolution of this parameter by matching the EFTofLSS prediction to simulation output at different redshifts, and find that it is well-described by a fitting function that only includes one additional parameter. After the two free parameters are fixed, the prediction agrees with nonlinear data within approximately~2\% up to at least $k\sim 1\invMpc$ at $z\geq 1$, and also within approximately~5\% up to $k\sim 1.2\invMpc$ at $z=1$ and $k\sim 2.3\invMpc$ at $z=3$, a major improvement with respect to other perturbative techniques. We also develop an accurate way to estimate where the EFTofLSS predictions at different loop orders should fail, based on the sizes of the next-order terms that are neglected, and find agreement with the actual comparisons to data. Finally, we use our matter power spectrum results to perform analytical calculations of lensing potential power spectra corresponding to both CMB and galaxy lensing. This opens the door to future direct applications of the EFTofLSS to observations of gravitational clustering on cosmic scales.
\noindent 
}

 \vspace{0.3cm}
\hrule
\def\thefootnote{\arabic{footnote}}
\setcounter{footnote}{0}

\vspace{.8cm}

\newpage
\tableofcontents
%\newpage

%--------------------------------------------------------------------------------------
% SECTION: INTRODUCTION
%--------------------------------------------------------------------------------------
\section{Introduction}

The WMAP and Planck satellites, which have measured the properties of the CMB with exquisite precision, have enormously improved our understanding of the primordial universe. But they have also made any further progress in our knowledge of the initial conditions of the universe terribly hard.  The next leading source of cosmological information is most probably going to be the Large Scale Structures (LSS) of the universe. But in order for them to be able to improve our knowledge of the primordial universe, a better understanding of their behavior is required. The Effective Field Theory of Large Scale Structures (EFTofLSS)~\cite{Baumann:2010tm,Carrasco:2012cv,Carrasco:2013mua,Pajer:2013jj,Carrasco:2013sva,Mercolli:2013bsa,Carroll:2013oxa,Porto:2013qua,Senatore:2014via,Angulo:2014tfa,Baldauf:2014qfa,Senatore:2014eva,Senatore:2014vja,Lewandowski:2014rca,Mirbabayi:2014zca} is a research program that aims at developing an analytic understanding of the clustering of dark matter and galaxies at large distances, in a perturbative expansion in the smallness of the size of the nonlinearities. 

In LSS, short distance nonlinearities are very large, and therefore not under perturbative control. Mode-mode coupling implies that short distance physics {\it does} affect long distances, where fluctuations are small and should naively be amenable to a perturbative treatment. In order to correctly encode the effect of short distance nonlinearities, in the EFTofLSS additional terms are present in the equations of motion. The form of these terms is highly constrained by symmetries, mainly the equivalence principle that controls the interactions with gravity~\cite{Baumann:2010tm,Carrasco:2012cv,Porto:2013qua}, and by the nature of the interactions between different species, such as baryons and dark matter~\cite{Lewandowski:2014rca}. At a given order in perturbation theory, only a finite number of terms is necessary, each one with its own numerical prefactor, that we can call ``coupling constant"~\cite{Carrasco:2012cv,Pajer:2013jj,Carrasco:2013sva}. Therefore, since the functional form of each term is fixed, and we are free to adjust a finite number of coupling constants,  the EFTofLSS {\it is} a predictive theory. In fact, it falls in the same class of predictivity of general relativity, the Chiral Lagrangian, or dielectric materials. All of these theories are valid in a specific energy range and make predictions, at a given level of accuracy, that depend on some unknown coupling constant that needs to be measured in some specific observations, after which all the remaining observations are predicted. In a modern language, all of these theories are called Effective Field Theories (EFTs).  

To make the analogy even more concrete, let us focus on the case of dielectric materials. When we consider propagation of electromagnetic waves in a dielectric material, we do not solve the vacuum Maxwell equations with an Avogadro number of atoms; instead, we solve the dielectric Maxwell equations with no sources. The dielectric equations are valid in the range of scales from the size of the dielectric material to the atomic scale. Therefore, waves propagating in the medium must have a wavelength that falls into this regime. Similarly, the EFTofLSS describes fluctuations in LSS quantities with wavelengths that are longer than the nonlinear scale.

In the dielectric material, at a given order in perturbation theory, the effect of the  detailed structure of the materials can be encapsulated in a few parameters: conductivity, polarizability, etc. Similarly, in the EFTofLSS, at a given order in perturbation theory we have to specify a few parameters, such as the speed of sound of the species (or more generally the coefficients of the various terms in the effective stress tensor)~\cite{Carrasco:2012cv,Carrasco:2013sva,Lewandowski:2014rca}, the bias parameters~\cite{Senatore:2014eva}, and the short-distance velocity contribution for redshift distortions~\cite{Senatore:2014vja}. Back to the case of dielectric materials, the actual numerical value of the conductivity can be measured directly in an experiment, or can be measured {\em a priori} by using a computer simulation.  Similarly in the EFTofLSS, the various parameters that are necessary to match observations can be measured directly in observations (that is without any need of simulations), or, even better, they can be anticipated by measurements in simulations. Finally, as the theory of dielectric materials is predictive and can be verified (and it has been verified) experimentally, similarly, the EFTofLSS is predictive and can be verified by observations.

So far, the EFTofLSS has been used to predict  the equal-time dark-matter  power spectrum at one-~\cite{Carrasco:2012cv} and at two-loop orders~\cite{Carrasco:2013sva,Senatore:2014via}, the equal-time momentum power spectrum at one loop~\cite{Senatore:2014via}, the equal-time dark matter bispectrum at one loop~\cite{Senatore:2014via}, the slope of the equal-time velocity vorticity power spectrum at leading order~\cite{Carrasco:2013sva,Mercolli:2013bsa}, the equal-time baryon power spectrum at one loop~\cite{Lewandowski:2014rca}, and the effect of baryons relative to dark matter effectively at two loops~\cite{Lewandowski:2014rca}. All of these quantities have been computed at redshift $z=0$. Only one parameter, the so-called effective speed of sound of dark matter, has been used for the predictions involving just dark matter, and an additional one, the so-called effective speed of sound of baryons, has been introduced when dealing with baryons. These parameters have been chosen to match numerical observations of power spectra and bispectra, and the EFTofLSS has then been able to match the numerical data to percent accuracy at $k\simeq 0.3\hinvMpc$ for calculations done at one-loop order, and at $k\simeq 0.6\hinvMpc$ for two-loop calculations. In the case of the dark matter speed of sound, this parameter has been also measured directly using the dark matter particles as degrees of freedom (this is the analogous procedure as predicting the conductivity of a material by solving the quantum mechanical equations of atoms), and the results agree with the number obtained by directly matching the long wavelength observables such as the power spectrum, within error bars~\cite{Carrasco:2012cv}. 

These results form a growing body of evidence for the correctness and viability of the EFT approach to structure formation. They represent a major improvement for the number of modes that are amenable to analytic control. Former analytic techniques, when implemented without the addition of extra ad-hoc (that is,  physically unjustified) fitting parameters, fail at about $k\simeq 0.1\hinvMpc$. Since the number of modes scales as the cube of the maximum wavenumber, these results suggest that there is a factor of 200 more modes amenable to analytic techniques than previously believed. This could have drastic  consequences for the  amount of physical information that LSS surveys can deliver.  

The above statement about the number of modes is made by extrapolating to higher redshifts the results obtained at redshift zero, a procedure that is, strictly speaking, unjustified. The purpose of this paper is to explore the predictions of the EFTofLSS at all redshifts. Since in the universe time-translations are spontaneously broken, the coefficients of the EFTofLSS are time-dependent. In fact, since the short modes evolve on the same time scale as the long modes, the EFTofLSS is actually local in space, but non-local in time~\cite{Carrasco:2013sva,Carroll:2013oxa}. Therefore, the coefficients of the counterterms of the EFTofLSS are integrals over time of some unknown function. Thanks to the fact that the linear evolution of the modes is $k$-independent, the ignorance of this function of time reduces to having a different time-dependent coefficient for each order in perturbation theory at which the counterterm is evaluated~\cite{Carrasco:2013sva}. In the local in time approximation, these different coefficients become equal, and it was shown in~\cite{Carrasco:2013sva,Senatore:2014eva} that this approximation appeared to be preferred by the data. By studying the prediction of the theory as a function of redshift, we will explore this time dependence as well.

Our main avenue for this exploration will be the equal-time matter power spectrum. After a brief review of the formalism behind the EFTofLSS, we will describe a method for estimating the contributions of various loop orders to the power spectrum prediction, properly accounting for the strength of mode couplings at different scales and for important combinations of loop diagrams that occur at different orders. We will then embark on a comparison of the EFTofLSS prediction to power spectra measured from $N$-body simulations at different redshifts, finding that the prediction's range of validity is consistent with our estimates at all times. Motivated by the picture of a scaling universe with a running slope, we present a two-parameter fitting function for the time-dependence of the effective speed of sound, and demonstrate that it can be used to reproduce the results obtained by fitting the speed of sound separately at each redshift.

With these results in hand, it is then possible to perform an analytical calculation of the lensing potential power spectrum $C_\ell^\psi$, and by extension many other observable quantities related to gravitational lensing. We present computations corresponding both to lensing of CMB photons and photons emitted by galaxies at lower redshifts, and quantify the accuracy of these computations: for CMB lensing, the two-loop EFT prediction for $C_\ell^\psi$ has less than 5\% theoretical uncertainty for $\ell\lesssim 1000$, while for photons emitted from sources at $z=2$ or $z=1$, similar accuracy is achieved for $\ell\lesssim 600$ and 350 respectively. We also estimate the effects of baryonic physics on these predictions, finding them to be at percent level in the above multipole ranges.

%--------------------------------------------------------------------------------------
% SECTION: REVIEW OF EFTOFLSS
%--------------------------------------------------------------------------------------
\section{Review of the EFTofLSS for Dark Matter}
\label{sec:review}

In the EFTofLSS, the dynamics of collisionless dark matter on large scales are described by the following Fourier-space equations of motion for the overdensity ($\delta \equiv \delta\rho/\rho$) and velocity divergence ($\theta\equiv\d_i v^i$) fields~\cite{Carrasco:2013mua}%
%%%%% FOOTNOTE %%%%%
~\footnote{Vorticity is generated in the EFTofLSS~\cite{Baumann:2010tm}, but it contributes only at high order~\cite{Carrasco:2013mua,Mercolli:2013bsa}, and can be neglected at the perturbative level at which we work in this paper.}%
%%%% END FOOTNOTE %%%%
:
\begin{align}
\label{eq:contapp}
&a\H \, \d_a \delta(\vk,a)+\theta(\vk,a)=
	 - \! \int_{\vq}\alpha(\vkp,\vk-\vkp)\delta(\vk-\vkp,a)\theta(\vkp,a)\ , \\
& a\H \, \d_a \theta(\vk,a)+\H \theta(\vk,a)+\frac{3}{2} \H^2(a)  \Omm(a)  \delta(\vk,a)
	= - \! \int_{\vq} \beta(\vkp,\vkkp)\theta(\vk-\vkp,a)\theta(\vkp,a) 
	- i k_i \gammai{}^i(\vk)\ ,
\label{eq:eulerwithdtau}
\end{align}
where
\beq
\Omm(a)= \Ommnow \frac{\H_0^2}{a\H^2(a)}\ ,
\qquad
\alpha(\vk_1,\vk_2)=\frac{\lp \vk_1+\vk_2\rp \cdot \vk_1}{k_1^2}\ ,\qquad
\beta(\vk_1,\vk_2)=\frac{\lp\vk_1+\vk_2\rp^2 \vk_1\cdot\vk_2}{2 k_1^2 k_2^2}\ ,
\eeq
and we have used the shorthand notation $\int_{\vq} \equiv \int \frac{d^3 \vq}{(2\pi)^3}$.
The operator $\gammai{}^i$ incorporates nonlinear, short-distance physics whose effects on the long-distance modes we would like to include in the theory. It is defined by $\gammai{}^i \equiv \rho^{-1} \d_j \tau^{ij}$, where $\tau^{ij}$ is the stress tensor, but from now on we will use the term ``stress tensor" to refer directly to the composite operator $\gammai{}^i$. We write this quantity as an expansion in derivatives and fields, consistent with the equivalence principle, which is the relevant symmetry at the scales we wish to describe.

One may worry that $\gammai{}^i$, which is generated by short modes of the fields that have been ``integrated out" of the theory %
%%%%% FOOTNOTE %%%%%
\footnote{In the EFTofLSS, modes are ``integrated out" by writing the theory in terms of fields that are smoothed over some scale, taking the expectation value over the short modes, and then taking the smoothing scale to zero within the long-wavelength theory---see~\cite{Carrasco:2012cv} for a precise description of the smoothing procedure and the construction of the EFTofLSS.}%
%%%% END FOOTNOTE %%%%
, could exhibit some degree of non-locality in time, since the (nonlinear) short modes will evolve on roughly the same timescales as the (perturbative) long modes~\cite{Carrasco:2013mua,Carroll:2013oxa}. However, it was recently argued in~\cite{Senatore:2014eva} that this non-locality might only affect the power spectrum through a correction of order $(t_\text{short}/t_\text{long}) \times P_\text{2-loop} \sim P_\text{3-loop}$, where in fact the short modes that contribute most strongly are expected to have $t_\text{short}/t_\text{long}\sim\mathcal{O}(1/10)$. While this argument is approximate and needs to be checked further, it agrees with the fact that, in~\cite{Carrasco:2013mua}, it was found that nonlinear power spectrum data at $z=0$ are better described by predictions derived from the time-local limit of $\gammai{}^i$ than by a simple parametrization of non-locality. Therefore, after reviewing how non-locality was handled in~\cite{Carrasco:2013mua}, the body of this paper will focus on the time-local case. This will be further justified by App.~\ref{app:nonlocal}, which shows that non-locality in time does not significantly alter the EFTofLSS predictions for the power spectrum at any redshift we consider.

In~\cite{Carrasco:2013mua} it was found that for a two-loop calculation of the power spectrum to describe nonlinear data with percent-level accuracy, it is sufficient to consider the lowest-order term in the expansion of $\gammai{}^i$, neglecting terms of quadratic or higher order in fields, and also neglecting higher derivative terms:
\beq
\label{eq:stresstensor}
\gammai{}^i = \int d\tau' \kappa_1(\tau,\tau') \d^i \d^2 \phi(\tau',\vx_{\rm fl}[\tau,\tau';\vx])
	= \int \frac{da'}{a' \H(a')} K(a,a') \d^i \delta(a',\vx_{\rm fl}[a,a';\vx])\ ,
\eeq
where we have used Poisson's equation in the second equality to trade $\d^2\phi$ for $\delta$, and $\kappa_1$ and $K$ are unknown kernels with support of order one Hubble time. Using $\vx_{\rm fl}$ as the path of a fluid element, defined by
\beq
\vx_{\rm fl}[a,a';\vx] \equiv \vx - \int_{a'}^{a} \frac{da''}{a''\H(a'')} \vv(a'',\vx_{\rm fl}[a,a'';\vx])\ ,
\eeq
ensures that the equations of motion are diffeomorphism-invariant (i.e.\ invariant under generalized Galilean transformations).
With this form of~$\gammai{}^i$,  the Euler equation~(\ref{eq:eulerwithdtau}) becomes 
\begin{align} \nn
&a\H \, \d_a \theta(\vk,a)+\H \theta(\vk,a)+\frac{3}{2} \H^2(a)  \Omm(a)  \delta(\vk,a)
= - \! \int_{\vq} \beta(\vkp,\vkkp)\theta(\vk-\vkp,a)\theta(\vkp,a) \\
\label{eq:eulerapp}
&\qquad\qquad\qquad\qquad\qquad\qquad\qquad\qquad\qquad\qquad\qquad
+  k^2 \int \! \frac{da'}{a'\H(a')} K(a,a') [ \delta(a',\vx_{\rm fl})]_{\vk}\ ,
\end{align}
where $\lb f(\vx_{\rm fl}) \rb_{\vk} \equiv \int d^3 x \, e^{-i\vk\cdot\vx} f(\vx_{\rm fl})$. 
One convenient choice for the kernel $K(a,a')$ is %
%%%%% FOOTNOTE %%%%%
\footnote{In~\cite{Carrasco:2013mua}, the specific choice $\zeta=2$  was used in Eq.~(\ref{eq:Kansatzold}). This ensured that the tree-level counterterm in the matter power spectrum, $P_\text{tree}^{(c_{\rm s})}$, had the same time-dependence as the one-loop correction $P_\text{1-loop}$. This choice has a negligible effect on predictions made at a single redshift (such as those presented in~\cite{Carrasco:2013mua} at $z=0$), but for the purpose of relating predictions at different times, it is important to allow for other choices of $\zeta$. See~\cite{Angulo:2014tfa} for the details of how to perform calculations in the $\zeta\neq 2$ case.}%
%%%% END FOOTNOTE %%%%
\beq
\label{eq:Kansatzold}
K(a,a') =\frac{(2 \pi )\co}{\knl^2}  \left [\zeta(\zeta+5/2) (p+1) \H(a)^2 f(a)^2 \, a'\H(a')\frac{d D_1(a')}{da'} \frac{D_1(a')^{p-1}}{D_1(a)^{p-\zeta}} \right] \ ,
\eeq
which parametrizes the non-locality in time by a fixed power $p$. Here, $D_1(a)$ is the linear growth factor, normalized to unity at $a=1$:
\beq
D_1(a) \equiv \frac{D_+(a)}{D_+(1)}\ , \qquad 
	D_+(a) \equiv \frac{5}{2} \Omm(a) \H^3(a) \int_0^a \frac{da'}{\H^3(a')}\ .
\eeq

The equations~(\ref{eq:contapp}) and~(\ref{eq:eulerapp}) admit solutions of the form
\begin{align} \nn
\delta(a,\vk) &= \sum_{n=1}^\infty [D_1(a)]^{n} \delta^{(n)}(\vk) 
	+ \sum_{n=1}^\infty [D_1(a)]^{n+\zeta} \tilde{\delta}^{(n)}(\vk)\ , \\
\theta(a,\vk) &= -\H(a) f(a) \sum_{n=1}^\infty [D_1(a)]^{n} \theta^{(n)}(\vk) 
	-\H(a) f(a) \sum_{n=1}^\infty [D_1(a)]^{n+\zeta} \tilde{\theta}^{(n)}(\vk)\ ,
\label{eq:deltathetaansatz}
\end{align}
under the assumption that $\Omm(a)=f(a)^2$, where $f(a)\equiv \d\log D_1 / \d\log a$. The error incurred using these approximate solutions instead of the exact solutions has been shown numerically to be very small in the one-loop power spectrum (e.g.~\cite{Carrasco:2012cv}), so we will make use of these solutions in the remainder of the paper. In an upcoming paper~\cite{formalism}, we will re-examine this approximation in detail.

Plugging the solutions~(\ref{eq:deltathetaansatz}) back into the equations of motion and writing
\beq
\delta^{(n)}(\vk) = \int_{\vq_1} \dots \int_{\vq_n} (2\pi)^3 \delta_{\rm D}(\vk-\vq_1-\cdots-\vq_n)
	F_n^{\rm (s)}(\vq_1,\dots,\vq_n) \delta^{(1)}(\vq_1) \dots \delta^{(1)}(\vq_n)\ ,
\eeq
and analogously writing $\theta^{(n)}(\vk)$, $\tilde{\delta}^{(n)}(\vk)$, and $\tilde{\theta}^{(n)}(\vk)$ in terms of $G_n$, $\tilde{F}_n$, and $\tilde{G}_n$, one obtains recurrence relations for the $F$ and $G$ kernels, which can be found, e.g., in~\cite{Angulo:2014tfa}. From there, any correlation function of $\delta$ and $\theta$ can be written down up to a specified order. The basic building block of these correlation functions is the linear solution $\delta^{(1)}(\vk)$, assumed to have Gaussian statistics (in the absence of primordial non-Gaussianity, which we will assume in this work) and power spectrum
\beq
\left\la \delta^{(1)}(\vk) \delta^{(1)}(\vk') \right\ra = (2\pi)^3 \delta_{\rm D}(\vk+\vk') P_{11}(k)\ ,
\eeq
which can be obtained from a public code such as CAMB~\cite{Lewis:1999bs}. The presence of $\gammai{}^i$ in the equations of motion will lead to the appearance of counterterms in expressions for correlation functions. From a very practical standpoint, these counterterms are the main distinction between the EFTofLSS and other perturbative approaches to structure formation.

%--------------------------------------------------------------------------------------
% SECTION: MATTER POWER SPECTRUM
%--------------------------------------------------------------------------------------
\section{The Matter Power Spectrum at Different Redshifts}
\label{sec:mps}

In this section, we compare the one- and two-loop EFTofLSS prediction for the matter power spectrum to data from $N$-body simulations. At a single redshift, the one-loop prediction has one free parameter, and so does the two-loop prediction, provided that the two-loop terms are renormalized via the procedure we will describe below.  At multiple redshifts, it might seem that this parameter needs to be fit separately at each redshift, but in fact there is a convenient way to parametrize the time-dependence that involves only one additional parameter. Sec.~\ref{sec:mpsformulas} contains the formulas used and assumptions made in our predictions. Sec.~\ref{sec:mpsestimates} shows estimates for how far in $k$ the one- and two-loop predictions should be expected to reach. Sec.~\ref{sec:cs2fitmethod} describes a procedure for renormalizing the two-loop contribution, and Sec.~\ref{sec:fits} contains the actual comparisons to data.

For the nonlinear power spectrum, we use version 2 (``FrankenEmu") of the Coyote interpolator~\cite{Heitmann:2008eq} with WMAP7 cosmological parameters %
%%%%% FOOTNOTE %%%%%
\footnote{The most accurate settings of the Coyote interpolator do not accept $h$ as an independent parameter, but rather fix its value based on the CMB distance to last scattering. Therefore, the nonlinear power spectra we compare to in this section actually have $h=0.7013$ instead of $h=0.704$, but this difference ($\sim$0.5\% of the value of $h$) is insignificant for the purpose of comparing to EFT predictions.}%
%%%% END FOOTNOTE %%%%
: $h=0.704$, $\Omega_{\rm  m}=0.272$, $\Omega_{\rm b}=0.0455$, $\Omega_\Lambda=1-\Omega_{\rm m}$, $n_{\rm s}=0.967$, and $\sigma_8=0.81$. As a conservative estimate of the error in the output of the interpolator, we take the nonlinear spectrum to have uniform 2\% errorbars for $z\leq 1$ and $k\leq 1\invMpc$, 3\% errorbars for $z>1$ and $k\leq 1\invMpc$, and 6\% errorbars for $z>1$ and $k>1\invMpc$%
%%%%% FOOTNOTE %%%%%
~\footnote{Our errorbars are determined in the following way. From our understanding, the output of the Coyote emulator has uncertainties that are quoted as $1\%$ for $k<1\invMpc$ and $z\leq 1$, and $5\%$ in all the other regimes of interest~\cite{Heitmann:2008eq}. From  Figs.~8--9 of the fourth paper in Refs.~\cite{Heitmann:2008eq}, it seems to us that the region $k<1\hinvMpc$, $z>1$ has a smaller error bar than $5\%$, and we take it to be about $2\%$. In this paper we are interested in understanding when the EFTofLSS fails in the UV to about $1\%$. So, to the error bars of Coyote, we add a $1\%$ error to represent when the theory is within $1\%$ of the numerical data. This error should also account for potential unknown systematics in the simulations and in the comparison between simulations and the EFTofLSS. At last, we point out that when the EFTofLSS begins to fail in the UV, it does so in a very steep way, so that the UV reach for the EFTofLSS that we establish later is expected to be quite insensitive to the precise choice of error bars. Similarly, the statement that the UV reach of the EFTofLSS is compatible with one from the estimated theoretical error that we derive later is quite unaffected by this.}%
%%%% END FOOTNOTE %%%%
. App.~\ref{app:wmap3} contains a consistency check of the EFT prediction against a different cosmology (WMAP3). We find results that are comparable to what we present in Sec.~\ref{sec:fits}. 

Our primary metric for quantifying the performance of various perturbation theory predictions will be the wavenumber at which the prediction begins to deviate from the measured nonlinear power spectrum by a certain percentage. In particular, we will choose this percentage to be our estimation of the uncertainty on the nonlinear spectrum we use, and define $\kfail[L]$ to be the wavenumber at which the $L$-loop EFT prediction exceeds this uncertainty~\footnote{We wish to emphasize that our chosen failure criterion is just one of many possibilities. In this work, we are comparing to numerical measurements with a certain uncertainty, and so we find it to be natural to use this uncertainty as a benchmark for our comparisons with theory. However, for more detailed applications of the EFTofLSS to observational data, or to other simulations with different uncertainties, there can certainly be other criteria which could be more suitable. Indeed, if predictions of very high precision are needed at some point in the future, the associated failure wavenumber of the EFTofLSS will be smaller than what we quote in this work, since the effects of loops and counterterms will be larger relative to the required precision.}. For the convenience of the reader, the uncertainty on the data is summarized in Table~\ref{tab:pnlerror}. 

\begin{table}[ht]
\begin{center}
\begin{tabular}{c|c||c}
	   $z$ & $k$  & failure threshold  \\ \hline
	$\leq 1$ & $\leq 1\invMpc$ & 2\%	  \\
	$>1$ & $\leq 1\invMpc$ & 3\%  \\
	$>1$ & $> 1\invMpc$ & 6\%	 
\end{tabular}
\caption{\label{tab:pnlerror} \small\it A summary of the uncertainty on the numerical data, and therefore of the criteria we use to define the ``failure" of a given prediction in this work: when the deviation of the $L$-loop EFT power spectrum prediction from the nonlinear measurements reaches the threshold in the rightmost column, we consider the corresponding wavenumber, $\kfail[L]$, to be where the prediction fails.}
\end{center}
\end{table}

%--------------------------------------------------------------------------------------
\subsection{Formulas}
\label{sec:mpsformulas}

The one-loop EFT formula for the power spectrum is given by
\beq
P_\text{EFT-1-loop}(k,z) = [D_1(z)]^2 P_{11}(k) + [D_1(z)]^4 \poneloop(k) + \ptreecs(k,z)\ ,
\eeq
where $\ptreecs(k,z) = -2(2\pi) \co(z) [D_1(z)]^2 (k^2/\knl^2) P_{11}(k)$, while the two-loop formula is
\begin{align} \nn
&P_\text{EFT-2-loop}(k,z) = P_\text{EFT-1-loop}(k,z) + [D_1(z)]^6 \ptwoloop(k)
	-2(2\pi) \ct(z) \frac{k^2}{\knl^2} P_{11}(k) \\
&\quad + (2\pi) \co(z) [D_1(z)]^4 \poneloopcs(k) 
	+ (2\pi)^2 \lp 1 + \frac{\zeta+\frac{5}{2}}{2(\zeta+\frac{5}{4})} \rp [\co(z)]^2 [D_1(z)]^2 \frac{k^4}{\knl^4} P_{11}(k)\ .
\label{eq:peft2loop}
\end{align}
Expressions for $\poneloop(k)$ and $\ptwoloop(k)$ are given in~\cite{Carrasco:2013sva}, and also in~\cite{Carrasco:2013mua}. Expressions for $\poneloopcs(k)$ are given in~\cite{Carrasco:2013mua} for the $\zeta=2$ case (see Eq.~(\ref{eq:Kansatzold})); the extension to $\zeta\neq 2$ is straightforward using the recurrence relations from~\cite{Angulo:2014tfa}. The coefficient of the $(k/\knl)^4 P_{11}$ term is different from that given in~\cite{Carrasco:2013mua,Senatore:2014via}, owing to the inclusion of a tree-level counterterm that has previously been omitted from two-loop computations; the derivation of this term (along with the other tree-level counterterms shown above) can be found in App.~\ref{app:treelevelcts}.

The time-dependence of $\poneloopcs$ comes from the fact that the diagrams in $\poneloopcs$ are built in one of two ways: by multiplying $\ptreecs/P_{11}$, which scales like $\co(z)$, by a diagram contained in $\poneloop$, which scales like $[D_1(z)]^4$; or by inserting $\ptreecs/P_{11}$ into a one-loop diagram. This second type of diagram may have time-dependence that could differ slightly from $\co(z)[D_1(z)]^4$, determined by the detailed form of the integrals over Green's functions in time. However, just as the Green's function integrals in $\poneloop$ collapse to $\sim [D_1(z)]^4$ to very good accuracy in {\lcdm}, we expect that the integrals in $\poneloopcs$ will also collapse to $\co(z) [D_1(z)]^4$ at a similar level of accuracy, and therefore we will use this expression in our calculations.
Furthermore, the $\poneloopcs$ and $(k/\knl)^4 P_{11}$ terms depend slightly on the value of~$\zeta$ we choose, so a particular value must be chosen before the prediction is computed; we use $\zeta=3$, which is very roughly motivated by scaling universe arguments%
%%%%% FOOTNOTE %%%%%
~\footnote{In principle, the time-dependence of $\co$ and therefore the value of $\zeta$ can be measured from fitting to simulations. However, as just mentioned, it is useful for the calculation to make an initial assumption about the time dependence. The fitting to the data that we will explain and perform later will be such that the residual dependence on this initial choice is very small.}%
%%%% END FOOTNOTE %%%%
. 
Finally, we use the form of $\poneloopcs$ corresponding to the time-local case (the $p\to\infty$ limit of Eq.~(\ref{eq:Kansatzold})). In App.~\ref{app:nonlocal}, we show that relaxing the assumption of locality in time does not significantly affect the comparison to data.

For numerical computations, we use CAMB~\cite{Lewis:1999bs} to generate a linear power spectrum, and a modification of the Copter library~\cite{Carlson:2009it} that makes use of the IR-safe integrands from~\cite{Carrasco:2013sva} and Monte Carlo integration routines from the CUBA library~\cite{CUBA} to compute the loop integrals.

%--------------------------------------------------------------------------------------
\subsection{Estimates and expectations}
\label{sec:mpsestimates}

\begin{figure}[t]
\begin{center}
\includegraphics[width=0.3\textwidth]{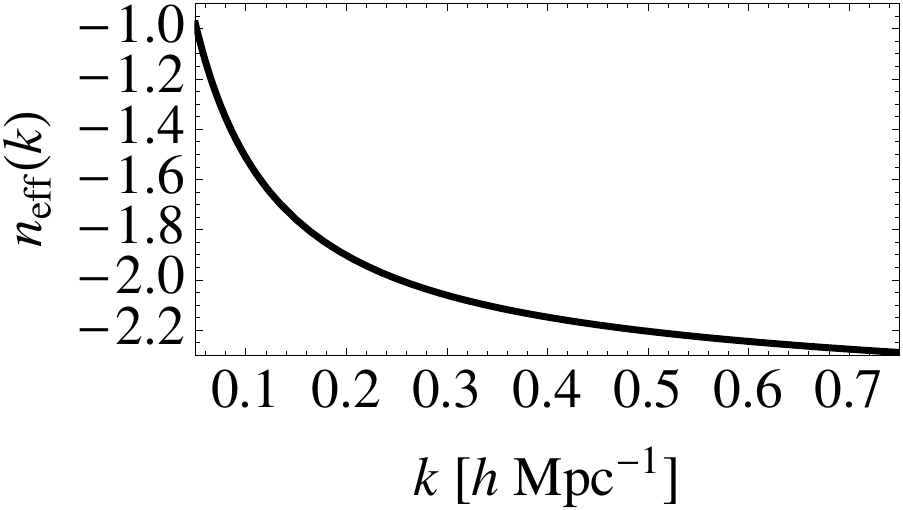}
\includegraphics[width=0.32\textwidth]{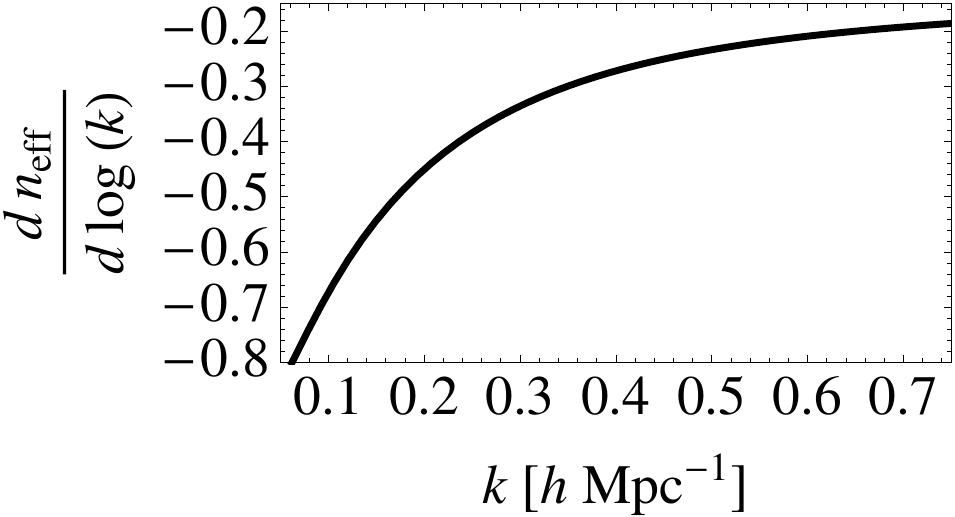}
\includegraphics[width=0.32\textwidth]{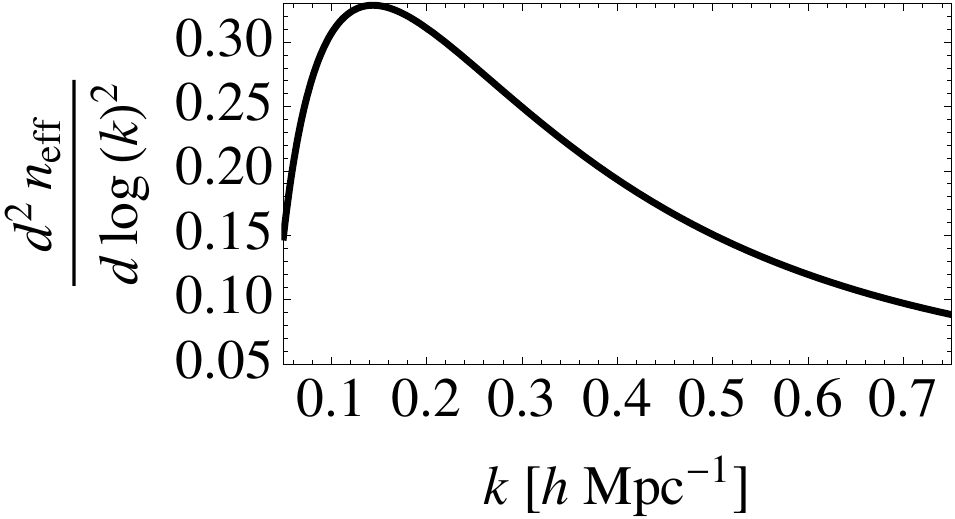}
\caption{\label{fig:neffk} \small\it Left: effective slope $n_{\rm eff}(k) = d\log P_{\rm nw}(k)/d\log k$, where $P_{\rm nw}$ is the linear matter power spectrum without BAO wiggles. $P_{\rm nw}(k)$ is given by Eq.~(\ref{eq:ehps}). Center: running of this slope, $dn_{\rm eff}/d\log k$. Right: running of the running, $d^2 n_{\rm eff}/d\log k^2$. The large running of the slope implies that a simple analogy with the case of a pure scaling universe will be insufficient to approximate the behavior of various terms in the power spectrum prediction.}
\end{center}
\end{figure}

An essential feature of the EFTofLSS is that it is possible to estimate at what scale a given prediction should fail, based on the size of the next-order terms that are neglected. In~\cite{Carrasco:2013mua,Angulo:2014tfa}, these estimates were performed by treating the predictions as taking place in a scaling universe ($P_{11}(k) \propto k^n$) with slope $n$ determined from the slope of the {\lcdm} linear power spectrum around the scales where the two-loop $z=0$ prediction should be valid ($k \lesssim 0.6\invMpc$). However, if one is interested in a wide range of scales and higher order contributions, where errors are given the possibility of piling up, the scaling-universe analogy can no longer be simply applied.  In fact, in Figure~\ref{fig:neffk}, we plot the effective slope $n_{\rm eff}(k)$ of the linear power spectrum without BAO wiggles, as found from the fitting formula in~\cite{Eisenstein:1997ik}%
%%%%% FOOTNOTE %%%%%
~\footnote{For reference, this formula is given by 
\beq
\label{eq:ehps}
P_{\rm nw}(k) \propto k^{n_{\rm s}} \!
	\lb \frac{\log(2e+1.8q)}{\log(2e+1.8q)+q^2 (14.2+731/(1+62.5q))} \rb^2\ ,
	\quad q \equiv  \lp \frac{k}{\invMpc} \rp \lp \frac{T_{\rm CMB}}{2.7\, {\rm K}} \rp^2
	\Gamma_{\rm eff}^{-1} \ ,
\eeq
where
\beq
\Gamma_{\rm eff} \equiv \Omm h \lp \alpha_\Gamma + \frac{1-\alpha_\Gamma}{1+(0.43ks)^4} \rp\ ,
\eeq
which itself depends on
\beq
s \equiv \frac{44.5 \log \lp 9.83/[\Omm h^2] \rp}{\lb 1+10(\Omega_{\rm b}h^2)^{3/4} \rb^{1/2}} \text{ Mpc}\ ,
	\quad \alpha_\Gamma \equiv 1 - 0.328\log(331\Omm h^2) \frac{\Omega_{\rm b}}{\Omm}
	+ 0.38\log(22.3\Omm h^2) \lp \frac{\Omega_{\rm b}}{\Omm} \rp^2\ .
\eeq
(The prefactors of $P_{\rm nw}$ are not needed to calculate $n_{\rm eff}(k) = d\log P_{\rm nw}(k)/d\log k$.)}%
%%%% END FOOTNOTE %%%%
, as well as the running of this slope, $dn_{\rm eff}/d\log k$, and the running of the running, $d^2n_{\rm eff}/d\log k^2$. While $d^2n_{\rm eff}/d\log k^2$ is quite small, the running is not small, at least for $k\lesssim 0.6\hinvMpc$, which means that the universe is not well approximated by a scaling one. Furthermore, and maybe more importantly, there can be important cancellations among diagrams that are not captured by the simple scaling argument, as we will now describe. This pushes us to develop a method to estimate the various contributions in a way that mirrors more closely the perturbative structure of the theory.

\subsubsection{One loop}

\begin{figure}[t]
\begin{center}
\includegraphics[width=0.6\textwidth]{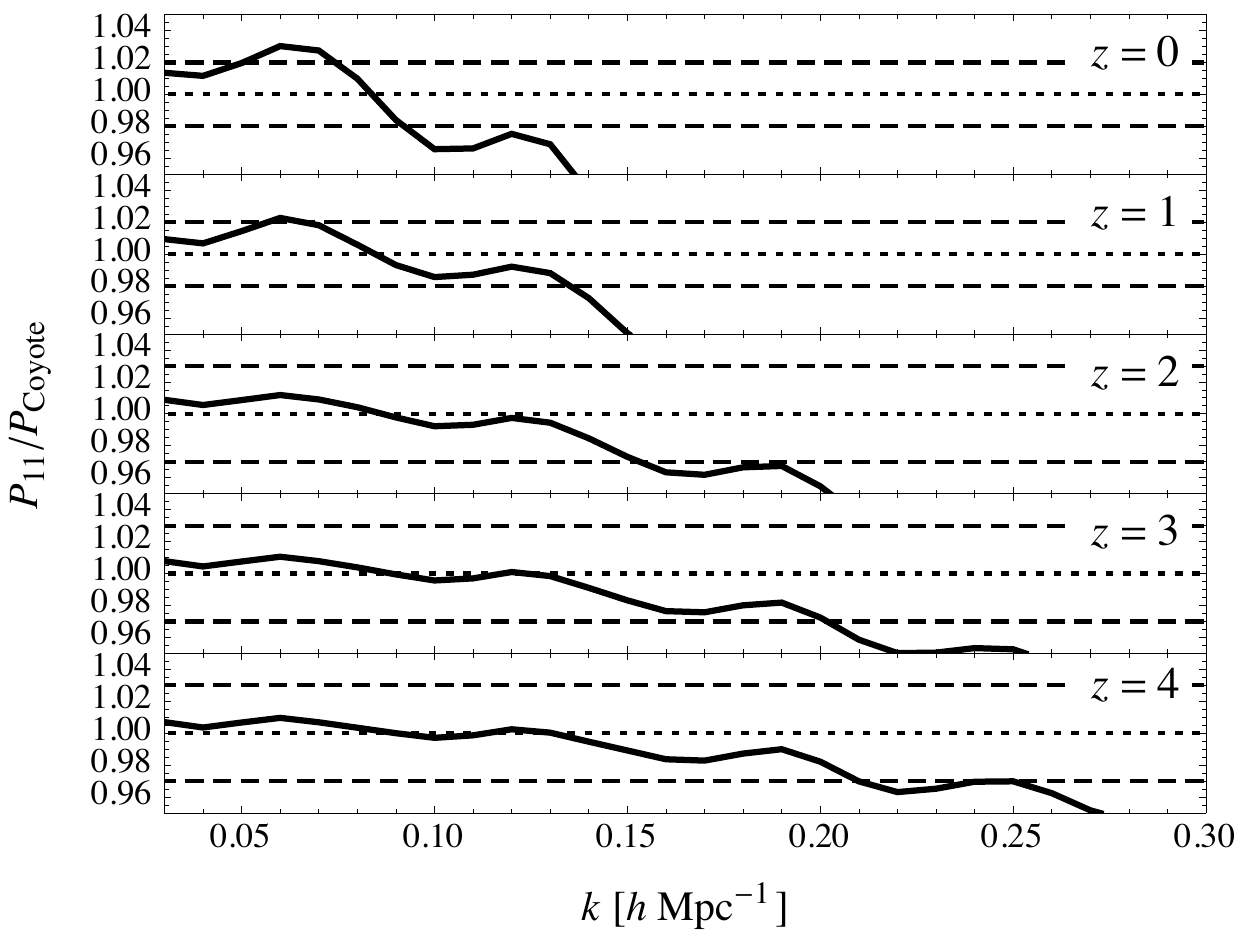}
\hspace{1cm}
\begin{tabular}[b]{c|c}
\multicolumn{2}{c}{\textbf{Linear theory}} \\ \hline
	   $z$ & estimated failure   \\
	    &  $[\hinvMpc]$   \\ \hline\hline
	0 & 0.04 -- 0.07  \\
	1 & 0.06 -- 0.09   \\
	2 & 0.08 -- 0.17   \\
	3 & 0.11 -- 0.23   \\
	4 & \, 0.13 -- 0.31  \vspace{3cm}
\end{tabular} 
\caption{\label{fig:linratios} \small\it Left: Linear theory prediction for the power spectrum, normalized to the nonlinear spectrum from the Coyote emulator at different redshifts. The dashed lines show the estimated error on the nonlinear data, while the dotted line at the value $P_{11}/P_\text{Coyote}=1$ is provided as a visual aid. Right: Estimates for when linear theory should fail, based on when the ratio $\poneloop^\text{(total)}/P_{11}$ exceeds the error on the nonlinear data. We quote a range of wavenumbers based on dividing and multiplying the estimate from Eq.~(\ref{eq:p1looptotest}) by a factor of~2. By comparing these estimates with the plots on the left, we find that linear theory fails roughly where it should.}
\end{center}
\end{figure}

Let us begin by formulating such an estimate for $\poneloop$. To do so, we recall that $\poneloop$ primarily captures the effects of density fluctuations from modes with $q<k$ and displacements from modes with $q>k$ on a density mode of wavenumber $k$~\cite{Porto:2013qua,Senatore:2014via}. These effects are respectively encapsulated in the parameters $\epsdl$ and $\epssg$, defined by%
%%%%% FOOTNOTE %%%%%
~\footnote{More in detail, the parameters upon which the Eulerian-space EFT Taylor expands are~\cite{Porto:2013qua,Senatore:2014via}
\bea\label{eq:epsparameters}
&&\epsilon_{\delta <} \sim \int^k_0 {d^3k' \over (2 \pi)^3} P_{11}(k')\ , 
\qquad \epsilon_{s >}  \sim k^2  \int_k^\infty {d^3k' \over (2 \pi)^3}  {P_{11}(k') \over k'^2}\ , 
\qquad \epsilon_{s_<}  \sim k^2  \int_0^k {d^3k' \over (2 \pi)^3}  {P_{11}(k') \over k'^2}\ .
\eea
They represent the effect of long-wavelength overdensities, the effect of short-wavelength displacements, and the effect of long-wavelength displacements. Here long and short are relative to a specific wavenumber $k$ that we are interested in. In the case of a system with two species, such as dark matter plus baryons~\cite{Lewandowski:2014rca}, we also have
\bea\label{eq:epsrelparameters}
\epsilon_{s < }^{\rm rel}(k) \sim k^2  \int_0^k \frac{ d^3 k' }{(2\pi)^3} \frac{\tilde P_{11}(k')}{k'{}^2} \ ,
\qquad \epsilon_{s > }^{\rm rel}(k) \sim  k^2 \int_k^\infty \frac{ d^3 k' }{(2\pi)^3} \frac{\tilde P_{11}(k')}{k'{}^2} \ ,
\eea
where $\tilde{P}_{11}$ is the power spectrum of $\d(\delta_{\rm c}-\delta_{\rm b})/\d\log a$, and $\delta_{\rm c}$, $\delta_{\rm b}$ are the overdensities of the dark matter and baryons respectively.
After IR-resummation~\cite{Senatore:2014via,Lewandowski:2014rca}, the parameters $\epsilon_{s_<}$ and $\epsilon_{s < }^{\rm rel}$ are not expanded upon.
}%
%%%% END FOOTNOTE %%%%
\begin{align}
\label{eq:epsdldef}
\epsdl(k) P_{11}(k) &\equiv \int_0^k dq\, p_\text{1-loop}(\vk,\vq)|_{q\ll k} 
	\approx P_{11}(k) \frac{2.3}{(2\pi)^2}  \int_0^k dq \, q^2 P_{11}(q)   \ , \\
%\end{align}
%\begin{align}
\label{eq:epssgdef}
- \epssg(k) P_{11}(k) &\equiv \int_k^\Lambda dq\, p_\text{1-loop}(\vk,\vq)|_{q\gg k}
	\approx  P_{11}(k) \frac{1}{(2\pi)^2} \int_k^\Lambda dq \, q^2 
	\lp -\frac{122k^2}{315q^2} \rp P_{11}(q)\ .
\end{align}
We therefore use
\beq
\label{eq:p1looptotest}
\poneloop^\text{(total)}(k) \approx \lp \epsdl - \epssg \rp P_{11}(k)
\eeq
as our estimate for $\poneloop^\text{(total)}(k) \equiv \poneloop(k) + \ptreecs(k)$. (This estimate assumes not more than an order one cancellation between $\poneloop$ and $\ptreecs$ at all redshifts.) In order to use this estimate without knowledge of the exact value of the $\co$ parameter that we will find by comparing to data, as well as of order one numerical factors resulting from the loop integrals, we multiply and divide the right-hand side of~(\ref{eq:p1looptotest}) by 2 and use the results as a range for the estimate of $\poneloop^\text{(total)}$. As an indicator of where linear theory should fail (i.e.\ $\kfail[0]$, as defined at the beginning of Sec.~\ref{sec:mps}), we find where our estimate for the ratio $\poneloop^\text{(total)}(k)/P_{11}(k)$ equals the uncertainty on the nonlinear spectrum. The result is that we expect the failure to occur somewhere in the range $0.04 \invMpc < k_{\rm fail}^{(0)} < 0.07 \invMpc$ at $z=0$, while the actual failure occurs at $k_{\rm fail}^{(0)}=0.05\invMpc$. Fig.~\ref{fig:linratios} displays the estimated and exact failures at various redshifts from 0 to~4.

For a more precise estimate, instead of multiplying and dividing Eq.~(\ref{eq:p1looptotest}) by 2, we could add the $\ptreecs(k)$ term, using the exact value of $\co$ obtained from one-loop fits to the data. Doing this gives $k_{\rm fail}^{(0)}=0.06\invMpc$ at $z=0$, and  also gives estimates at other redshifts that are similarly close to the actual failure of linear theory. We will follow this kind of strategy when estimating $\ptwoloop^\text{(total)}$ and $\pthreeloop^\text{(total)}$ (the sums of all terms entering at two- and three-loop order, respectively), in order to obtain the most precise picture of when the one- and two-loop EFT predictions should fail.

\subsubsection{Two loops}

\begin{figure}[t]
\begin{center}
\includegraphics[width=0.6\textwidth]{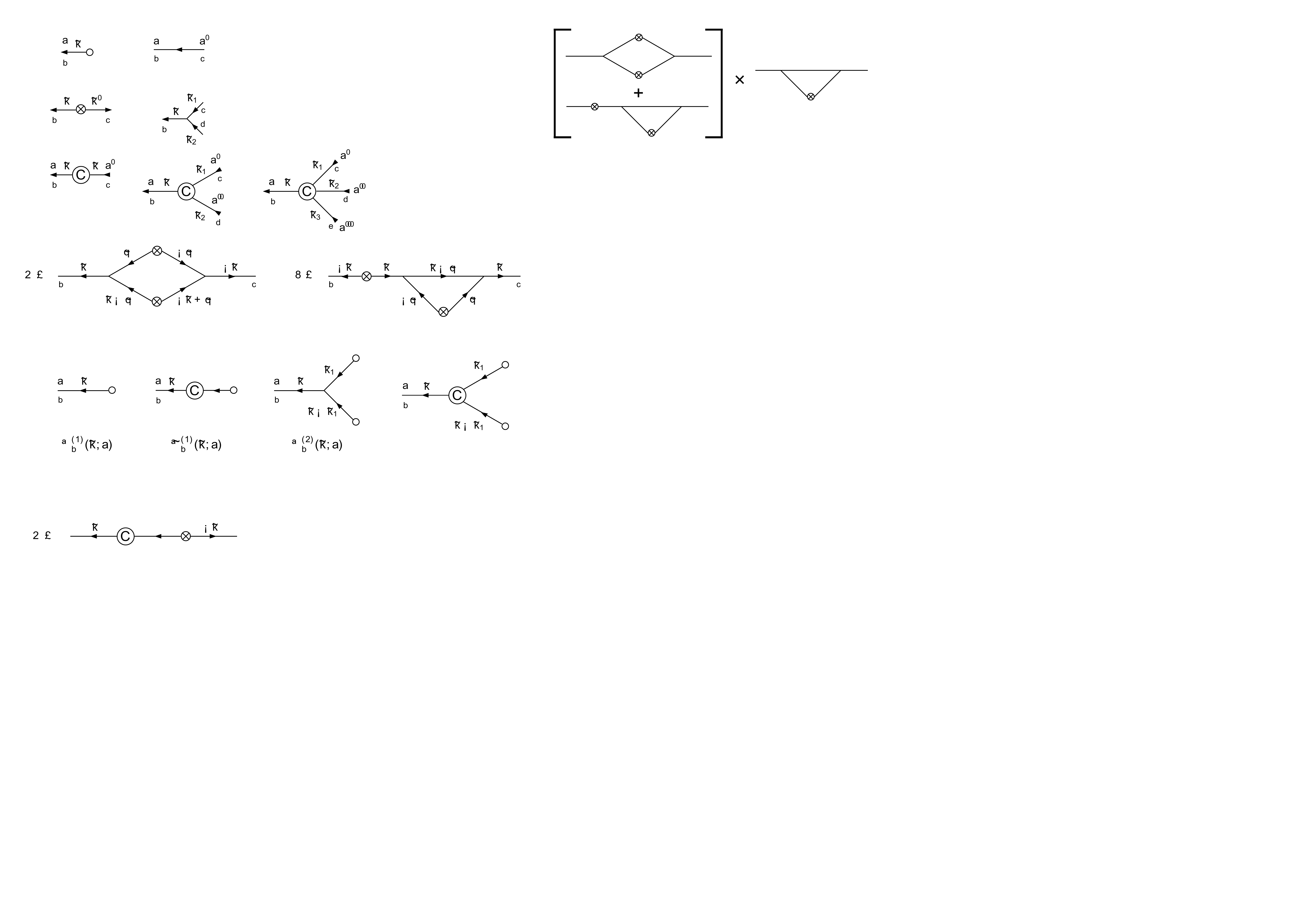}
\caption{\label{fig:p2loop-schematic} \small\it Schematic depiction of the dominant two-loop diagrams in the power spectrum: $\lb P_{22}+P_{13} \rb \times P_{13}^{\text{(no-IR)}}/P_{11}$.}
\end{center}
\end{figure}

To estimate the size of $\ptwoloop^\text{(total)}$, we make use of the fact that $\ptwoloop$ will be dominated by diagrams containing two independent trivial angular integrations, which each contribute a factor of $(2\pi)$. The sum of these diagrams can be approximated, up to combinatorics, as a product of the standard one-loop diagrams, $\poneloop$, and a $P_{13}$ diagram with the $P_{11}(k)$ factor removed (see Fig.~\ref{fig:p2loop-schematic}).%
%%%%% FOOTNOTE %%%%%
\footnote{When the diagrams are written using only cubic vertices instead of symmetrized $n$-point vertices (e.g.~\cite{formalism}), there will be also diagrams with two independent angular integrations but whose loops are nested. However, these diagrams will contain more integrations over Green's functions in time. Recall that in the Einstein de-Sitter case, these Green's functions are roughly the scale factor $a$, so $m$ integrals over Green's functions will give an overall prefactor of $\sim 1/m!$. This factor will suppress diagrams with nested loops relative to those with non-nested loops, so we only consider the latter type of diagram in our estimates.} %
%%%% END FOOTNOTE %%%%
  However, an exact calculation of $P_{13}(k)$ will include a contribution from modes with momentum $q\ll k$, which will cancel with the corresponding contribution from other diagrams (at one-loop order the cancellation is provided by $P_{22}(k)$).  We should therefore include only the part of $P_{13}$, that we call $P_{13}^\text{(no-IR)}$, that is not going to cancel against other diagrams in the final result. This contribution cannot be calculated precisely, but a rough estimate can be obtained by finding where the $P_{13}$ and $P_{22}$ integrals exhibit strong cancellations when each integral in wavenumber $q$ is cut off at some $q_{\rm max}$, and then taking $P_{13}^\text{(IR)}$ as the contribution to $P_{13}$ from $q<q_{\rm max}$. We find that $q_{\rm max}$ is somewhere between $k/5$ and $2k/5$ for $k\lesssim 0.4\invMpc$, and this leads to $P_{13}^\text{(no-IR)}= P_{13}(k)-P_{13}^{\rm (IR)}(k) = \alpha' P_{13}(k)$ with $1/10 \lesssim \alpha' \lesssim 1/3$. Also, the quantity that appears in $L$-loop contributions will not be $\poneloop$ or $\alpha' P_{13}$ on their own, but rather the combinations $\poneloop+\ptreecs$ or $\alpha' P_{13}+\ptreecs$. For convenience, we can use the fact that $P_{13} \sim 3\poneloop$ in the range we care about to replace $\alpha' P_{13}$ by $\alpha \poneloop$ in our estimates, so that $3/10 \lesssim \alpha \lesssim 1$.

In summary, then, we estimate
\beq
\label{eq:p2looptotest}
\ptwoloop^\text{(total)}(k) \approx \beta \lb \frac{\alpha \poneloop(k)+\ptreecs(k)}{P_{11}(k)} \rb 
	\lb \poneloop(k) +\ptreecs(k) \rb\ ,
\eeq
with $\beta$ ranging between $1/2$ and $2$ to account for unknown combinatorics and other order one numerical coefficients.  This expression highlights the possibility that there may be rather large cancellations between loops and counterterms at two-loop (and higher) order, because loops beyond the first contribute factors of $\alpha \poneloop + \ptreecs$ instead of $\poneloop+\ptreecs$ (which was assumed, and found, not to exhibit a cancellation beyond order one).

\begin{figure}[t]
\begin{center}
\includegraphics[width=0.8\textwidth]{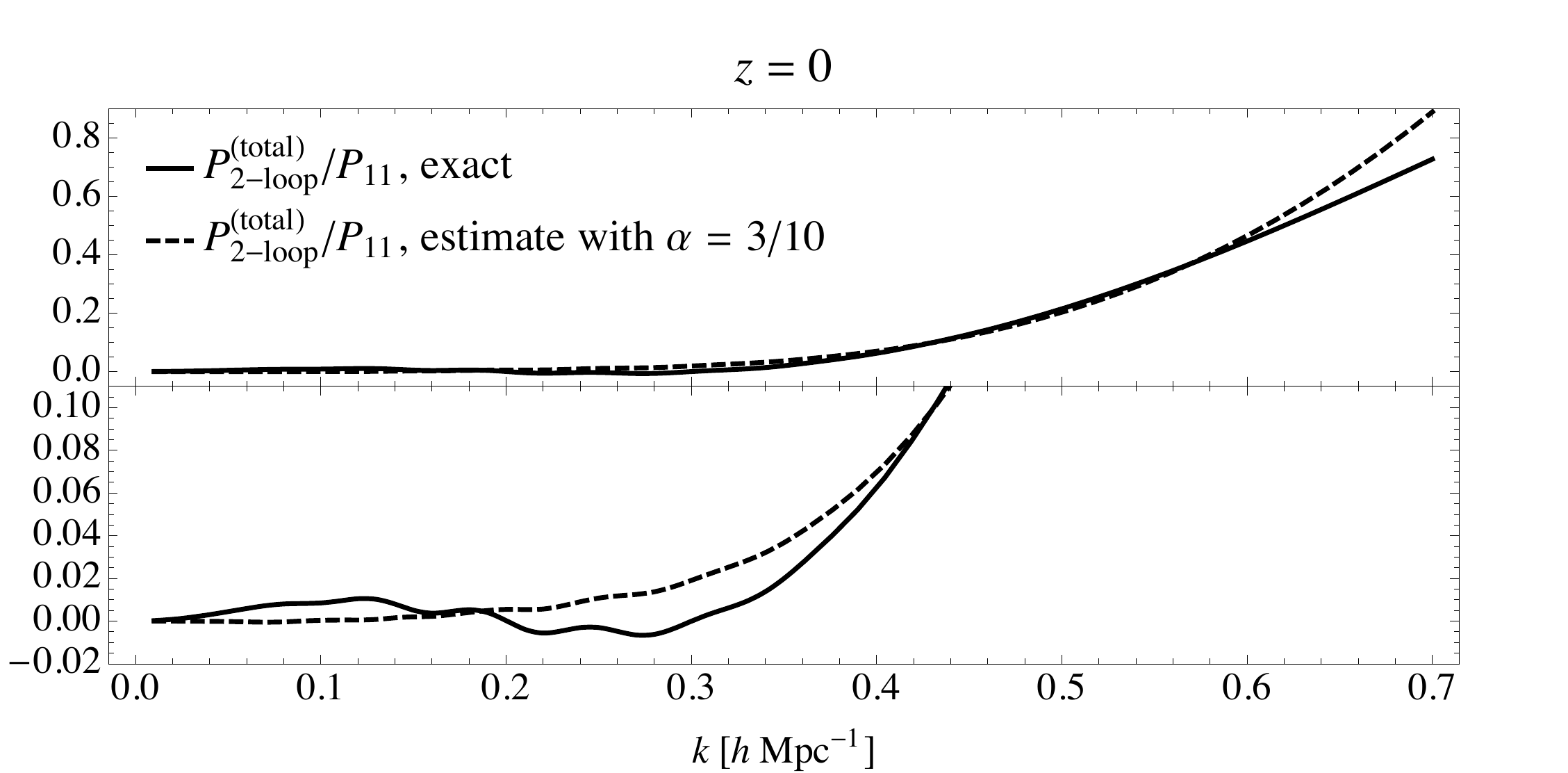}
\caption{\label{fig:p2loopestalphaz0} \small\it  Exact and approximate versions of $\ptwoloop^\text{(total)}/P_{11}$ at $z=0$, the latter plotted using Eq.~(\ref{eq:p2looptotest}) with $\alpha\simeq3/10$ and $\beta=1$. The bottom panel is a zoomed-in version of the top panel. Overall, the shapes of both curves match very well up to $k\sim 0.6\invMpc$. }
\end{center}
\end{figure}

We can explicitly see such a cancellation occurring if we try to match our estimate for $\ptwoloop^\text{(total)}$ with the exact calculation. With $\beta$ fixed to 1, our estimate matches the shape of the exact $\ptwoloop^\text{(total)}$ if $\alpha\simeq 3/10$ (see Fig.~\ref{fig:p2loopestalphaz0})%
%%%%% FOOTNOTE %%%%%
~\footnote{A similar estimate for the contribution of the two-loop finite term is presented in App.~\ref{app:sptestimate}, while comparisons between the estimated and exact versions of $\ptwoloop^\text{(total)}$ at different redshifts can be found in App.~\ref{app:estimatecomparisons}. }%
%%%% END FOOTNOTE %%%%
. Examination of $\ptreecs$ after $\co$  has been matched to data reveals that $\ptreecs \sim -(1/3) \poneloop$ at $z=0$ and $k\lesssim 0.4\invMpc$. Therefore, at these values of $\co$ and $\alpha$, the factor $(\alpha\poneloop+\ptreecs)$ in~(\ref{eq:p2looptotest}) heavily suppresses the magnitude of $\ptwoloop^\text{(total)}$ in this range.
This suppression was first noticed in~\cite{Carrasco:2013mua} as a cancellation between the different terms entering the power spectrum prediction at two loops, but it is more naturally understood as arising from the relationship between loops and counterterms when they are present as subdiagrams within higher-loop diagrams.

\begin{figure}[t]
\begin{center}
\includegraphics[width=0.7\textwidth]{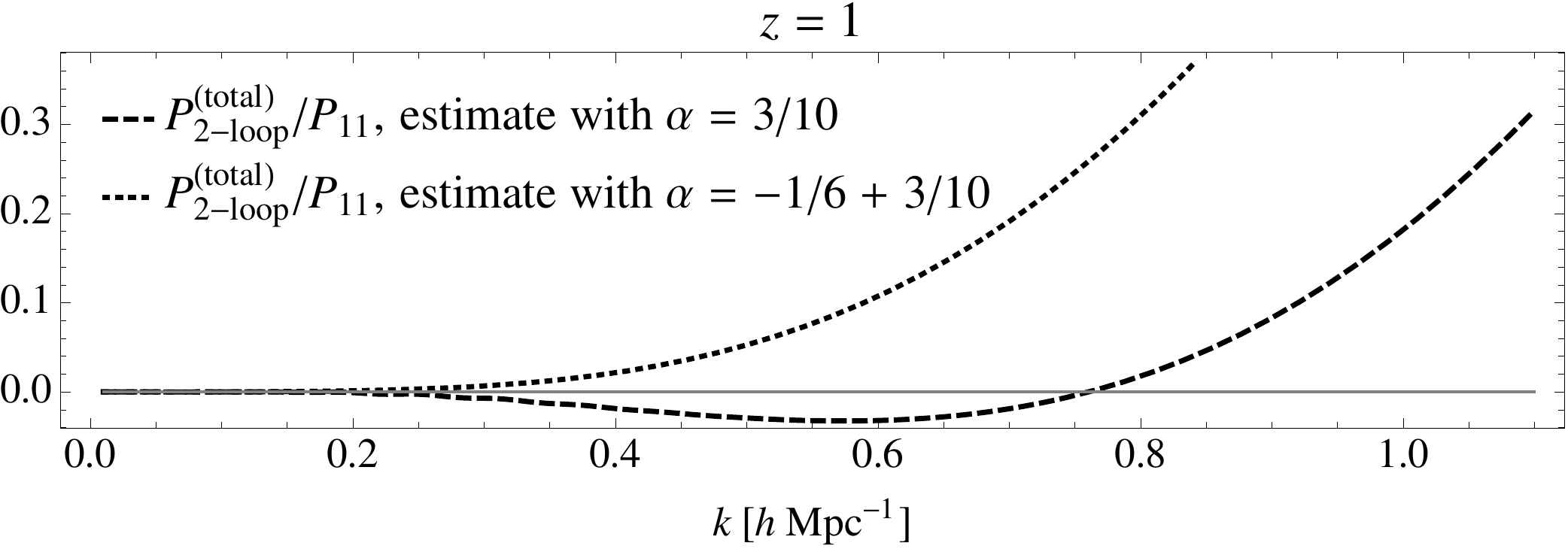}
\caption{\label{fig:p2loopestalphaz1} \small\it  Approximate versions of $\ptwoloop^\text{(total)}/P_{11}$ at $z=1$, plotted using Eq.~(\ref{eq:p2looptotest}) with $\alpha\simeq3/10$~(dashed curve) and $-1/6+3/10$~(dotted cruve). Both curves use $\beta=1$. The dashed curve has zero crossings that are not expected from the exact calculation, while the dotted curve has the same slope as the dashed one at low $k$, but reflected across the $k$-axis, eliminating any zero crossings. This shift in the value of $\alpha$ is the minimal way to modify our estimate to remove the zero crossings; we also implement similar shifts at $z=2$ and 3.}
\end{center}
\end{figure}

When attempting to estimate $\ptwoloop^\text{(total)}$ at higher redshifts, we find that some choices for $\alpha$ imply that the expression~(\ref{eq:p2looptotest}) will have zero crossings that are not expected to be present in the full calculation, or, if present, clearly cannot be exactly reproduced by our estimates (see Fig.~\ref{fig:p2loopestalphaz1} for an example at $z=1$). This is particularly problematic because we would like to use the prefactor $\beta$ to incorporate the uncertainty in the overall amplitudes of the estimates, but points close to a zero crossing will not be rescaled in the desired way if $\beta$ is rescaled. Therefore, at redshifts where these zero crossings appear in Eq.~(\ref{eq:p2looptotest}), we will modify the estimate to preserve its $k$-dependence at low $k$ when $\alpha\simeq 3/10$, but reflect it across the $k$-axis so that the estimate no longer changes sign (see Fig.~\ref{fig:p2loopestalphaz1}). Shifting the value of $\alpha$ is the minimal way to accomplish this goal. At the redshifts where there are zero crossings, namely $z=1,2,3$, we find that the requisite shifts are $\Delta\alpha \simeq -1/6, -1/3,$ and $-1/2$ respectively.

To obtain a rough value for $k_{\rm fail}^{(1)}$, the wavenumber at which $P_\text{EFT-1-loop}(k)$ will fail, we will use the scale at which the estimate for the ratio $\ptwoloop^\text{(total)}(k)/P_\text{EFT-1-loop}(k)$ equals the error on the nonlinear data. In our estimate for $\ptwoloop^\text{(total)}(k)$, we allow $\alpha$ and $\beta$ to vary within the ranges specified above: $3/10 \leq \alpha \leq 1$ (appropriately shifted at redshifts where there is a zero crossing in the estimate) and $1/2 \leq \beta \leq 2$. We also allow for $\co$ to deviate by 20\% from its best-fit value (taken from our fits to the Coyote emulator in Sec.~\ref{sec:fits}). At $z=0$, this results in an expectation that $k_{\rm fail}^{(1)}$ should fall somewhere within the range $0.14 \invMpc < k_{\rm fail}^{(1)} < 0.58 \invMpc$, encompassing the actual value of $k_{\rm fail}^{(1)}=0.36 \invMpc$. Table~\ref{tab:eftloopestimates} shows estimates for $k_{\rm fail}^{(1)}$ for redshifts from 0 to 4, along with exact $k_{\rm fail}^{(1)}$ values obtained from the exact fits in Sec.~\ref{sec:fits}.

\begin{table}[t]
\begin{center}
\begin{tabular}{c|c|c||c|c}
	& \multicolumn{2}{c||}{\textbf{One-loop EFT}} & \multicolumn{2}{c}{\textbf{Two-loop EFT}} \\ \hline
	   $z$ & estimated failure & actual failure & estimated failure & actual failure  \\
	    &  $[\hinvMpc]$ & $[\hinvMpc]$ &  $[\hinvMpc]$ & $[\hinvMpc]$  \\ \hline\hline
	0 & 0.14 -- 0.58 & 0.36	& 0.26 -- 0.78 	& 0.65  \\
	1 & 0.20 -- 0.60 & 0.45	& 0.42 -- 0.85  	& 1.2  \\
	2 & 0.27 -- 0.85 & 0.68	& 0.65 -- 1.50	& 1.6  \\
	3 & 0.5 -- 1.4 & 1.15		& 1.2 -- 3.0	& 2.3  \\
	4 & 0.5 -- 2.3 & 1.65 		& 1.3 -- 5.0	& 3.3
\end{tabular}
\caption{\label{tab:eftloopestimates} \small\it Estimates for where the one- and two-loop EFT predictions should fail, based on where the ratios $\ptwoloop^\text{(total)}/P_\text{EFT-1-loop}$ and $\pthreeloop^\text{(total)}/P_\text{EFT-2-loop}$ respectively exceed the error on the nonlinear data we use. We also state the scales at which the exact one- and two-loop EFT calculations fail, as seen in the comparison to data in Sec.~\ref{sec:fits}. The details of the estimates can be found in the main text. The exact two-loop calculation sometimes reaches slightly farther than the estimates suggest, but this is not cause for alarm given the $\orderone$ uncertainty in the estimates. 
Note also that there is some sizable uncertainty in the determination of the actual failure points, stemming from the uncertainty in $\co$ and the range of reasonable choices for $\kren$; see the main text for further discussion.
}
\end{center}
\end{table}

Before moving on, we remark that $P_{13}$ not only has an IR contribution that cancels with $P_{22}$, but it also has a UV contribution (from modes with $q\gtrsim \knl$) that is not under perturbative control and degenerate with a contribution from a counterterm. Estimating this UV contribution at $z=0$ for $k\sim 0.4\invMpc$ by integrating $P_{13}$ from $q=0.7 \invMpc$ to $\infty$, we find that it can be as much as 50\% of $\alpha\poneloop$, indicating that an unrenormalized calculation of a one-loop subdiagram can contain a large uncontrolled contribution. In this light, it is not surprising that the $\ptreecs$ term turns out to be a sizable fraction of the size of $\poneloop$ after $\co$ is fit to data, since $\ptreecs$ must correct for UV contributions to $\poneloop$ that should not be included.

\subsubsection{Three loops}

For $\pthreeloop^\text{(total)}$, we follow the same strategy as for $\ptwoloop^\text{(total)}$, multiplying the lower-loop contribution by a factor representing the additional loop:
\beq
\label{eq:p3looptotest}
\pthreeloop^\text{(total)}(k) \approx \beta \lb \frac{\alpha \poneloop(k)+\ptreecs(k)}{P_{11}(k)} \rb 
	\ptwoloop^\text{(total)}(k)\ .
\eeq

We use the exact calculation for $\ptwoloop^\text{(total)}$, and also use the same ranges for $\alpha$, $\beta$, and $\co$ as before. Similarly, the estimate for where $P_\text{EFT-2-loop}$ should fail, denoted by $k_{\rm fail}^{(2)}$, is taken to be where $\pthreeloop^\text{(total)}/P_\text{EFT-2-loop}$ exceeds the error on the nonlinear data. At $z=0$, Eq.~(\ref{eq:p3looptotest}) gives $0.26 \invMpc < k_{\rm fail}^{(2)} < 0.78 \invMpc$, which includes the exact value of $k_{\rm fail}^{(2)} = 0.65 \invMpc$. Values for other redshifts are shown in Table~\ref{tab:eftloopestimates}. At some redshifts these estimates fall slightly short of where the exact calculation fails, but this is not cause for alarm, given the fact that these estimates are only intended to be accurate up to $\orderone$ factors. The fact that the estimated and exact reach of $P_\text{EFT-2-loop}$ show the same trend indicates that the suppression due to the $(\alpha\poneloop+\ptreecs)$ factor that we discussed at two loops indeed persists at higher loops, since otherwise the prediction would fail earlier.

Note that there is some sizable uncertainty in the determination of the actual failure points listed in Table~\ref{tab:eftloopestimates}, stemming from the uncertainty in $\co$ and the range of reasonable choices for $\kren$; this is shown by the blue bands in Fig.~\ref{fig:datacomparison-wmap7}. At $z=0$, this uncertainty could result in an actual failure point of around $k\sim 0.2\invMpc$ for $P_\text{EFT-2-loop}$, because of the sensitivity of the low-redshift calculation to accidental cancellations between different terms. $k\sim 0.2\invMpc$ is slightly below the lower bound of our $z=0$ estimate in Tab.~\ref{tab:eftloopestimates}, but this is not cause for concern since the estimates themselves have $\mathcal{O}(1)$ uncertainties. Meanwhile, at higher redshifts, the uncertainty in the failure point of the prediction is much less severe, and the range of possible failure points indicated by Fig.~\ref{fig:datacomparison-wmap7} lies comfortably within our estimates.

%--------------------------------------------------------------------------------------
\subsection{Determining $\ct$}
\label{sec:cs2fitmethod}

In moving from the one-loop to the two-loop prediction for the power spectrum, we must include several new terms, the most important being the term containing the two-loop diagrams themselves, $\ptwoloop$. This term will have a contribution that scales like $(k/\knl)^2 P_{11}$, but this contribution arises from UV modes that are not under perturbative control, and whose impact on the power spectrum is degenerate with the choice of $\co$. Therefore, this contribution must be cancelled by adding a counterterm $\sim \ct (k/\knl)^2 P_{11}$ and adjusting $\ct$ appropriately. We emphasize that adjusting the coefficient of the $(k/\knl)^2 P_{11}$ term in this way does not change the value of the physical parameter $\co$; rather, the only role of $\ct$ is to cancel the unphysical part of $\ptwoloop$ that is erroneously included in the calculation.

In a scaling universe, one could analytically calculate the form of $\ct$ by Taylor-expanding the integrand of $\ptwoloop$ for large internal momenta (or alternately for small $k$): integrating the result will give a constant (dependent on the method used to regulate the integral, such as a hard UV cutoff~$\Lambda$) times $k^2 P_{11}$, and $\ct$ can then be set to to cancel off this term. In Appendix~\ref{app:cs2log}, we argue that in {\lcdm}, it is dangerous to use this procedure if one does not have access to nonlinear" power spectrum data that are very precise at low $k$. The data we currently use do not have the desired precision, and therefore we must use a different strategy for determining $\ct$ in our case.

\begin{figure}[t]
\begin{center}
\includegraphics[width=0.6\textwidth]{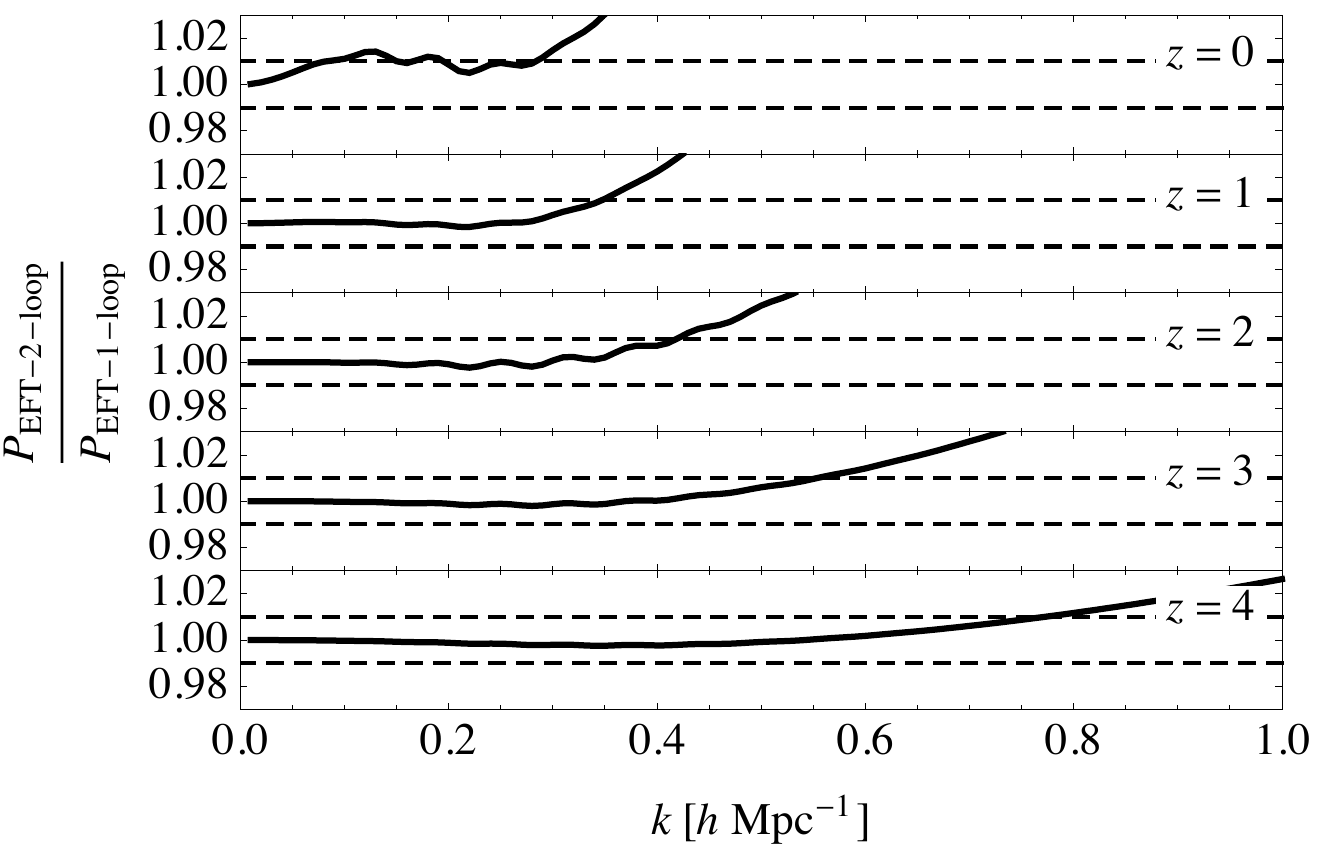}
\caption{\label{fig:21ratio} \small\it Ratio of $P_\text{EFT-2-loop}/P_\text{EFT-1-loop}$ at various redshifts, with $\co(z)$ set to what one would get from a fit of $P_\text{EFT-1-loop}$ to nonlinear data, and $\ct(z)$ set to a value that keeps the ratio flat for as large a range as possible at low $k$. (At $z\geq 1$, this translates into having the ratio as close to unity as possible, while at $z=0$, the flat region is offset from unity by 1\% ---see the main text for further discussion.) The highest scale where the ratio exceeds a 1\% deviation from unity (shown in dashed lines) determines $\kren(z)$. }
\end{center}
\end{figure}

In~\cite{Carrasco:2013mua} and~\cite{Senatore:2014via}, the prediction was only considered at a single redshift, and $\ct$ was fixed by simply requiring that the one-loop and two-loop EFT predictions matched either over a certain range of $k$ or at one specific renormalization scale $\kren$. We will adopt the latter approach here, but to do so, we must first define a prescription for choosing $\kren$ at each redshift. In general, $\kren$ should be chosen in the range where $\ptwoloop^\text{(finite)}$ is small, a fact that pushes $\kren\to0$, but also where the $k^2P_{11}(k)$ contribution of $\ptwoloop$ is much larger than the  error in the numerical data, pushing $\kren\to\infty$. From a practical standpoint, this will occur where the one-loop and two-loop EFT predictions begin to depart from each other; more specifically, in this work we set $\kren$ to be the scale at which this departure exceeds~1\%, when $\co$ is fixed by a fit of $P_\text{EFT-1-loop}$ to the nonlinear data and $\ct(z)$ is set to a value that keeps the ratio  $P_\text{EFT-2-loop}/P_\text{EFT-1-loop}$  flat for as large a range as possible at low $k$. Several examples of the ratio $P_\text{EFT-2-loop}/P_\text{EFT-1-loop}$ are shown in Fig.~\ref{fig:21ratio}, from which $\kren$ at each redshift can simply be read off. 

Note that at $z\geq 1$, when we apply the procedure described above, $P_\text{EFT-2-loop}/P_\text{EFT-1-loop}$ is very close to unity at low $k$, while at $z=0$ the ratio is systematically offset from unity before deviating at greater than 1\%. We allow for this deviation at low $z$ to account for possible numerical error in our loop integrals or IR-resummation. Alternatively, if we are very strict with the low-$k$ flatness criterion, we instead find that the ratio deviates at 1\% starting around $k \sim 0.1\invMpc$, leading to $\kren = 0.1\invMpc$ instead of $0.3\invMpc$ at $z=0$, where however the measurement of $\co$ would be challenging as its effect is dangerously closer to the numerical errors. We defer a more thorough investigation of these two options to future work. In this paper, we use $\kren=0.3\invMpc$ at $z=0$, but incorporate the uncertainty in $\kren$ into our final results, as described in the next section.

%--------------------------------------------------------------------------------------
\subsection{Comparison with simulations}
\label{sec:fits}

We are now ready to compare the two-loop EFT prediction for the matter power spectrum, Eq.~(\ref{eq:peft2loop}), to results from simulations at different redshifts. (In fact, we will use the IR-resummed version of~(\ref{eq:peft2loop}), the details of which can be found in~\cite{Senatore:2014via}.) There is only one free parameter, $\co(z)$, to be fit at each redshift, since the procedure described in Sec.~\ref{sec:cs2fitmethod} fixes $\ct(z)$ once $\co(z)$ has been determined. At each redshift, we determine a best-fit value of $\co$ by the value for which the ratio $P_\text{EFT-2-loop}/P_\text{Coyote}$ is constant over as large a range of wavenumbers as possible. As a conservative estimate of the uncertainty in this value, we vary $\co$ in either direction until the ratio $P_\text{EFT-2-loop}/P_\text{Coyote}$ deviates from its constant value by 3\% at the upper end of the constant region. We apply this procedure at each redshift used to build the Coyote emulator, $z \in \{0,1/9,1/4,3/7,2/3,1,3/2,2,5/2,3,4 \}$.
Furthermore, we incorporate our uncertainty in the optimal value of $\kren$ by repeating this procedure at various $\kren$ values that are smaller than the value determined in Sec.~\ref{sec:cs2fitmethod}%
%%%%% FOOTNOTE %%%%%
~\footnote{For $z\geq 2/3$, we consider $\kren$ values as much as 30\% below the value we choose to calculate the main results for the two-loop prediction (we do not allow $\kren$ to range above the value from Sec.~\ref{sec:cs2fitmethod}, since it is likely that this would lead to significant contamination of our matching procedure by finite two-loop terms). For $z\leq3/7$, we allow $\kren$ to range down to where the flatness criterion on the ratio $P_\text{EFT-2-loop}/P_\text{Coyote}$ is most strictly satisfied---for example, at $z=0$ we consider the range $0.1\invMpc<\kren<0.3\invMpc$. Of course, a lower $\kren$ will lead to higher contamination of the measurement of $\co$ by numerical error.}%
 %%%% END FOOTNOTE %%%%
.

\begin{figure}[t]
\begin{center}
\includegraphics[width=0.6\textwidth]{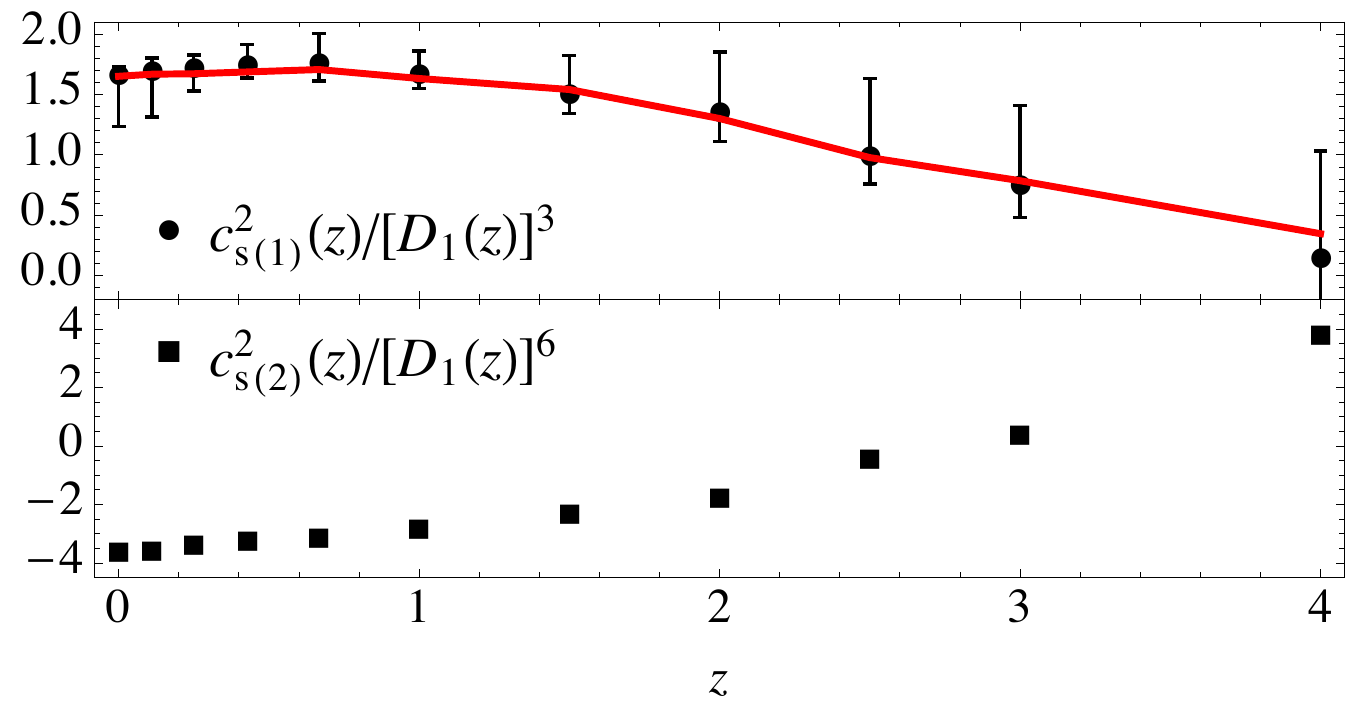}
\caption{\label{fig:cs1z-wmap7} \small\it Top: The values of $\co$, with associated errorbars, found by matching the two-loop EFT prediction for the power spectrum to the outputs of the Coyote emulator at different redshifts, in units of $(2\pi)^{-1} (\knl/\invMpc)^2$.  The specific matching procedure is described in the main text. The red curve shows the fitting function~(\ref{eq:cs1z}) with $\beta=-1.44$ (the best-fit value) and higher coefficients set to zero. The good match between the red curve and the points demonstrates that Eq.~(\ref{eq:cs1z}) is successful in describing the time-dependence of $\co(z)$ using only two parameters,~$\co(0)$ and~$\beta$. Bottom: The corresponding values of~$\ct$, again in units of $(2\pi)^{-1} (\knl/\invMpc)^2$, found from the procedure from Sec.~\ref{sec:cs2fitmethod}.}
\end{center}
\end{figure}

The resulting $\co(z)$ values are shown in Fig.~\ref{fig:cs1z-wmap7}, divided by $[D_1(z)]^3$ to decrease the range of the plot. The $z=0$ value,
\beq
\co(0) = (1.65^{+0.08}_{-0.42}) \times \frac{1}{2\pi} \lp \frac{\knl}{\invMpc} \rp^2\ ,
\eeq
is consistent with that found in~\cite{Carrasco:2013mua,Senatore:2014via} for a similar cosmology, a reassuring sign that minor variations in the background cosmology translate into miniscule variations in the value of $\co$%
%%%%% FOOTNOTE %%%%%
 \footnote{However, larger variations in $\co$ are noticeable when it is measured from cosmologies with different values of $\sigma_8$: in~\cite{Angulo:2014tfa}, we found $\co(0)\simeq 2.45 \times (2\pi)^{-1} (\knl/\invMpc)^2$ in a cosmology with $\sigma_8=0.9$, while in Appendix~\ref{app:wmap3} we find $\co(0)\simeq 1.31\times (2\pi)^{-1} (\knl/\invMpc)^2$ in a cosmology with $\sigma_8=0.74$. Taken together with the $\co$ measured in this section (from a cosmology with $\sigma_8=0.81$), these points suggest very roughly that $\co(0) \propto \sigma_8^{\sim 3.5}$, although further investigation will be required to verify this behavior. We leave this to future work.}%
 %%%% END FOOTNOTE %%%%
 . In a scaling universe, $\co(z)$ would scale with time like a power of $D_1(z)$~\cite{Mercolli:2013bsa,Baldauf:2014qfa}. As shown in Fig.~\ref{fig:neffk}, the effective tilt $n_{\rm eff}(k)$ of the linear power spectrum in {\lcdm} deviates significantly from the scaling form. Nevertheless, a useful starting point for parametrizing the time-dependence of $\co(z)$ is to suppose that it is determined by the effective tilt at $k=\kren$, along with corrections arising from the running of $n_{\rm eff}(k)$:
\beq
\label{eq:cs1z}
\co(z) = \co(0) [D_1(z)]^{\frac{4}{3+N(z)}}\ , \quad
	N(z) = \left. n_{\rm eff}(k) + \beta \frac{dn_{\rm eff}(k)}{d\log(k)} 
	+ \gamma \frac{d^2 n_{\rm eff}(k)}{d\log(k)^2} + \cdots \right|_{k=\kren(z)}\ ,
\eeq
where $\beta$ and $\gamma$ are unknown coefficients, and the ellipsis denotes higher logarithmic derivatives of $n_{\rm eff}(k)$. We use $k=\kren$ because it is the scale at which the effects encapsulated by $\co$ will affect the power spectrum quite strongly. Apart for the first derivative term in~(\ref{eq:cs1z}), the higher derivative terms are very small, so it should suffice to consider only the first correction to $n_{\rm eff}$ due to running. In fact, $n_{\rm eff}$ plus the first correction is able to match the fitted $\co(z)$ values very well, with a best-fit value for $\beta$ of $-1.44\pm 0.42$ (fitting $\beta$ and $\gamma$ simultaneously does not alter the results). The best-fit curve is shown in red in Fig.~\ref{fig:cs1z-wmap7} %
%%%%% FOOTNOTE %%%%%
 \footnote{Note that a very good approximation for the $\kren(z)$ values we have used in this section is $\kren(z)=(0.3+0.03z^2) \invMpc$. Using this formula, Eq.~(\ref{eq:cs1z}) can be computed as a smooth, analytical function, which may be useful for readers interested in using the EFTofLSS power spectrum prediction in other applications.}%
%%%% END FOOTNOTE %%%%
.

\begin{figure}[t!]
\begin{center}
\includegraphics[width=1.00\textwidth]{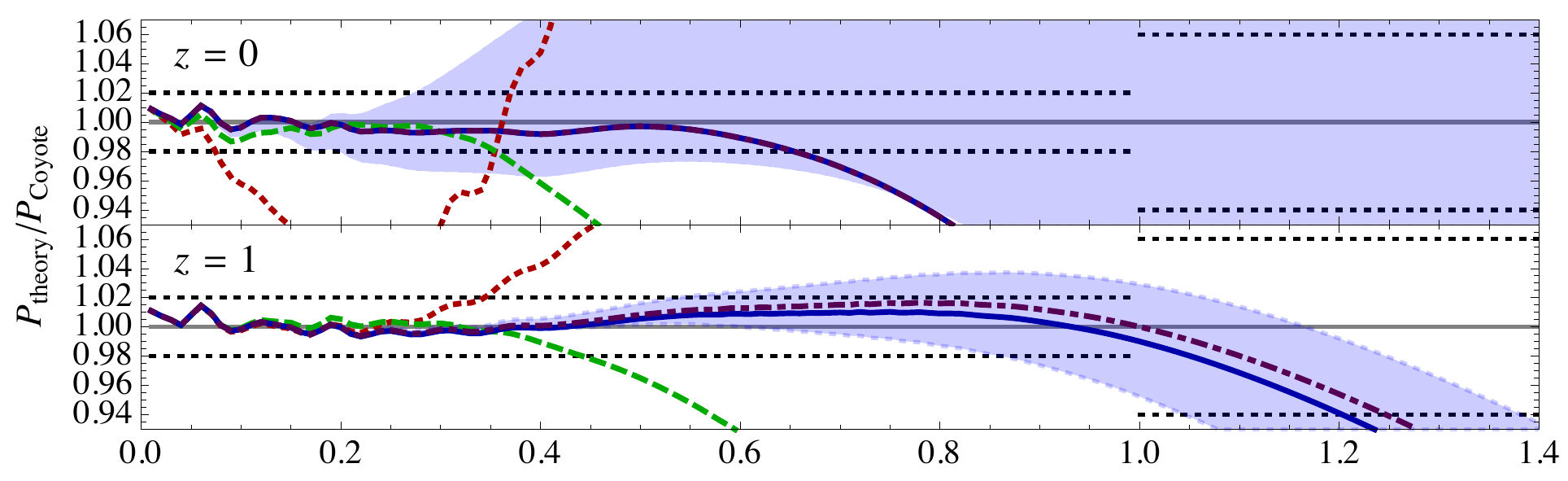}
\includegraphics[width=1.005\textwidth]{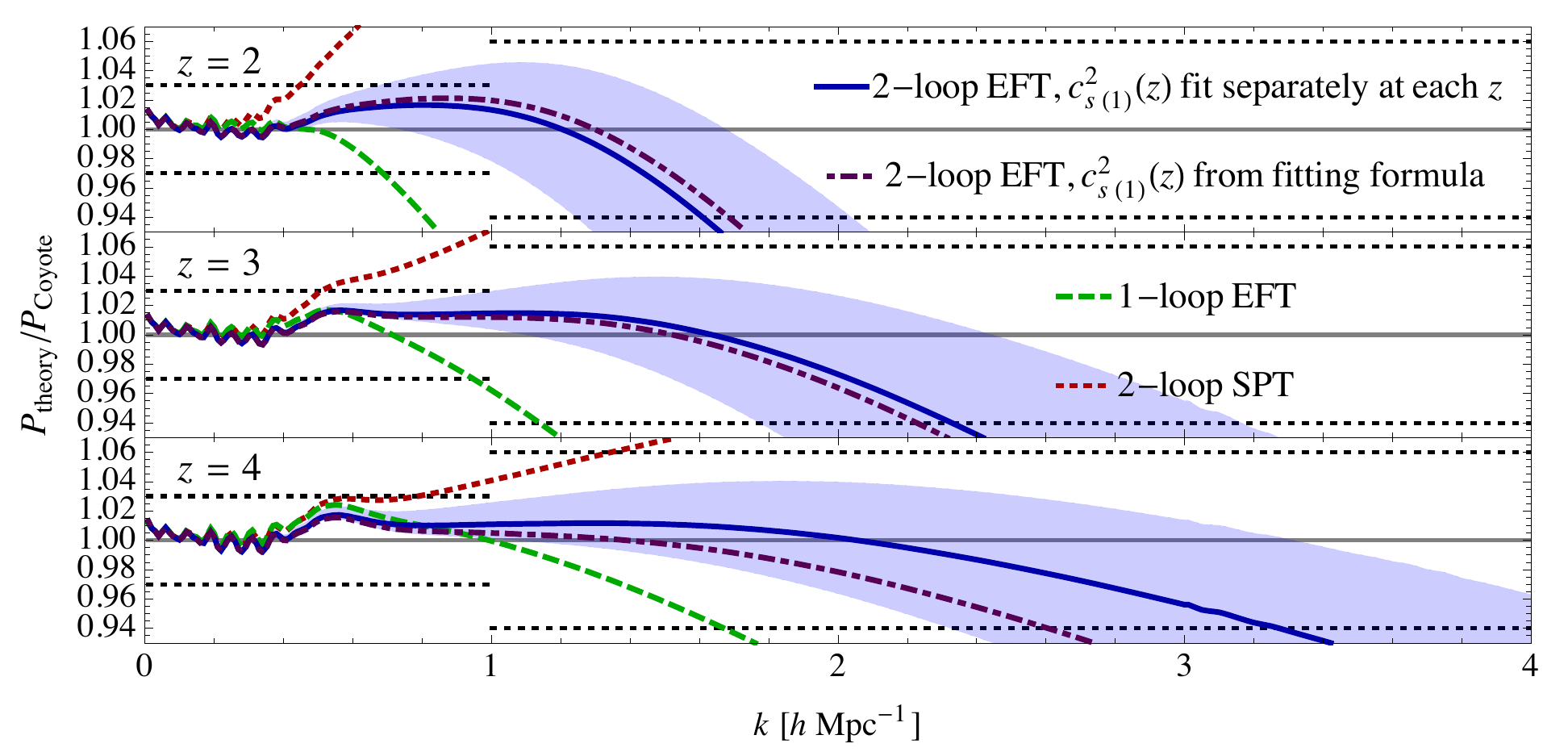}
\caption{\label{fig:datacomparison-wmap7} \small\it  Various theory curves, normalized to nonlinear power spectra at five different redshifts. The black dotted lines show the estimated uncertainty on the nonlinear spectra, while the blue band shows the uncertainty on the two-loop EFT prediction based on the uncertainty on $\co$ at that redshift. Whether $\co(z)$ is determined separately at each redshift (blue solid curves) or parametrized using Eq.~(\ref{eq:cs1z}) (purple dot-dashed curves), the two-loop EFT predictions agree with the data to much higher $k$ than two-loop SPT, in a manner consistent with the estimates of the theoretical error from Sec.~\ref{sec:mpsestimates}.  }
\end{center}
\end{figure}

Strictly speaking, the functional form in (\ref{eq:cs1z}) is in disagreement with the power-law time-dependence, $\co(z)\propto D_1(z)^\zeta$, with $\zeta\sim 3$, that has been repeatedly assumed during the paper (as for example in the assumption that the time-dependence of the terms proportional to $\co(z)$ is given by the EdS approximation, as well as in the derivation of the expressions for the higher order counterterms). It is possible to relax these assumptions by exactly solving for the time-dependence of the diagrams, by using the full Green's functions as for example done in~\cite{Carrasco:2012cv}. We have performed this calculation (which will be presented in an upcoming paper~\cite{formalism}), and found that the results are virtually unchanged.

\begin{figure}[t]
\begin{center}
\includegraphics[width=0.9\textwidth]{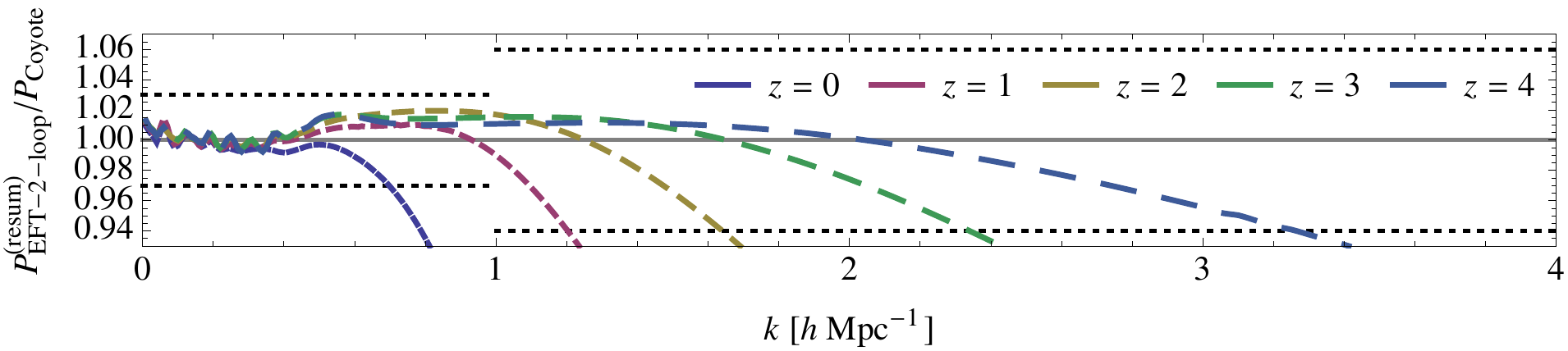}
\caption{\label{fig:datacomparison-wmap7-all2loop} \small\it  The (resummed) two-loop EFT predictions for the matter power spectrum at five different redshifts, displayed on the same set of axes and normalized to nonlinear spectra at the same redshifts. }
\end{center}
\end{figure}

Comparisons of various theoretical predictions to nonlinear spectra at different redshifts are shown in Fig.~\ref{fig:datacomparison-wmap7}. The solid blue curves show the two-loop EFT prediction with $\co(z)$ measured separately at each redshift, while the purple dot-dashed curves instead use Eq.~(\ref{eq:cs1z}) with $\beta=-1.44$  and setting all higher coefficients to zero. For reference, we also show the one-loop EFT and two-loop SPT predictions%
%%%%% FOOTNOTE %%%%%
~\footnote{We compare with SPT and not with other techniques, such as RPT or LPT, because these and all other techniques who differ from SPT only by a resummation of the IR-modes are such that, if  they are correctly formulated, they must give exactly the same UV reach for the dark matter power spectrum as in SPT~\cite{Senatore:2014via}. }%
%%%% END FOOTNOTE %%%%
. Fig.~\ref{fig:datacomparison-wmap7-all2loop} shows all two-loop EFT curves on the same plot. As previously found in~\cite{Senatore:2014via}, the resummed two-loop prediction matches the nonlinear power spectrum at $z=0$ within 2\% (the estimated uncertainty on the nonlinear data) for $k\lesssim 0.6\invMpc$ after fitting a single parameter, $\co$, that can be determined from data at a single redshift. By fitting $\co$ at each redshift, or alternatively making use of Eq.~(\ref{eq:cs1z}) with only one extra free parameter, it is possible to match the nonlinear spectrum at {\em all} redshifts up to $z=4$ and up to scales consistent with our theoretical expectations of how far the prediction should reach (see Sec.~\ref{sec:mpsestimates}). 

One can quantify the improvement of the two-loop EFT prediction with respect to two-loop SPT by the increase in the number of modes reliably described by each prediction, as determined by the scale $k_{\rm fail}$ at which each prediction deviates from the nonlinear data by an amount equal to the uncertainty on the data. The number of modes scales roughly like $k_{\rm fail}^3$, so the improvement can be estimated by $\lp k_{\rm fail}^{\rm (EFT)}/k_{\rm fail}^{\rm (SPT)} \rp^3$. It is also possible to perform a more detailed mode-counting calculation, based on integrating the number of modes contained in thin shells between $z=0$ and the redshift of interest%
%%%%% FOOTNOTE %%%%%
~\footnote{The details of this calculation are as follows. Starting from the comoving distance between an observer and a point at scale factor $a$,
\beq
\chi(a) = \int_a^1 \frac{da'}{(a')^2 H(a')}\ ,
\eeq
the volume of a single spatial shell between scale factors $a$ and $a+da$ is given by
\beq
V_\text{shell}(a) = 4\pi \chi^2(a) \, d\chi = 4\pi \chi^2(a) \frac{d\chi}{da} da
	= -\frac{4\pi \chi^2(a)}{a^2 H(a)} da\ .
\eeq
The number of modes within the volume between redshifts zero and $z$ is found by integrating the number of modes within each shell, estimated by the volume of the shell divided by the ``fundamental cell" corresponding to $k_{\rm fail}$ at the shell's redshift:
\beq
N_\text{modes}(z) \simeq \int_{a(z=0)}^{a(z)} \frac{V_\text{shell}(\ta)}{ \lb 2\pi/k_{\rm fail}(a) \rb^3 }
	= \int_{1/(1+z)}^1 \frac{d\ta}{\ta^2 H(\ta)} \frac{4\pi \chi^2(\ta)}{ \lb 2\pi/k_{\rm fail}(a) \rb^3 }\ .
\eeq }%
%%%% END FOOTNOTE %%%%
.  The results of both calculations are shown in Table~\ref{tab:modecounting} for several redshifts. The gain in the number of modes is necessarily smaller at higher redshift, since the power spectrum around the quasi-linear scale approaches the slope of $n=-3$ where loops become more important than counterterms. Nevertheless, the EFT allows one to describe a factor of $\sim$44 more modes out to $z=2$, or $\sim 70$ at $z=4$, which can in principle lead to a huge gain in the information we can extract from ongoing and upcoming galaxy surveys.

\begin{table}[t]
\begin{center}
\begin{tabular}{c|c|c|c}
	   $z$ & $\lp k_{\rm fail}^\text{(EFT)}/k_{\rm fail}^\text{(SPT)} \rp^3$ 
	   	& $N_\text{modes}^\text{(EFT)}/N_\text{modes}^\text{(SPT)}$
		& $N_\text{modes}^\text{(EFT)}$  \\
		 \hline\hline
	1/9 & $(0.66/0.08)^3=$ 562 & 610 & $1.5\times10^5$ \\
	3/7 & $(0.88/0.15)^3=$ 202 & 246 & $1.1\times10^7$ \\
	1 &  $(1.2/0.35)^3=$ 40 & 34 &  $2.4\times10^8$ \\
	2 & $(1.6/0.45)^3=$ 45 & 44 & $2.2\times10^9$ \\
	3 &  $(2.3/0.5)^3=$ 97 & 59 & $8.2\times10^{9}$ \\
	4 &  $(3.3/0.8)^3=$ 70 & 71 & $2.4\times10^{10}$
\end{tabular}
\vspace{8pt}
\caption{\label{tab:modecounting} \small\it The ratio of the number of modes reliably described by the two-loop EFT and two-loop SPT predictions for the matter power spectrum, calculated in two ways: a simple estimate using the fact that $N_{\rm modes} \propto k_{\rm fail}^3$, and a more detailed estimate based on integrating $N_{\rm modes}$ over thin redshift shells. We also show the absolute number of modes described by the two-loop EFT prediction. Depending on the redshift ranges of interest, the EFT allows a gain of anywhere between $\sim$30 and $\sim$600 times as many modes as SPT. }
\end{center}
\end{table}

Finally, let us briefly comment on our procedure for determining $\co$ at each redshift. An alternative procedure would be performing a least-$\chi^2$ fit over a certain range of $k$, but there are a few obstacles to implementing this procedure in practice. One is that the range over which one should fit at each redshift is quite uncertain---the estimates in Sec.~\ref{sec:mpsestimates} provide a rough guideline, but the resulting best-fit value for $\co$ will strongly depend on the extent of the chosen range. Another obstacle is that there may be systematic errors in the output of the Coyote emulator, which would bias our fits even if these errors were within the tolerances quoted in~\cite{Heitmann:2008eq}. For example, we expect that the ``bump" seen at $z\geq 3$ in Fig.~\ref{fig:datacomparison-wmap7} is a property of the emulator rather than the EFT predictions, since it is also present in $P_\text{SPT-2-loop}/P_\text{Coyote}$. A least-$\chi^2$ fit would not be able to incorporate this piece of information without an extra ad hoc prescription, such as inflating the errorbars on the nonlinear spectrum in the neighborhood of the bump. The method we use allows us to account for the presence of the bump, at the possible expense of overtuning the fit, but we have chosen conservative errors  (which incorporate our uncertainty in the renormalization scale $\kren$) and consistent $k$-ranges on our determination of $\co(z)$ to allow for this possibility. Better methods can in principle be designed to measure $\co$ directly from the power spectrum, depending on the quality of available numerical data. Alternatively, $\co$ can be measured directly in simulations using the dark matter particles as degrees of freedom, as done in~\cite{Carrasco:2012cv}.

%--------------------------------------------------------------------------------------
% SECTION: LENSING POTENTIAL POWER SPECTRUM
%--------------------------------------------------------------------------------------
\section{Analytical Calculation of Lensing Potential}

Now that we have the prediction of the EFTofLSS at all redshifts, we are able to calculate various observables related to gravitational lensing in a purely analytical way, without relying on simulations%
%%%%% FOOTNOTE %%%%%
~\footnote{As already stressed, the parameter $\co$ can in principle be extracted directly from observations, rather than from simulations.}%
%%%% END FOOTNOTE %%%%
. In particular, we will focus on the lensing potential~$\psi$, defined so that its gradient along the line of sight gives the total angle through which a photon is deflected due to the gravitational potential along the photon's trajectory. If one can reliably compute the statistics of~$\psi$, one can then straightforwardly compute various observables (such as shear correlation functions) that are relevant to weak lensing surveys (see, e.g.,~\cite{Munshi:2006fn} for a review). Alternatively, a calculation of the angular power spectrum of~$\psi$ can be directly compared to reconstructions from CMB measurements.

To calculate the statistics of~$\psi$, we first need the power spectrum of the gravitational potential~$\phi$, defined by
\beq
\left\langle \phi(\vk,a) \phi(\vk',a) \right\rangle = \frac{2\pi^2}{k^3} \delta_{\rm D}(\vk+\vk') P_\phi(k,a)\ ;
\eeq
this can be related to the matter power spectrum via Poisson's equation, yielding
\beq
\label{eq:phips}
P_\phi(k,a) = \frac{9\, \Omm(a)^2 \H(a)^4}{8\pi^2} \frac{P_\delta(k;a)}{k}\ .
\eeq
After expanding the lensing potential in spherical harmonics, $\psi(\hn) = \sum_{\ell m} \psi_{\ell m} Y_{\ell m}(\hn)$, we can define its angular power spectrum by
\beq
\left\langle \psi_{\ell m} \psi_{\ell' m'}^* \right\rangle = \delta_{\ell\ell'} \delta_{mm'} C_\ell^{\psi}\ .
\eeq
In the Limber approximation and assuming a flat universe, $C_\ell^\psi$ is given by a line-of-sight integral over $P_\phi$ times a geometric factor:
\beq
\label{eq:clpsi}
C_\ell^\psi = \frac{8\pi^2}{\ell^3} \int_0^{\chi_*} d\chi \, \chi\,  P_\phi \! \lp k=\frac{\ell}{\chi}; a(\chi) \rp \lp \frac{\chi_*-\chi}{\chi_* \chi} \rp^2\ ,
\eeq
where $\chi$ is the comoving distance with respect to a chosen observer, running from zero to $\chi_*$, the distance to the emitting source (for CMB lensing, the source is located at the last scattering surface). The derivation of this formula, as well as a thorough discussion of gravitational lensing, can be found in~\cite{Lewis:2006fu}. The EFT prediction for $C_\ell^\psi$ can be calculated by using $P_\text{EFT-1-loop}$ or $P_\text{EFT-2-loop}$ for $P_\delta$ in Eq.~(\ref{eq:phips}), and then inserting~(\ref{eq:phips}) into~(\ref{eq:clpsi}).

\begin{figure}[t]
\begin{center}
\includegraphics[width=0.65\textwidth]{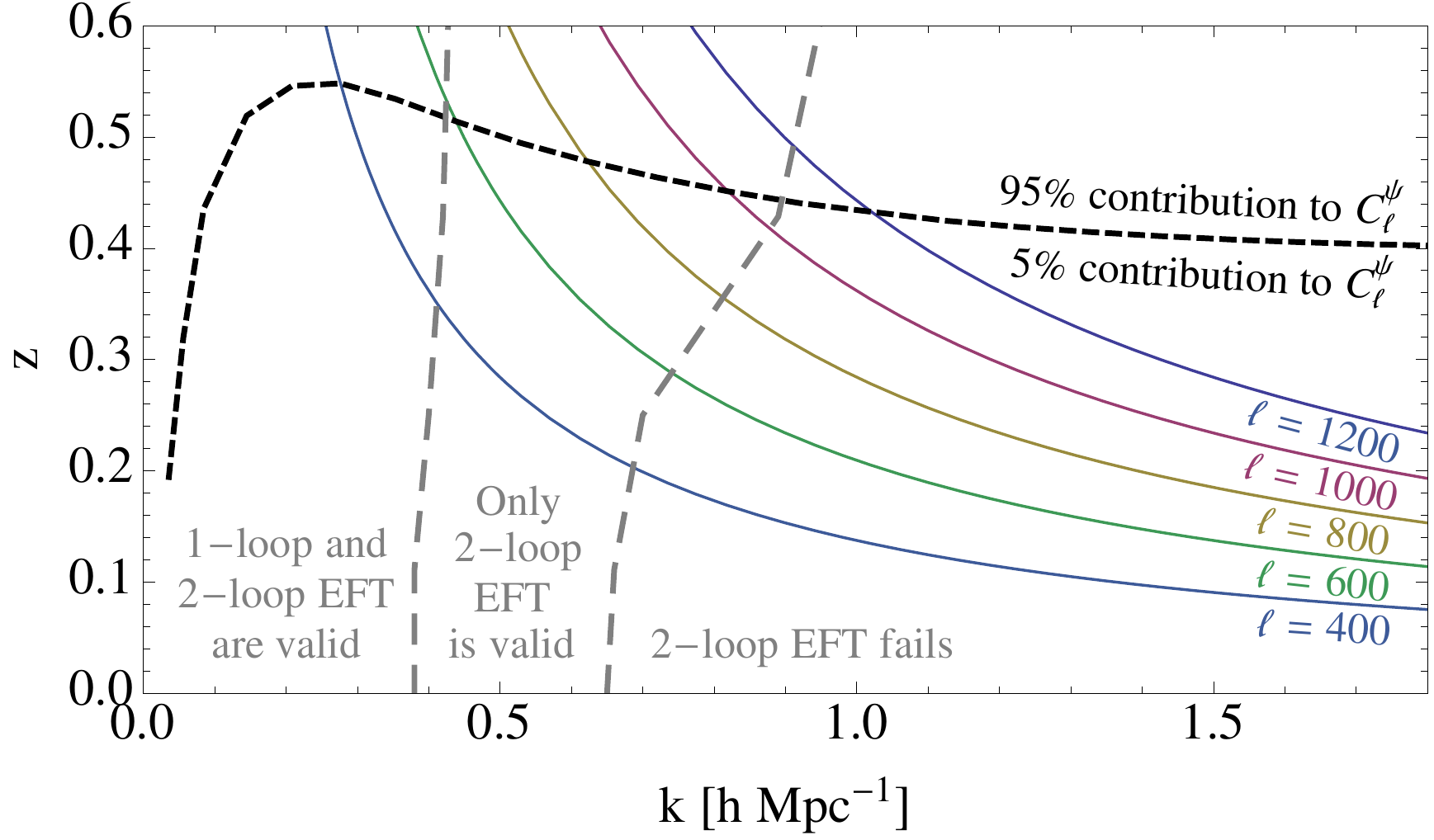}
\caption{\label{fig:lensingranges} \small\it  A graphical depiction of the regions in the $(k,z)$ plane corresponding to the validity of one- and two-loop EFT, the paths along which the integral for the lensing potential power spectrum $C_\ell^\psi$ is evaluated, and the scales and redshifts that contribute 95\% of the value of $C_\ell^\psi$ for selected $\ell$ values. We find that one-loop EFT can be used to calculate $C_\ell^\psi$ with less than 5\% theoretical error for $\ell\lesssim 600$, while the corresponding limit for two-loop EFT is $\ell\lesssim 1000$. See the main text for further discussion.}
\end{center}
\end{figure}

To investigate what ranges of $\ell$ can be accurately described using the EFT at one or two loops, we plot several sets of curves in Fig.~\ref{fig:lensingranges} for lensing of the CMB. The solid lines trace the paths in the $(k,z)$ plane along which the integral~(\ref{eq:clpsi}) is evaluated for various values of $\ell$, while the black short-dashed line separates each path into the region contributing 95\% (above the black line) or 5\% (below the black line) of the value of the integral%
%%%%% FOOTNOTE %%%%%
 \footnote{We use Halofit~\cite{Takahashi:2012em}  for this calculation, since it provides a rough indication of the size of nonlinearities that will affect the boundary between 95\% and 5\% contributions to the integral.}%
 %%%% END FOOTNOTE %%%%
 . This shows, for example, that $C_{\ell=400}^\psi$ is dominated by the region of the integral corresponding to $k\lesssim 0.25\invMpc$ and $z\gtrsim 0.55$, while, as expected, higher-$k$ modes contribute more strongly at higher $\ell$. The grey long-dashed lines show where the one- and two-loop EFT predictions for the matter power spectrum fail, as found in Sec.~\ref{sec:fits} with WMAP7 cosmological parameters. For each $\ell$ curve, if the 95\% region overlaps with the region where the one- or two-loop EFT fails, then the theoretical error in the EFT calculation of $C_\ell^\psi$ will surpass 5\%, which we take as our threshold for a useful calculation%
%%%%% FOOTNOTE %%%%%
  \footnote{Observational uncertainties on future measurements of $C_\ell^\psi$ will precisely determine the required theoretical precision. For now, we choose 5\% as an optimistic forecast of these future requirements. As presented in recent publications, current measurements of $C_\ell^\psi$ bandpowers have uncertainties that are well above this level~\cite{vanEngelen:2012va,Das:2013zf,Ade:2015zua,Story:2014dwa}, although these uncertainties could be reduced by changing how the measurements of individual multipoles are binned.}%
%%%% END FOOTNOTE %%%%  
. This occurs at $\ell \sim 600$ for one loop and $\ell\sim 1000$ for two loops.

\begin{figure}[t]
\begin{center}
\includegraphics[width=0.65\textwidth]{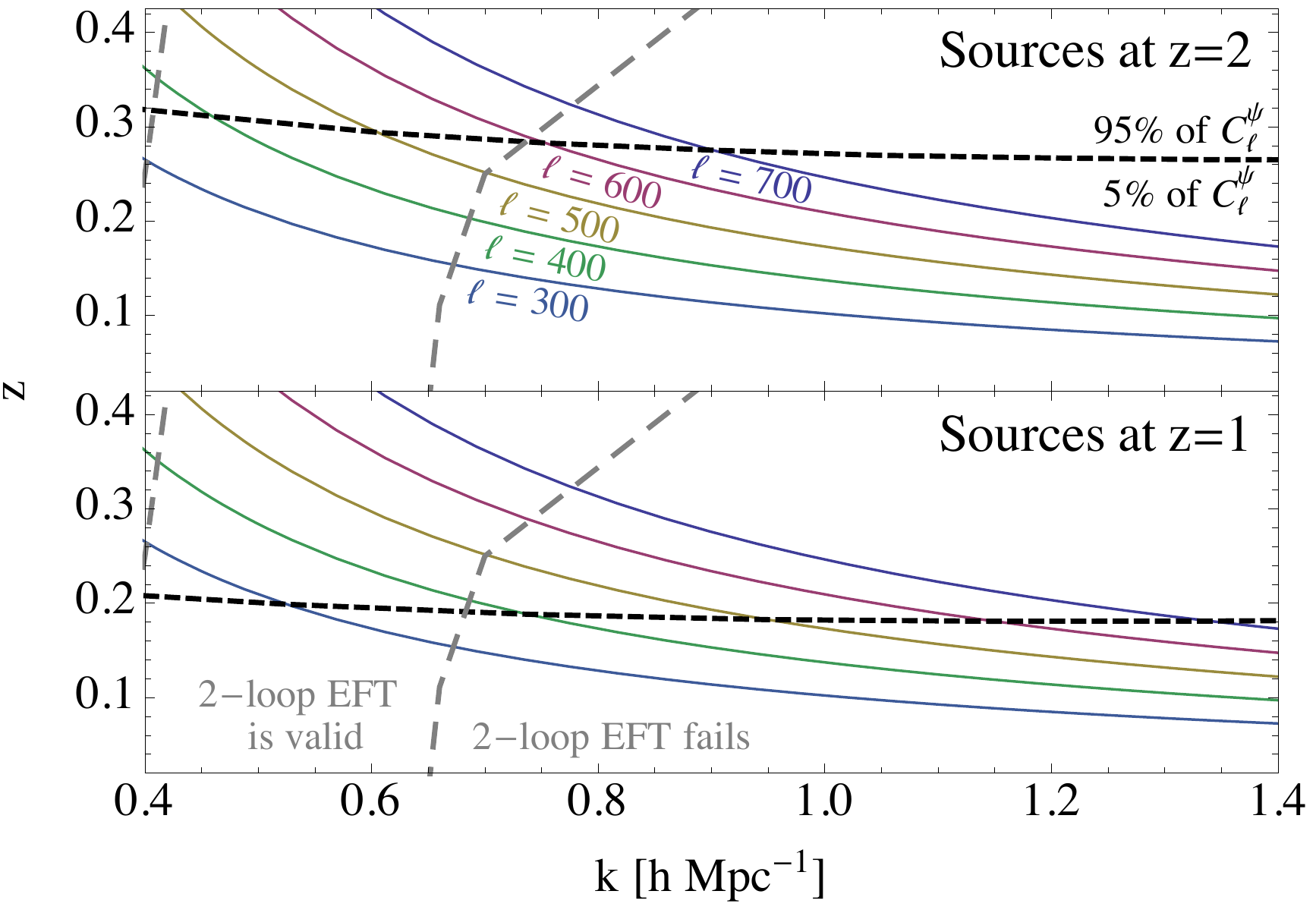}
\caption{\label{fig:lensingrangesz12} \small\it  As Fig.~\ref{fig:lensingranges}, but for the lensing potential power spectrum corresponding to sources at $z=2$ (top panel) and $z=1$ (bottom panel). The two-loop EFT prediction for $C_\ell^\psi$ has less than 5\% theoretical error for $\ell\lesssim 600$ for sources at $z=2$ and $\ell\lesssim 350$ for $z=1$. }
\end{center}
\end{figure}

In Fig.~\ref{fig:lensingrangesz12} we repeat Fig.~\ref{fig:lensingranges} but for sources at redshift $z=2$ and $z=1$ corresponding roughly to the reach of ongoing lensing surveys. For photons originating at these redshifts (as opposed to the last scattering surface), the effects of late-time nonlinearities are more important. We see that the EFTofLSS allows us to predict $C_\ell^\psi$ to $5\%$ accuracy up to $\ell\sim 600$ for sources at $z=2$ and $\ell \sim 350$ for $z=1$. This is a large improvement with respect to former analytical techniques.

\begin{figure}[t]
\begin{center}
\includegraphics[width=0.95\textwidth]{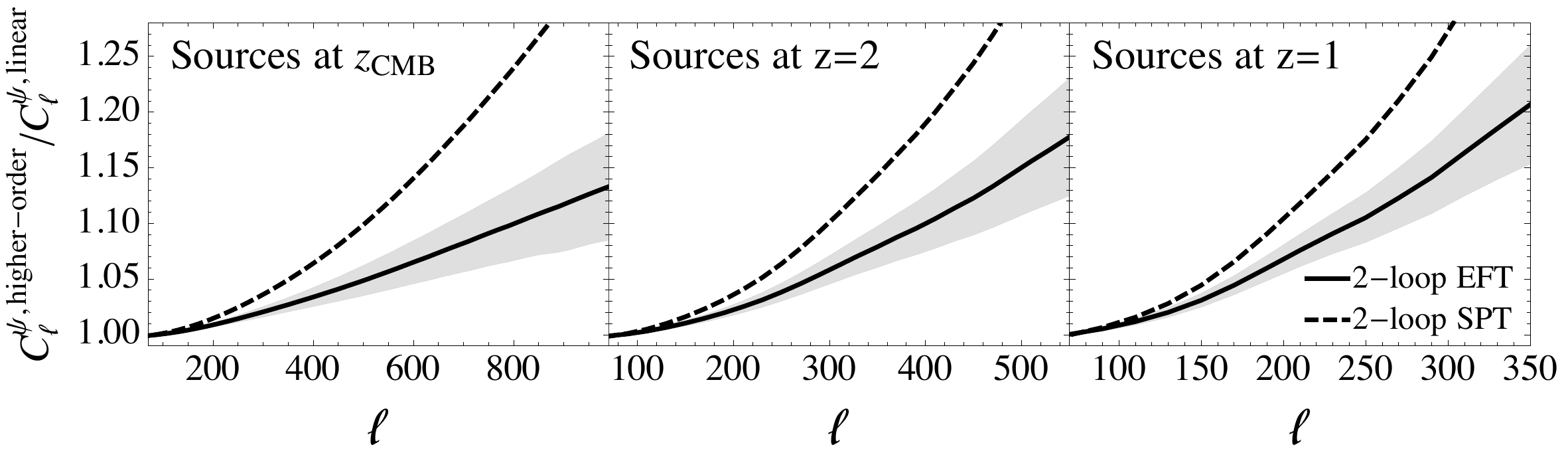}
\caption{\label{fig:clpsi_ratios} \small\it  The ratio of the two-loop EFT and SPT predictions for $C_\ell^\psi$ to the prediction from linear theory. The grey bands show the theoretical error on the EFT prediction, based on the contribution to the $C_\ell^\psi$ integral from scales beyond the reach of the EFT. It is likely that future data will become sensitive to the nonlinearities in the lensing potential that are encoded in the differences between these curves (which may also be accessible to current measurements of CMB lensing, after an appropriate choice of binning scheme). We furthermore see that, for $\ell\gtrsim 200$, SPT at two loops is as bad an approximation to the EFTofLSS (which we showed is the correct result) as linear theory. }
\end{center}
\end{figure}

In Fig.~\ref{fig:clpsi_ratios}, we plot the ratio of the two-loop EFT and SPT predictions for $C_\ell^\psi$ over the predictions from linear theory%
%%%%% FOOTNOTE %%%%%
~\footnote{In computing the two-loop EFT prediction for $C_\ell^\psi$, it is important to recall that $P_\text{EFT-2-loop}$ is written as an expansion in powers of $k/\knl$ times $P_{11}$, and at high $k$ these terms will cause the prediction to deviate much more strongly from $P_{\rm NL}$ than $P_{11}$ does. Such a deviation is guaranteed regardless of the loop order, since as $k\to\knl$ the entire perturbative expansion breaks down completely. To avoid using the EFT prediction in a regime where it does not apply, for the EFT curves in Fig.~\ref{fig:clpsi_ratios}, we use $P_\text{EFT-2-loop}$ for $P_\delta$ when $k<k_{\rm fail}^{(2)}(z)$, and $P_{11}$ when $k>k_{\rm fail}^{(2)}(z)$. This introduces a small error in the calculation of $C_\ell^\psi$, which should be no more than 5\%, since we have already established that for $\ell\lesssim 1000$ the region $k>k_{\rm fail}^{(2)}(z)$ contributes no more than 5\% of the value of the integral corresponding to lensing of the CMB. Using $P_{\rm EFT}$ instead of $P_{11}$ in this regime would probably make the error induced by the contribution from high wavenumbers more important.}%
%%%% END FOOTNOTE %%%%
, along with grey bands indicating the theoretical error on the EFT prediction, based on the contribution to the lensing integral from scales beyond the EFT's reach. As presented in recent publications, current data are unable to distinguish between these curves, since, with the chosen binning schemes, the errorbars on $C_\ell^\psi$ are $\mathcal{O}(10\%)$~\cite{vanEngelen:2012va,Das:2013zf,Ade:2015zua} or slightly smaller~\cite{Story:2014dwa}. For example, the 2015 Planck data~\cite{Ade:2015zua} measure the overall amplitude of the lensing-potential power spectrum in the range $40\lesssim \ell\lesssim 400$, to a precision of about~2.5\%. Fig.~\ref{fig:clpsi_ratios} shows that this might allow us at most a very marginal measurement of gravitational nonlinearities%
%%%%% FOOTNOTE %%%%%
~\footnote{ Notice that one could design a procedure that would allow us to measure an integral over redshifts of the speed of sound, which would allow us to extract this information directly from data, without relying on simulations.}%
%%%% END FOOTNOTE %%%%
. Future data are expected to be precise enough to detect these effects with high significance, making the EFTofLSS an important tool for using lensing to tighten constraints on cosmological parameters and search for new physics.

\begin{figure}[t]
\begin{center}
\includegraphics[width=0.55\textwidth]{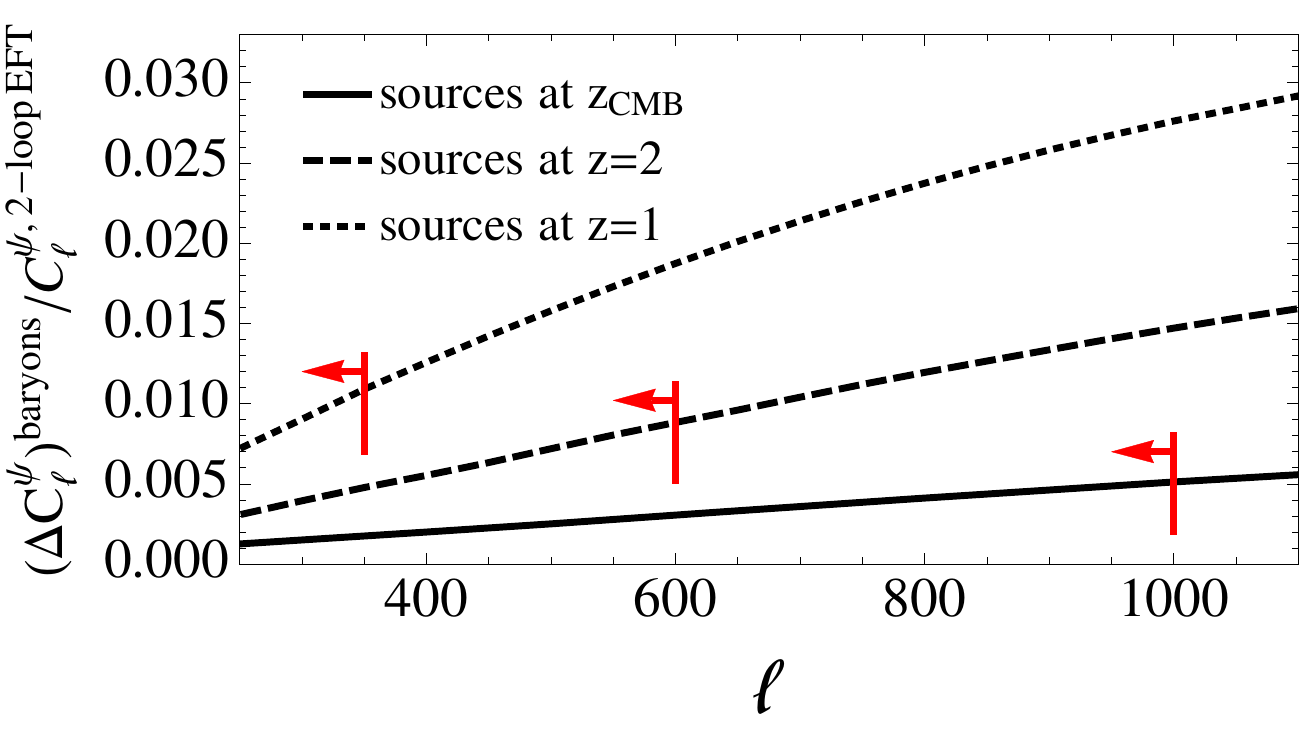}
\caption{\label{fig:clpsi_baryons} \small\it  The effect of baryonic physics on the two-loop EFT prediction for $C_\ell^\psi$, as approximately captured by a 15\% shift in $\co(z)$ in the tree-level counterterm $\ptreecs$~\cite{Lewandowski:2014rca}. The red arrows indicate the ranges in which each EFT prediction has less than 5\% theoretical error, as estimated from Figs.~\ref{fig:lensingranges} and~\ref{fig:lensingrangesz12}. We find that baryonic effects on~$C_\ell^\psi$ never exceed $\sim$1\% in these ranges and for these sources.}
\end{center}
\end{figure}

Thus far, we have presented results for dark matter only, so an important question is whether the effects of baryons could become important at the scales we are concerned with. At lowest order, these effects will manifest in the power spectrum as a shift of $\sim$15\% in the value of $\co$ in the tree-level counterterm $\ptreecs$ (see~\cite{Lewandowski:2014rca} for a detailed discussion). To quantify the impact on the lensing potential power spectrum, we calculate the change in $C_\ell^\psi$ resulting from a 15\% shift in $\co(z)$ at each $z$, and compare to the full two-loop $C_\ell^\psi$ calculation. Fig.~\ref{fig:clpsi_baryons} shows that, in the range of multipoles where the EFT prediction has better than 5\% theoretical error, and for the sources we consider, baryonic physics never alters the lensing potential power spectrum by more than $\sim$1\%.  

In summary, all of this is a promising indication that the EFT prediction could be applied to upcoming observational data, an exciting prospect that we plan to explore in future work.

%--------------------------------------------------------------------------------------
% SECTION: DISCUSSION
%--------------------------------------------------------------------------------------
\section{Discussion}

In this paper we have studied the prediction of the EFTofLSS as applied to dark matter at different redshifts. This has allowed us to explore to which high wavenumber the EFTofLSS can push itself as we increase redshift. We have found that the reach of the EFT in the UV grows with redshift, largely outperforming SPT and all former analytic techniques, all of which, when correctly implemented, have the same UV reach as SPT.  At two loops with IR-resummation, the EFTofLSS agrees within approximately 1\% with matter power spectra measured from simulations up to $k\simeq 0.6\invMpc$ at $z=0$. At redshifts $z\geq 1$, the agreement is within approximately 2\% up to at least $k\simeq1\invMpc$, and also within approximately 5\% up to $k\simeq 1.2\invMpc$ at $z=1$ and $k\simeq 2.3\invMpc$ at $z=3$. Uncertainties in the fitted value of $\co$ and the range of reasonable choices for $\kren$ can change where the predictions fail; these uncertainties are largest at low redshift, where the calculation is particularly sensitive to accidental cancellations that may occur between different terms that enter at two loops, and become milder with increasing redshift.

As we go to higher redshift, the slope of the power spectrum near the nonlinear scale becomes closer to $-3$. This makes the effect of loops more important relative to the effect of the counterterms, so that the gain of the EFT with respect to SPT decreases at higher redshifts. In fact, in the EFTofLSS it is possible to estimate the size of the higher-order contributions that have not been included, so that one can forecast the scale at which a calculation at a given order fails. After pointing out that approximating the universe as scaling is only marginally correct, we have developed a more accurate way of estimating the size of higher order terms, with the results that the UV reach of our two-loop calculation (along with the associated uncertainty in that reach) agrees with these estimates. 

We have studied how the parameters of the EFT depend on time, finding that, in the local in time approximation, the $\co$ term can be well described by a function close to a power-law that slightly runs with $z$,  and whose generic features can be anticipated by using an approximate scaling symmetry of the universe, and that ultimately requires a single parameter to be fit. In summary, only two coupling constants---$\co$ at redshift $z=0$, and the slope of the power law, controlled by a parameter $\beta$---are sufficient to extend the predictions of the power spectrum in the EFTofLSS to all redshifts far beyond the reach of SPT.

These power spectrum results allow us to perform analytical calculations related to gravitational lensing, such as the lensing potential power spectrum presented in this work. We have found that the EFTofLSS at two-loop order is able to predict $C_\ell^\psi$ corresponding to lensing of CMB photons with better than 5\% accuracy for $\ell\lesssim 1000$. For photons emitted by galaxies at $z=2$ or $z=1$, similar accuracy is achieved by the EFT prediction for $\ell\lesssim 600$ and 350, respectively. This opens up an incredibly interesting window to compare the EFTofLSS to currently available data as well as to data that will be available in the very near future. We plan to perform a detailed study in an upcoming paper.

We stress that the presence of coupling constants in the EFTofLSS is not a catastrophe for the theory. These parameters can be measured directly in observations (without resorting to simulations), or can be measured in  small (and therefore fast) numerical simulations%
%%%%% FOOTNOTE %%%%%
~\footnote{For early ideas and applications on how to use small $N$-body simulations to extract information about the properties of long wavelength fluctuations, see~\cite{Baldauf:2011bh,Carrasco:2012cv,Wagner:2014aka}.}%
%%%% END FOOTNOTE %%%%
. It would be interesting to explore how much cosmological information is lost if these coupling constants were to be measured directly in observations.  We guess not much, given that in the EFTofLSS the same coupling constant predicts several observables: fixing the speed of sound parameter using the matter power spectrum allows  us to predict the matter bispectrum~\cite{Angulo:2014tfa} and the momentum power spectrum~\cite{Senatore:2014via} at one loop up to about $k\simeq 0.3\hinvMpc$ without any remaining free parameter. We leave a more detailed study of these aspects to future work. As we keep exploring the predictions of the EFTofLSS for dark matter, biased tracers, lensing, and so on, the repeated successes of this novel analytic approach seem to indicate that we might be able to extract a much greater amount of cosmological information from large scale structures than previously believed.

%--------------------------------------------------------------------------------------
% SECTION: ACKNOWLEDGMENTS
%--------------------------------------------------------------------------------------
\section*{Acknowledgments}

We are very much indebted to Eiichiro Komatsu for strongly motivating us in the importance of this project. We also thank John Joseph Carrasco, Antony Lewis, and Matias Zaldarriaga for useful conversations. This research was supported by the Munich Institute for Astro- and Particle Physics (MIAPP) of the DFG cluster of excellence ``Origin and Structure of the Universe." S.F.~is partially supported by the Natural Sciences and Engineering Research Council of Canada.  L.S. is supported by DOE Early Career Award DE-FG02-12ER41854 and by NSF grant PHY-1068380.

%--------------------------------------------------------------------------------------
% APPENDICES
%--------------------------------------------------------------------------------------
\appendix
\section*{Appendix}

%--------------------------------------------------------------------------------------
% APPENDIX: NEW TREE-LEVEL COUNTERTERM
%--------------------------------------------------------------------------------------
\section{Derivation of Tree-Level Counterterms}
\label{app:treelevelcts}

We will now present the derivation of the tree-level counterterms involved in one- and two-loop EFT predictions, since one of these terms was erroneously omitted from prior work~\cite{Carrasco:2013mua,Senatore:2014via}. For convenience, we will repeat some of the equations from Sec.~\ref{sec:review}.

We begin with the Eulerian-space EFTofLSS equations of motion, with the stress tensor $\gammai{}^i$ given by Eqs.~(\ref{eq:stresstensor}) and~(\ref{eq:Kansatzold}):
\begin{align}
\label{eq:contApp}
&a\H \, \d_a \delta(\vk,a)+\theta(\vk,a)=
	 - \! \int_{\vq}\alpha(\vkp,\vk-\vkp)\delta(\vk-\vkp,a)\theta(\vkp,a)\ , \\ \nn
&a\H \, \d_a \theta(\vk,a)+\H \theta(\vk,a)+\frac{3}{2} \H^2(a)  \Omm(a)  \delta(\vk,a)
= - \! \int_{\vq} \beta(\vkp,\vkkp)\theta(\vk-\vkp,a)\theta(\vkp,a) \\
&\qquad\qquad\qquad\qquad\qquad\qquad\qquad\qquad\qquad\qquad
+  \epsilon k^2 \int \! \frac{da'}{a'\H(a')} K(a,a') [ \delta(a',\vx_{\rm fl})]_{\vk}\ ,
\label{eq:eulerApp}
\end{align}
where $\epsilon$ is a parameter inserted to organize the power of $K(a,a')$ appearing in the perturbative solutions we will obtain. To solve these equations, we make use of two assumptions:
\begin{itemize}
\item $\Omm(a) \approx f(a)^2$, and
\item $c_n(a)  = \bar c_n ( \xi D_1(a)^\zeta \H^2 f^2)$, where
\beq
c_n(a) \equiv \int \frac{da'}{a'\H(a')} K(a,a') \frac{D_1(a')^n}{D_1(a)^n}
\eeq
and $\xi$ is a constant that we are free to set to a convenient value, since it simply rescales each~$\bar{c}_n$.
\end{itemize}
Solutions to~(\ref{eq:contApp}) and~(\ref{eq:eulerApp}) can also be found without making use of either of these assumptions (e.g.~\cite{formalism}), but we use them here because they greatly simplify the algebra involved, and we have checked that the exact implementation of the time-dependence does  not make a relevant difference~\cite{formalism}. Under these assumptions, we can use the following solutions:
\begin{align} \nn
\delta(a,\vk) &= \sum_{n=1}^\infty [D_1(a)]^{n} \delta^{(n)}(\vk) 
	+ \epsilon \sum_{n=1}^\infty [D_1(a)]^{n+\zeta} \tilde{\delta}^{(n)}(\vk)
	+ \epsilon^2 \sum_{n=1}^\infty [D_1(a)]^{n+2\zeta} \hat{\delta}^{(n)}(\vk)\ , \\
\theta(a,\vk) &= -\H(a) f(a) \left\{ \sum_{n=1}^\infty [D_1(a)]^{n} \theta^{(n)}(\vk) 
	+ \epsilon \sum_{n=1}^\infty [D_1(a)]^{n+\zeta} \tilde{\theta}^{(n)}(\vk)
	+ \epsilon^2 \sum_{n=1}^\infty [D_1(a)]^{n+2\zeta} \hat{\theta}^{(n)}(\vk) \right\}\ .
\label{eq:dtansatzwitheps2}
\end{align}
Since in this Appendix we are only interested in tree-level counterterms, we will only need the $n=1$ terms in each sum above. Plugging these back into~(\ref{eq:contApp}) and~(\ref{eq:eulerApp}) and collecting terms of order~$\epsilon^1$, we find:
\beq
(1+\zeta) \tilde{\delta}^{(1)}(\vk) - \tilde{\theta}^{(1)}(\vk) = 0, \quad 
	-\lp \frac{3}{2}+\zeta \rp \tilde{\theta}^{(1)}(\vk) + \frac{3}{2} \tilde{\delta}^{(1)}(\vk) = 
	k^2 \bar{c}_1 \xi \, \delta^{(1)}(\vk)\ .
\label{eq:deltilde1eqs}
\eeq
Solving for $\tilde{\delta}^{(1)}(\vk)$ and conveniently setting $\xi=\zeta(\zeta+5/2)$ to cancel the resulting denominator, we find 
\beq
\tilde{\delta}^{(1)}(\vk) = -  \bar{c}_1 k^2 \delta^{(1)}(\vk)\ .
\eeq
In~\cite{Carrasco:2013mua}, $\bar{c}_1$ was identified with $(2\pi) \co/\knl^2$, since the calculation was only performed at a single redshift. In this paper, it will be convenient to absorb a factor of $[D_1(z)]^\zeta$ into a time-dependent function $\co(z)$, coinciding with the convention from~\cite{Carrasco:2013mua} at $z=0$. 
Therefore, the counterterm corresponding to $2 \langle \delta^{(1)} \tilde{\delta}^{(1)} \rangle$ (which enters the one-loop EFT prediction) is written as 
\beq
\ptreecs(k,z)(k,z) = -2 (2\pi) \co(z) [D_1(z)]^2 \frac{k^2}{\knl^2} P_{11}(k)\ ,
\eeq
while the counterterm corresponding to $\langle \tilde{\delta}^{(1)} \tilde{\delta}^{(1)} \rangle$ takes the form
\beq
P_\text{tree}^{(k^4,1)}(k,z) = (2\pi)^2 [\co(z)]^2 [D_1(z)]^2 \frac{k^4}{\knl^4} P_{11}(k)\ .
\eeq

The second counterterm that scales like $(k/\knl)^4 P_{11}$ arises from correlating $\hat{\delta}^{(1)}$ with $\delta^{(1)}$. To find the expression for $\hat{\delta}^{(1)}$, we follow the same steps that lead to Eq.~(\ref{eq:deltilde1eqs}), but collect terms of order $\epsilon^2$ instead of $\epsilon^1$; this gives
\beq
(1+2\zeta) \hat{\delta}^{(1)}(\vk) - \hat{\theta}^{(1)}(\vk) = 0, \quad 
	-\lp \frac{3}{2}+2\zeta \rp \hat{\theta}^{(1)}(\vk) + \frac{3}{2} \hat{\delta}^{(1)}(\vk) = 
	k^2 \bar{c}_{1+\zeta} \xi \, \tilde{\delta}^{(1)}(\vk)\ .
\eeq
Solving for $\hat{\delta}^{(1)}(\vk)$ and using our previous choice for $\xi$, we find
\beq
\hat{\delta}^{(1)}(\vk) = \frac{\zeta+\frac{5}{2}}{4(\zeta+\frac{5}{4})} 
	\bar{c}_1 \bar{c}_{1+\zeta} k^4 \delta^{(1)}(\vk)\ .
\eeq
At this point, we will assume locality in time, for which $\bar{c}_{1+\zeta}=\bar{c}_1$~\cite{Carrasco:2013mua}. Then, taking $2 \langle \delta^{(1)} \hat{\delta}^{(1)} \rangle$ gives us our second counterterm:
\beq
P_\text{tree}^{(k^4,2)}(k,z) = (2\pi)^2  \frac{\zeta+\frac{5}{2}}{2(\zeta+\frac{5}{4})} [\co(z)]^2 [D_1(z)]^2 \frac{k^4}{\knl^4} P_{11}(k)\ .
\eeq
The $\zeta$-dependent factor in $P_\text{tree}^{(k^4,2)}$ is a slowly-varying function of $\zeta$ for $1<\zeta<5$, so deviations from our initial choice of $\zeta=3$ will have only a weak effect on this term.

%--------------------------------------------------------------------------------------
% APPENDIX: SPT ESTIMATES
%--------------------------------------------------------------------------------------
\section{Notes on Estimates}

\subsection{Estimating Finite Parts of SPT Loop Corrections}
\label{app:sptestimate}

\begin{figure}[t]
\begin{center}
\includegraphics[width=0.7\textwidth]{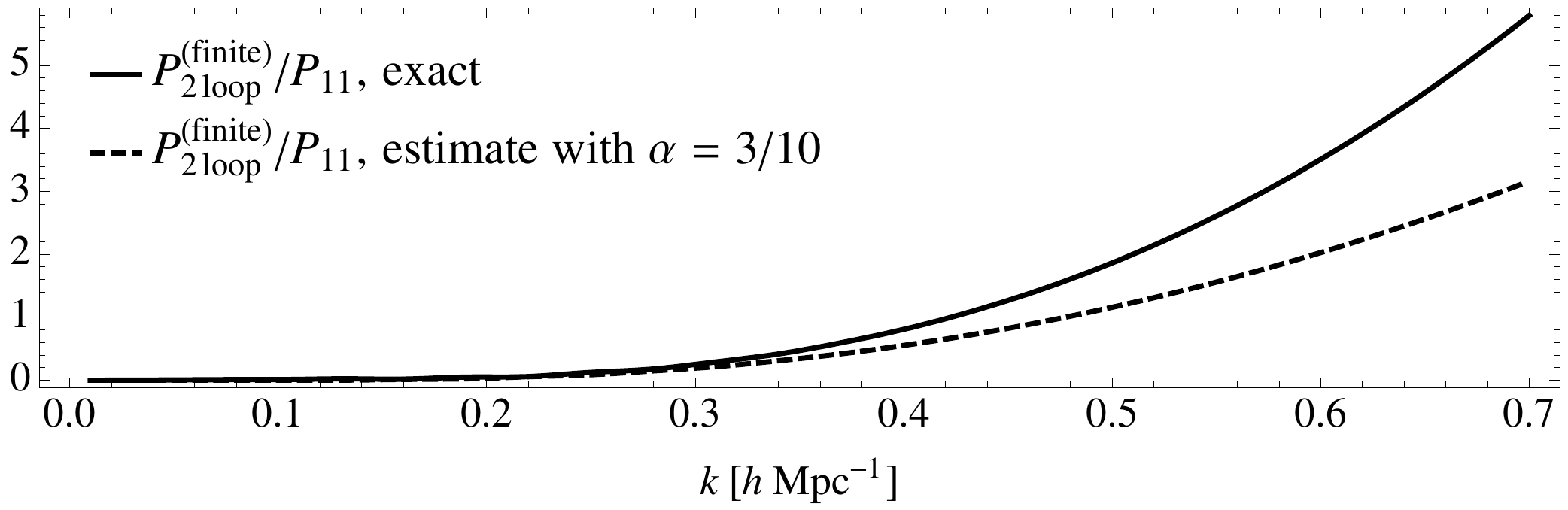}
\caption{\label{fig:p2loopfiniteestalpha}\small\it Exact and approximate versions of $\ptwoloop^\text{(finite)}/P_{11}$ at $z=0$, where $\ptwoloop^\text{(finite)}(k,0) \equiv \ptwoloop(k) - 2(2\pi)\ct(0) (k/\knl)^2 P_{11}(k)$. The estimate is plotted using Eq.~(\ref{eq:p2looptotest}) with $\co(0)=0$, $\alpha\simeq3/10$ and $\beta=1$. The two curves match to within a factor of $\sim$2 up to $k\sim0.7\invMpc$; an even better match can be achieved by letting $\alpha$ or $\beta$ float freely instead of fixing their values to those from Sec.~\ref{sec:mpsestimates}. }
\end{center}
\end{figure}

In this short section, we simply highlight the fact that the method of estimating loop corrections from Sec.~\ref{sec:mpsestimates} also applies to the finite parts of SPT terms alone. At two loops, this corresponds to the combination
\be
\ptwoloop^\text{(finite)}(k,0) \equiv \ptwoloop(k) - 2(2\pi)\ct(0) (k/\knl)^2 P_{11}(k)\ ,
\ee
where the $\ct$ term removes the UV contribution to $\ptwoloop$ (see Sec.~\ref{sec:cs2fitmethod}). Fig.~\ref{fig:p2loopfiniteestalpha} compares the exact calculation of $\ptwoloop^\text{(finite)}$ with the approximation from Eq.~(\ref{eq:p2looptotest}), with $\co(0)=0$ and $\alpha$ and $\beta$ fixed to the values used in Fig.~\ref{fig:p2loopestalphaz0} (3/10 and 1, respectively). The exact and approximate calculations match to within a factor of $\sim$2 up to $k\sim 0.7\invMpc$, with the discrepancy due entirely to the normalization of the estimate, set by the combination $\beta \times \alpha$ in this case. Letting $\beta$ or $\alpha$ float away from the values used in the plot enables an even better agreement. Since the SPT contribution is a part of the EFT one, it is important to be able to estimate that as well.

\subsection{Comparisons of Estimates and Exact Calculation of $\ptwoloop^\text{(total)}$}
\label{app:estimatecomparisons}

\begin{figure}[t]
\begin{center}
\includegraphics[width=0.8\textwidth]{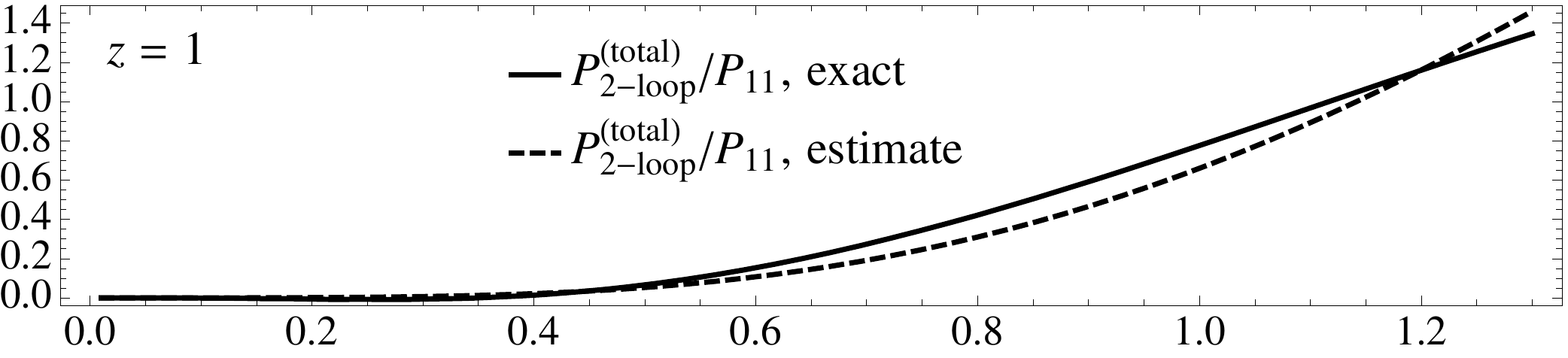}
\includegraphics[width=0.8\textwidth]{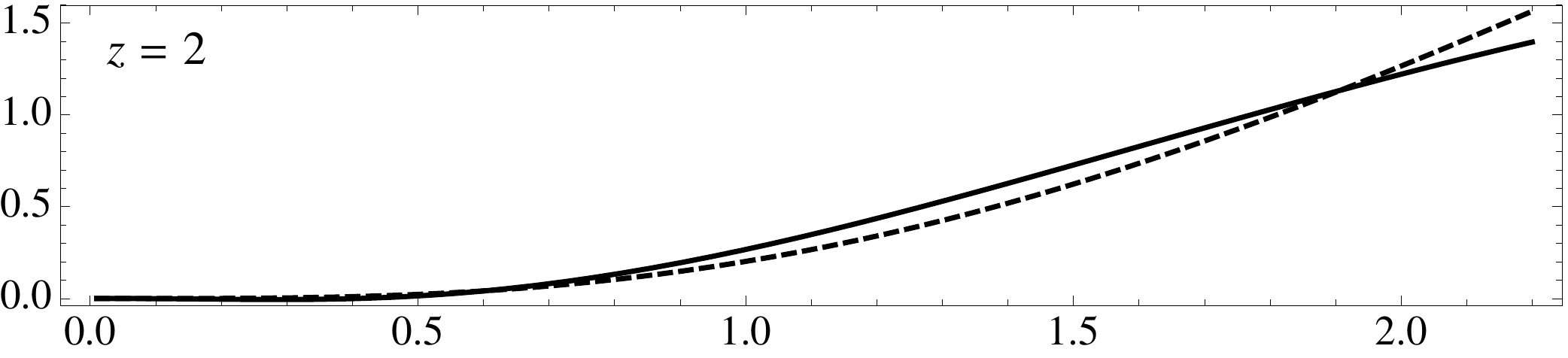}
\includegraphics[width=0.8\textwidth]{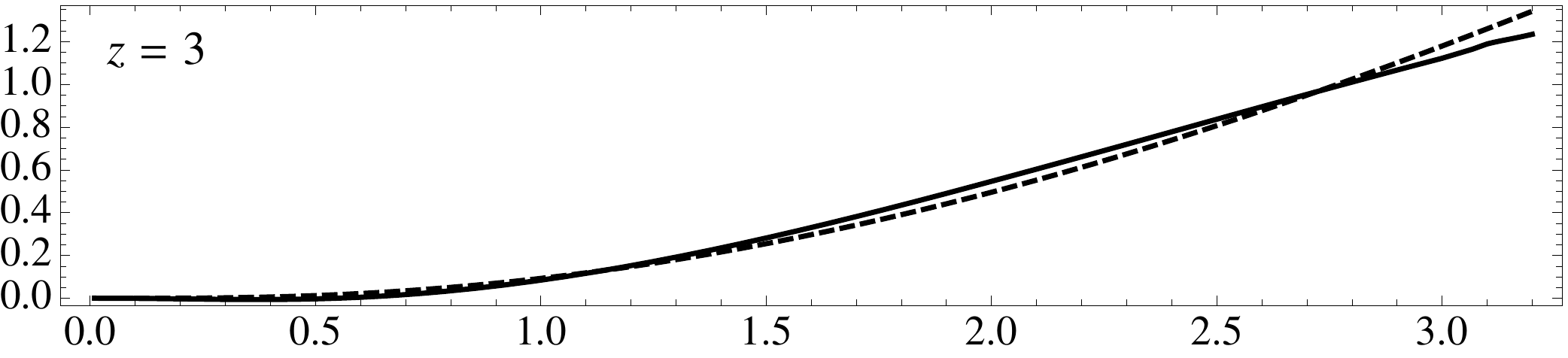}
\includegraphics[width=0.8\textwidth]{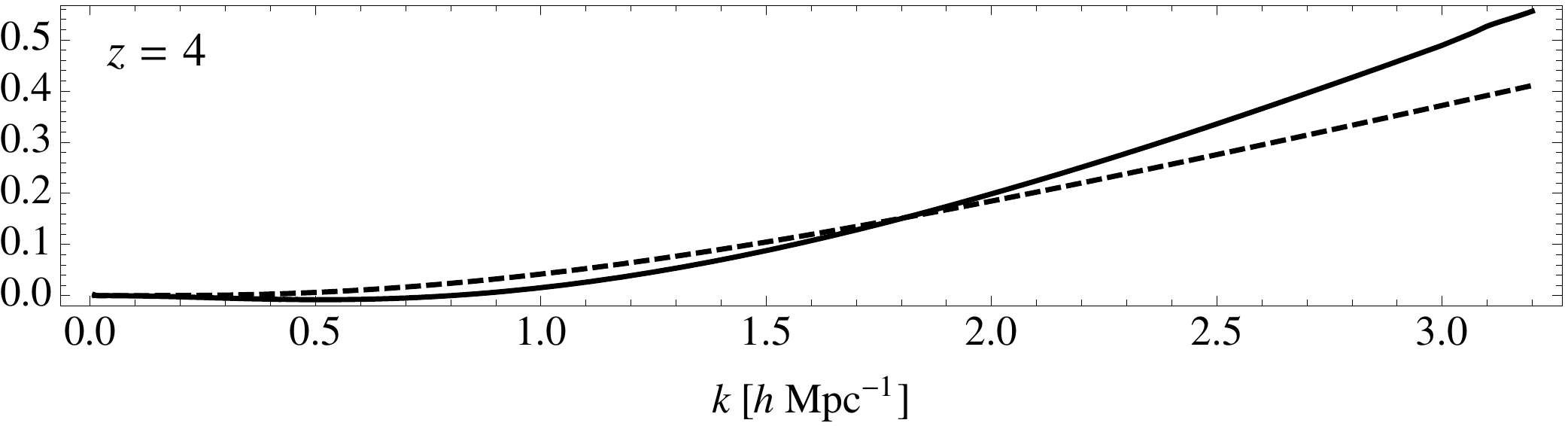}
\caption{\label{fig:highzestimates} \small\it  Comparisons of estimates and exact calculations of $\ptwoloop^\text{(total)}$ for $z=1$ to $4$. The estimates are computed as described in Sec.~\ref{sec:mpsestimates}.}
\end{center}
\end{figure}

For the interested reader, in this section we provide comparisons between the estimates for $\ptwoloop^\text{(total)}$ from Sec.~\ref{sec:mpsestimates} and the exact calculations from $z=1$ to $4$. As with Fig.~\ref{fig:p2loopestalphaz0}, in these plots we have fixed $\co$ to its best-fit value at each redshift, and have fixed $\alpha$ to $3/10$ plus any small shifts that are required to remove zero crossings from the estimated curves. One can see from Fig.~\ref{fig:highzestimates} that the estimates provide a good match to the exact calculations at all redshifts, though one should keep in mind the presence of the same uncertainties that we describe in Sec.~\ref{sec:mpsestimates}.

%--------------------------------------------------------------------------------------
% APPENDIX: CS2 AND LOGS IN P2LOOP
%--------------------------------------------------------------------------------------
\section{Additional Discussion of $\ct$}
\label{app:cs2log}

As mentioned in Sec.~\ref{sec:cs2fitmethod}, it is possible to calculate $\ct$ analytically in a scaling universe by taking the limit of $\ptwoloop$ in which the external momentum $k$ is much smaller than the internal momenta $q$ and $p$. Explicitly, if $P_{11}(k) \propto k^n$, then this limit of the integral will be dominated by the term containing the $F_5^{\rm (s)}$ kernel, which scales like $k^2/q^2$ in the $k\ll q$ limit. Therefore, the UV limit of $\ptwoloop(k)$ looks like
\beq
\label{eq:p2loopuv}
\ptwoloop^{\rm (UV)}(k) 
	\sim \lb  \int^\Lambda_{k/\epsilon} dq\, q^n \int^\Lambda_{k/\epsilon} dp \, 
	p^{2+n} \rb \frac{k^2}{\knl^2} P_{11}(k)
	= C \Lambda^{4+2n} \frac{k^2}{\knl^2} P_{11}(k)
		\lp1+{\cal O}\left(\frac{k}{\knl}\right)\rp\ ,
\eeq
where $C$ is a constant that can be calculated exactly by evaluating the $q$ and $p$ integrals, and $\epsilon$ is an order one number. (The $\mathcal{O}(k/\knl)$ factor comes from the lower limit of integration, but evaluating~(\ref{eq:p2loopuv}) with $k\ll \knl$ ensures that this factor will be highly suppressed, allowing for an accurate determination of $C$.) Setting $\ct \propto -C\Lambda^{4+2n}$ (after restoring the proper conventions) will precisely cancel this contribution at all values of $k$.  Alternatively, one could fix $\ct$ by matching to the full calculation of $\ptwoloop(k)$ at some small but finite value of $k$, where the $(k/\knl)^2 P_{11}(k)$ term in $\ptwoloop(k)$ is dominant. Note that this entire discussion is taking place at $z=0$, but the time-dependence of $\ct$ will be fixed to the time-dependence of $\ptwoloop$, so there is no need to distinguish between different times.

\begin{figure}[t]
\begin{center}
\includegraphics[width=0.7\textwidth]{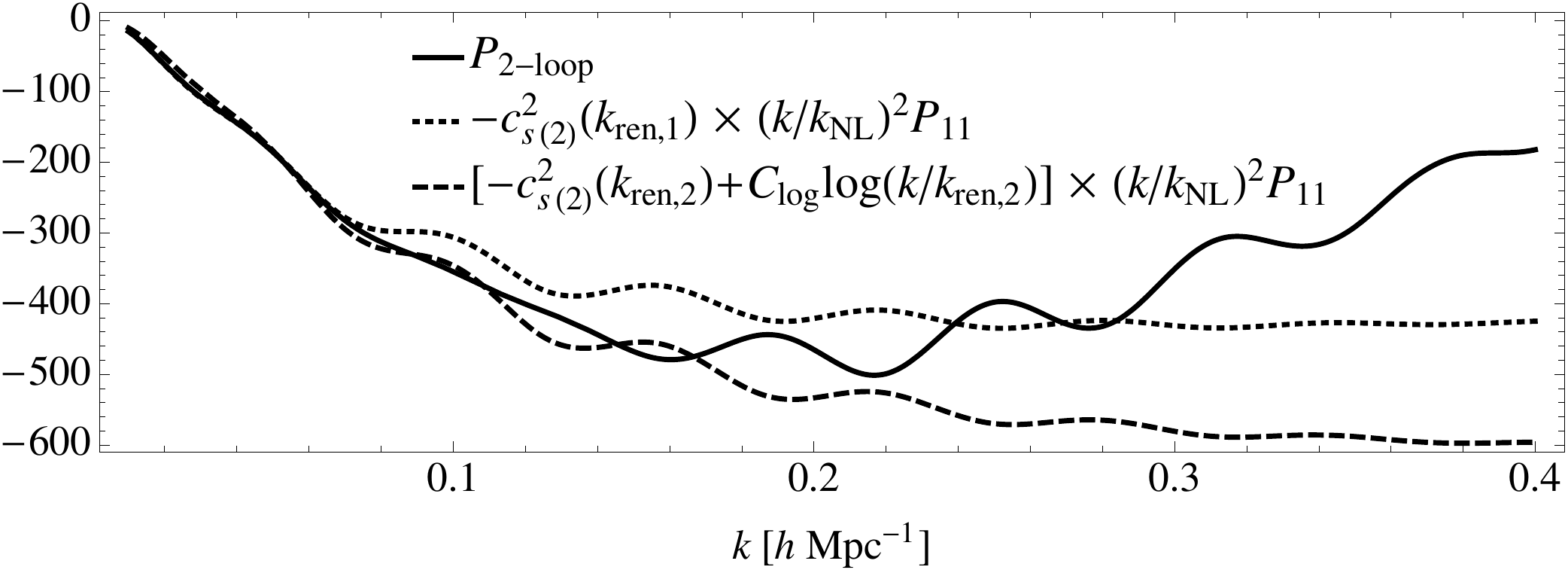}
\caption{\label{fig:p2looplog} \small\it We plot the full $\ptwoloop$ calculation in {\lcdm} (solid), a $\ct (k/\knl)^2 P_{11}$ term with $\ct$ fit at $\kren=0.05\invMpc$ (dotted), and the sum of two terms (dashed): a $\ct (k/\knl)^2 P_{11}$ term with $\ct$ fit at $\kren=0.3\invMpc$, and a $\log(k/\kren) (k/\knl)^2 P_{11}$ term. The dashed curve can be made to agree with $\ptwoloop$ at low $k$ if the coefficient of the log term is set to $C_{\rm log}\simeq -0.55$. This implies that we cannot exclude the possibility that $\ptwoloop$ contains a contribution that depends logarithmically on $k$, and therefore $\ct$ may depend on $\kren$. }
\end{center}
\end{figure}

The situation changes slightly at certain values of $n$, however. A case of interest is $n=-2$, which is quite close to the slope of the power spectrum on quasi-linear scales in our universe~\footnote{For this discussion, the relevant scales are the ones close to where we choose the renormalization scale.}, for which $\ptwoloop^{\rm (UV)}(k)$ depends logarithmically on $\Lambda$:
\beq
\ptwoloop^{({\rm UV},n=-2)}(k) 
	= \lb C_{\rm fin} + C_{\rm log} \log\!\lp \frac{k}{\Lambda} \rp \rb \frac{k^2}{\knl^2} P_{11}(k)\ .
\eeq
In order for $\ct$ to cancel this $\Lambda$-dependence, it will gain a dependence on a renormalization scale~$\kren$: $\ct(\kren) \propto \lb -C_{\rm fin} + C_{\rm log} \log(\Lambda/\kren) \rb$. The same phenomenon occurs when loops are renormalized in quantum field theory, where the renormalization scale is often denoted by $\mu$ and $C_{\rm log}$ is the analogue of a $\beta$-function. If we were to attempt to determine $\ct$ by matching to $\ptwoloop$ at low $k$, the value we would find would depend on the scale $\kren$ at which we perform the matching, with the difference between matching at two scales $k_{{\rm ren},1}$ and $k_{{\rm ren},2}$ given by $C_{\rm log} \log(k_{{\rm ren},1}/k_{{\rm ren},2})$.

In {\lcdm}, the effective slope $n_{\rm eff}(k)$ of the linear power spectrum varies strongly with $k$, so that the UV part of $\ptwoloop^{\rm (UV)}$ cannot be determined as cleanly as in the scaling case. Since $n_{\rm eff}(k)$ passes through $-2$ around $k\sim 0.25\invMpc$ (see Fig.~\ref{fig:neffk}), we cannot exclude the possibility that $\ptwoloop^{\rm (UV)}$ contains some kind of logarithmic dependence on $k$ around this scale, implying that $\ct$ would depend on the scale at which it is fixed. To show this concretely, Fig.~\ref{fig:p2looplog} displays the full $\ptwoloop$ curve, along with two other calculations. The dotted curve corresponds to $\ct (k/\knl)^2 P_{11}$ where $\ct$ has been matched to $\ptwoloop$ at $k_{{\rm ren},1}=0.05\invMpc$. The dashed curve shows the sum of $\ct (k/\knl)^2 P_{11}$ with $\ct$ fit using the procedure from Sec.~\ref{sec:cs2fitmethod} (with an effective renormalization scale of $k_{{\rm ren},2}=0.3\invMpc$), and a log term corresponding to renormalization at $k_{{\rm ren},2}$. The fact that both curves match equally well with $\ptwoloop$ at low $k$ shows that we cannot distinguish whether or not there is a log term in $\ptwoloop$, and therefore we cannot exclude the possibility that $\ct$ depends on $\kren$.

Therefore, to be consistent, we must use (roughly) the same scale to determine both $\co$ and $\ct$. We cannot use data at very low $k$ to determine $\co$, due to the fact that at low $k$ the contribution of the $\co$ term becomes smaller than the numerical, and cosmic variance, error in the simulations. This implies that we should not match $\ct$ to $\ptwoloop$ at low $k$, but should instead use information at higher scales to determine $\ct$%
%%%%% FOOTNOTE %%%%%
~\footnote{Alternatively, one could measure $\ct$ at low $k$'s and then compute the running by numerically evaluating the two-loop integral, an approach that seems numerically quite challenging.}%
%%%% END FOOTNOTE %%%%
. The precise procedure we use, based on matching at a certain $z$-dependent $\kren$, is described in Sec.~\ref{sec:cs2fitmethod}. The fact that $\kren$ depends on $z$ also explains why the time-dependence of $\ct$ is not simply given by the time-dependence of $\ptwoloop$.

%--------------------------------------------------------------------------------------
% APPENDIX: CONSISTENCY CHECK
%--------------------------------------------------------------------------------------
\section{Consistency Check using Different Cosmology}
\label{app:wmap3}

\begin{figure}[t]
\begin{center}
\includegraphics[width=0.6\textwidth]{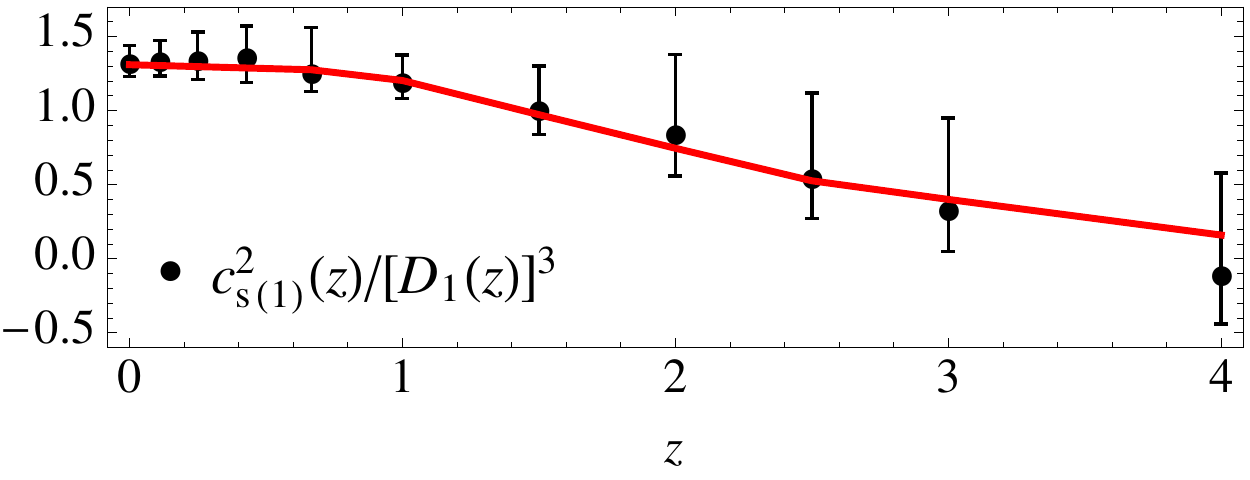}
\caption{\label{fig:cs1z-wmap3} \small\it As Fig.~\ref{fig:cs1z-wmap7}, but for WMAP3 instead of WMAP7 cosmological parameters. The red curve shows the fitting function~(\ref{eq:cs1z}) with $\beta=-1.20$, which exhibits excellent agreement with the values of $\co(z)$ obtained separately at each redshift.}
\end{center}
\end{figure}

\begin{figure}[t]
\begin{center}
\includegraphics[width=0.9\textwidth]{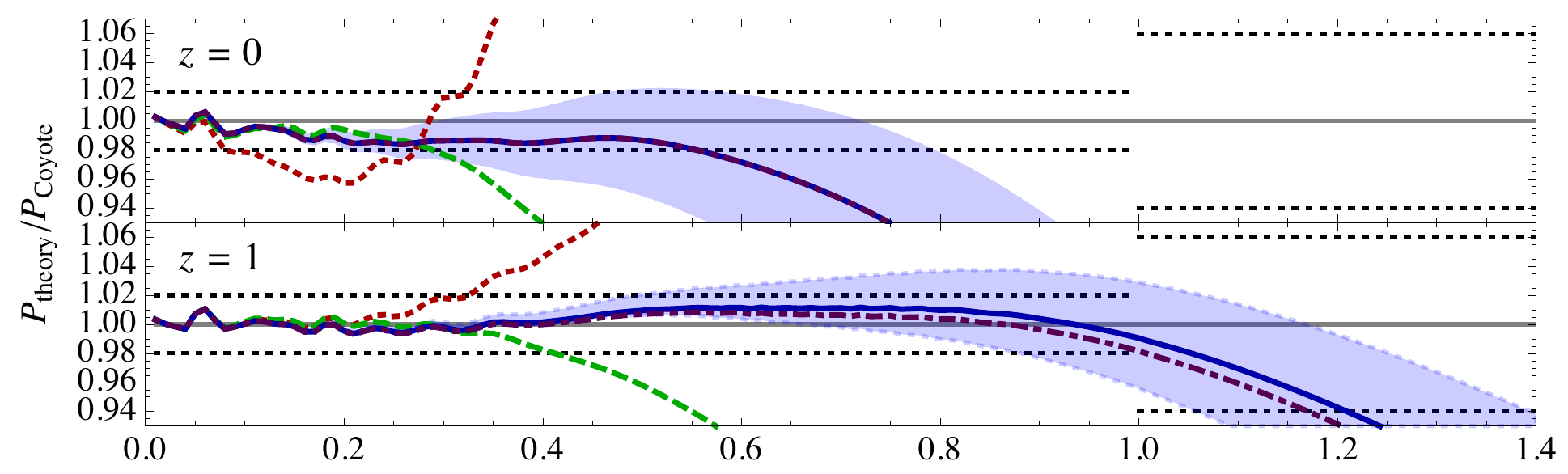}
\includegraphics[width=0.903\textwidth]{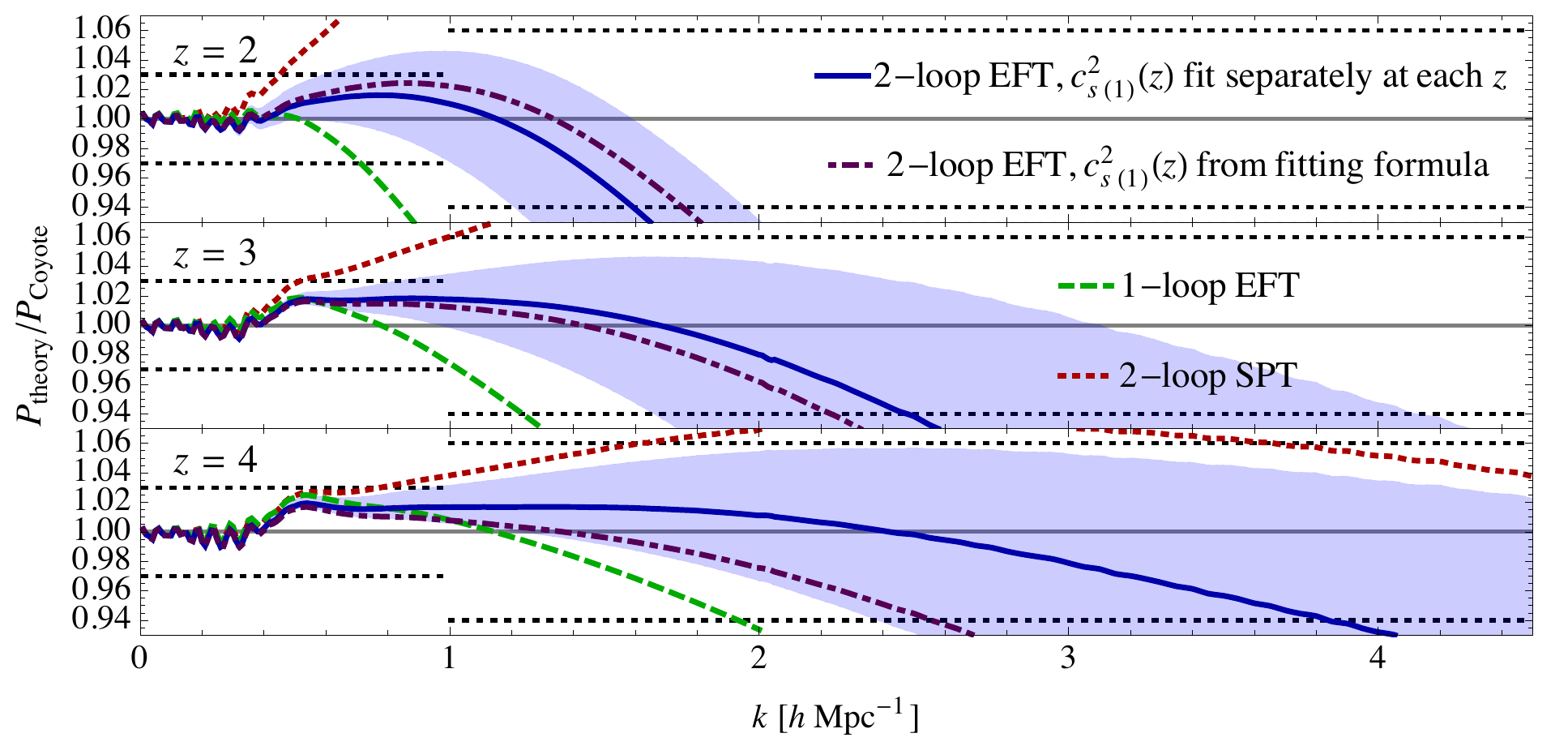}
\caption{\label{fig:datacomparison-wmap3} \small\it  As Fig.~\ref{fig:datacomparison-wmap7}, but for WMAP3 instead of WMAP7 cosmological parameters. The reach of the EFT predictions is comparable what we find in Sec.~\ref{sec:fits}. }
\end{center}
\end{figure}

In this appendix, we apply the procedures in Secs.~\ref{sec:cs2fitmethod} and~\ref{sec:fits} to the nonlinear power spectrum generated by Coyote using WMAP3 cosmological parameters: $h=0.73$, $\Omega_{\rm  m}=0.238$, $\Omega_{\rm b}=0.0418$, $\Omega_\Lambda=1-\Omega_{\rm m}$, $n_{\rm s}=0.951$, and $\sigma_8=0.74$. Fig.~\ref{fig:cs1z-wmap3} shows the values of $\co(z)$ determined separately at each redshift, along with the fitting formula~(\ref{eq:cs1z}) with the best-fit value $\beta=-1.20$, while Fig.~\ref{fig:datacomparison-wmap3} compares the EFT predictions with the nonlinear spectra at different redshifts.  Overall, we find results that are comparable to what we found for a WMAP7 cosmology, both for the reach of the EFT prediction at various redshifts and for the validity of Eq.~(\ref{eq:cs1z}) for describing the time-dependence of $\co(z)$.

%--------------------------------------------------------------------------------------
% APPENDIX: NON-LOCALITY IN TIME
%--------------------------------------------------------------------------------------
\section{Effect of Non-Locality in Time}
\label{app:nonlocal}

We have performed our main study using the assumption that the short modes that have been ``integrated out" of the EFTofLSS affect the long modes in a way that is purely local in time, an assumption that may not be strictly correct in the real universe~\cite{Carrasco:2013mua,Carroll:2013oxa,Senatore:2014eva}. This assumption affects the two-loop matter power spectrum through the terms $\poneloopcs$ and $P_\text{tree}^{(k^4,2)}$. The first of these can be written in terms of three separate functions $\tilde{P}_n(k)$ and three coefficients $\bar{c}_n$ (see~\cite{Carrasco:2013mua} for more details):
\beq
\label{eq:poneloopcssplit}
(2\pi) \co(z) \poneloopcs(k) 
	= \bar{c}_1(z) \tilde{P}_1(k) + \bar{c}_2(z) \tilde{P}_2(k) + \bar{c}_3(z) \tilde{P}_3(k)\ ,
\eeq
while the second can be written as
\beq
\label{eq:ptreek42cn}
P_\text{tree}^{(k^4,2)}(k,z) =  \frac{\zeta+\frac{5}{2}}{2(\zeta+\frac{5}{4})} 
	\bar{c}_1(z) \bar{c}_{1+\zeta}(z) [D_1(z)]^2 k^4 P_{11}(k)\ .
\eeq
The first coefficient, $\bar{c}_1$, is defined by $\bar{c}_1(z) = (2\pi) \co(z) /\knl^2$, while $\bar{c}_n = \frac{p+1}{p+n} \bar{c}_1$ if Eq.~(\ref{eq:Kansatzold}) is used for the kernel $K(a,a')$ appearing in the Euler equation. The parameter $p$ is a simple way to parametrize the severity of the non-locality in time: a lower value for $p$ corresponds to a broader kernel, while $p\to\infty$ corresponds to the local limit, effectively turning the kernel into a delta function at $a'=a$.

\begin{figure}[t]
\begin{center}
\includegraphics[width=0.9\textwidth]{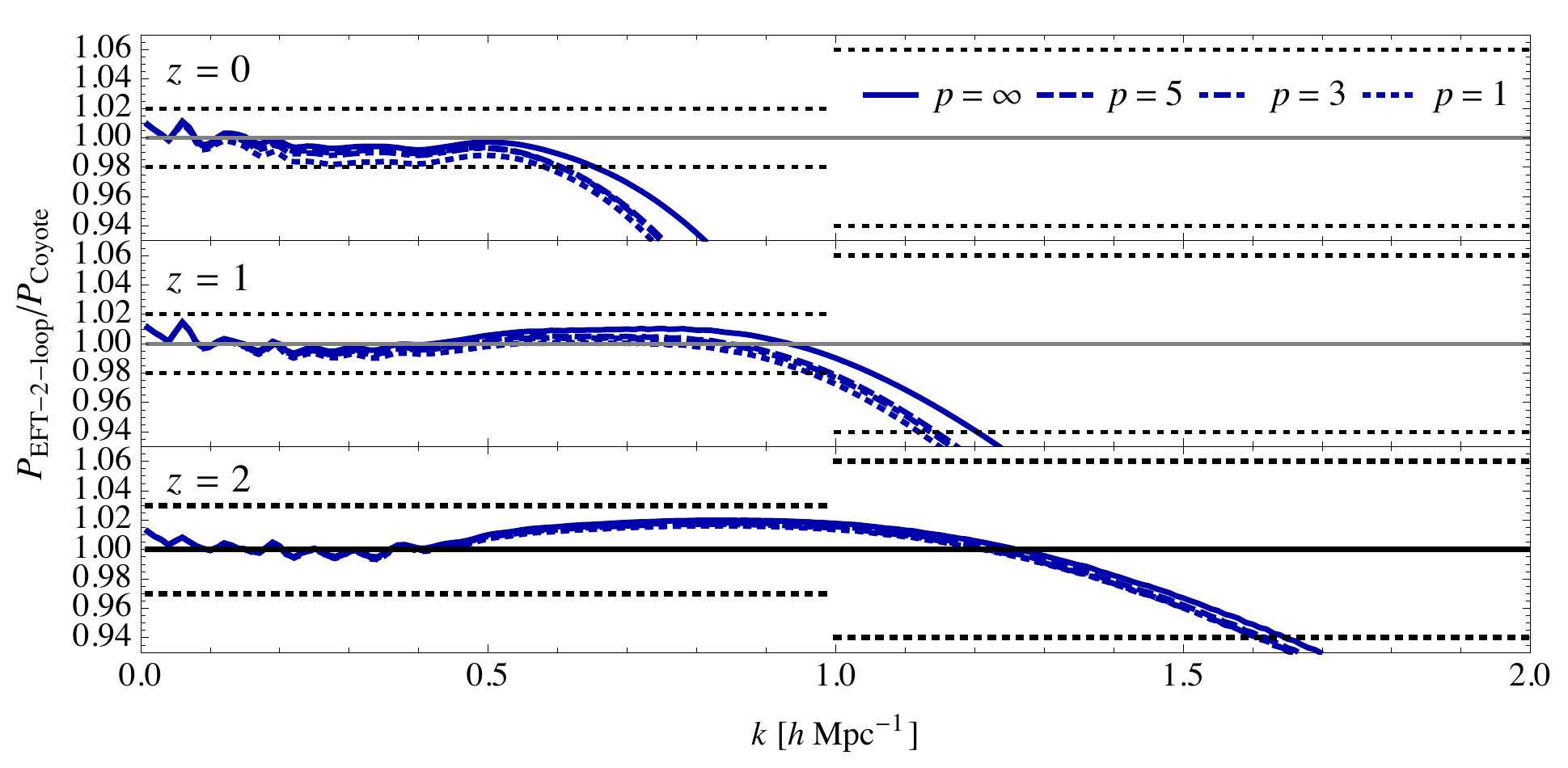}
\caption{\label{fig:nonlocal-p-wmap7} \small\it  Two-loop EFT predictions for the WMAP7 power spectrum, parametrizing the severity of the non-locality in time of the $\poneloopcs$ counterterm with a single parameter $p$. The speed of sound parameter $\co$ is fit separately at each redshift and $p$ value using the procedure described in Sec.~\ref{sec:fits}. Taking the uncertainties on the nonlinear data into account, we find that introducing non-locality in time does not improve the agreement of the prediction with the data at any redshift. }
\end{center}
\end{figure}

In Fig.~\ref{fig:nonlocal-p-wmap7}, we repeat the procedures in Secs.~\ref{sec:cs2fitmethod} and~\ref{sec:fits} using different values of $p$ in the two-loop EFT prediction, re-fitting $\co$ in each case. The effect of changing $p$ is most prominent at $z=0$, when the greatest amount of time evolution has taken place. For the most nonlocal case we consider, $p=1$, there is a $\sim$2\% shift in the power spectrum for $k \gtrsim 0.2\invMpc$, with a corresponding 6\% shift (upward) in the value of $\co(0)$. For higher, and maybe more reasonable~\cite{Senatore:2014eva}, values of $p$, the effect on the power spectrum is much less, especially when compared to the error induced by the uncertainty in $\co$ (see Fig.~\ref{fig:datacomparison-wmap7}). The non-locality has even less of an effect at higher redshift, and, importantly, cannot improve the reach of the prediction into the UV at any redshift.

\begin{figure}[t]
\begin{center}
\includegraphics[width=0.9\textwidth]{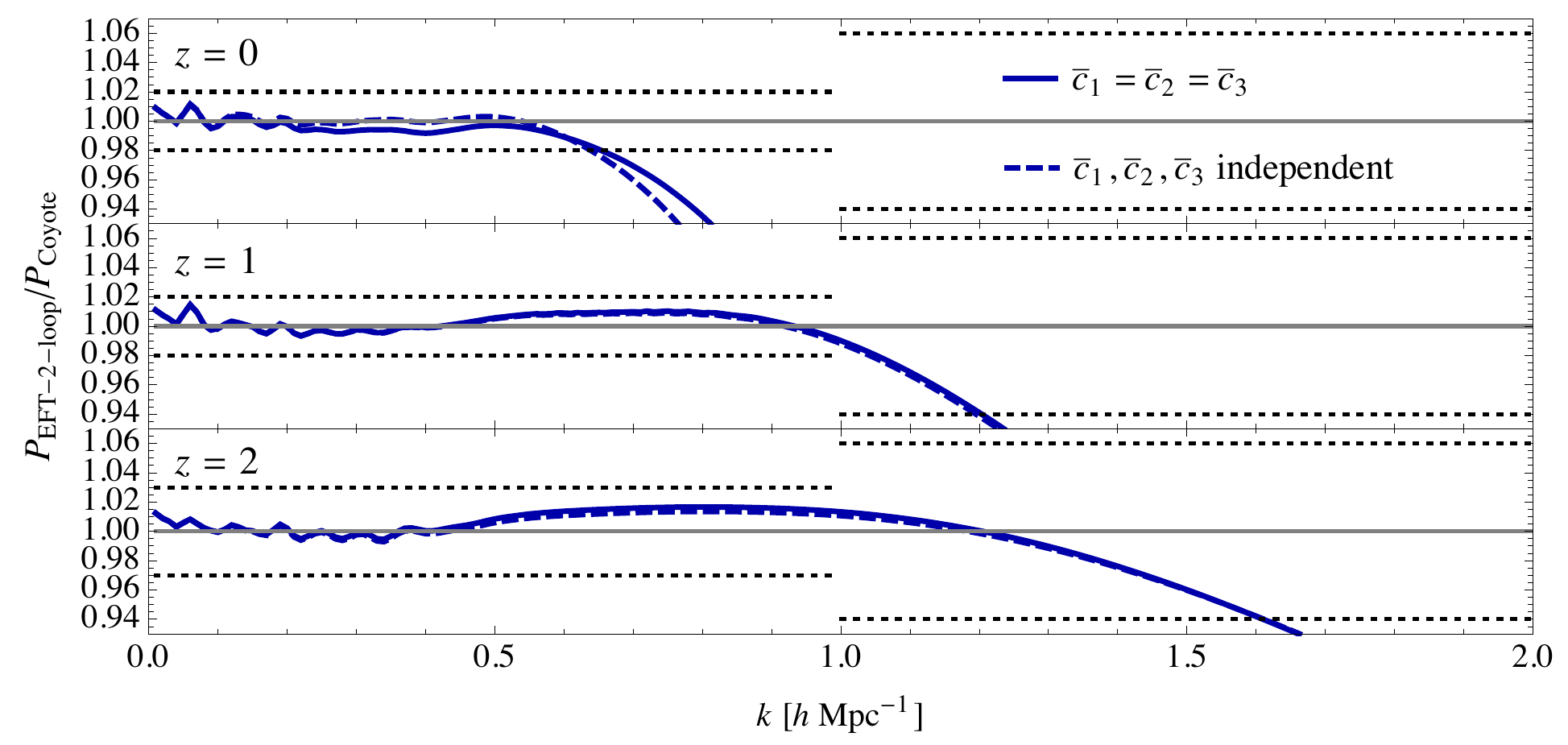}
\caption{\label{fig:nonlocal-arbcbar-wmap7} \small\it  Two-loop EFT predictions for the WMAP7 power spectrum, allowing for a generalized form of non-locality in time in which the four $\bar{c}_n$ coefficients appearing in $\poneloopcs$ and $P_\text{tree}^{(k^4,2)}$ (see Eqs.~(\ref{eq:poneloopcssplit}) and~(\ref{eq:ptreek42cn}))  are treated as independent, and are fit as described in the main text. This results in predictions that are not significantly different from the time-local case, motivating the adoption of the time-local approximation in practical computations. }
\end{center}
\end{figure}

Instead of using the relationship between the $\bar{c}_n$ coefficients implied by Eq.~(\ref{eq:Kansatzold}), we can also allow for a  completely general type of non-locality by treating each $\bar{c}_n$ as independent. The two-loop EFT prediction then has four free parameters at each redshift, which we can write as $\co(z)$ and $r_n(z) \equiv \bar{c}_n(z)/\bar{c}_1(z)$ for $n=2,3,1+\zeta$. We determine the four parameters at each redshift through a least-$\chi^2$ fit to the nonlinear power spectrum, subject to the constraints that $|r_{1+\zeta}| \leq  |r_3| \leq |r_2| \leq 1$ and $\co(z)$ is within 30\% of the value obtained from the time-local fit. The first constraint comes from the fact $c_n$ is equal to an integral over the $n^{\rm th}$ power of the growth factor $D_1(z)$, and $0\leq D_1(z) \leq 1$, so higher powers of $D_1(z)$ in the integrand will act to suppress the absolute value of the integral. The second constraint comes from the fact that $\co$ can be determined from the one-loop calculation, and the value for $\co$, when adding time non-locality at two loops, should not be drastically different from the one-loop value. 

At each redshift, we perform this fit between $k_{\rm min}=0.15\invMpc$ and a floating $k_{\rm max}$, taking the overall best fit to be one that gives a flat region in $P_\text{EFT-2-loop}/P_\text{Coyote}$ that reaches farthest into the UV. (For the fits at $z= \{ 0,1,2 \}$, the $k_{\rm max}$ values corresponding to these best fits are $\{ 0.6, 1.4, 1.9 \} \invMpc$.) The results are shown in Fig.~\ref{fig:nonlocal-arbcbar-wmap7}. We find only minor differences between these general predictions and those of the time-local approximation. In summary, Figs.~\ref{fig:nonlocal-p-wmap7} and~\ref{fig:nonlocal-arbcbar-wmap7} demonstrate that the matter power spectrum at $z\leq 2$  (and also at $z\leq 4$, although we do not plot redshifts greater than~2) does not provide any compelling evidence for non-locality in time, although such evidence may emerge from consideration of other observables. This is an interesting topic for future studies. 

We finally point out the following. Since, from comparison to data, we find no need to abandon the time-local approximation, the EFTofLSS at two-loop order has only two free parameters for all redshifts: $\co$ at $z=0$ and its `$z$-slope' $\beta$, as described. In the non-local in time case, additional parameters are present, but it is incorrect to consider these additional parameters as having been fitted to obtain the results presented in the main text of this paper. This is because the local-in-time case constitutes a special, highly symmetric, point in the theory space of the EFTofLSS. As is standard, when theories are enhanced by additional symmetries or properties, the number of free parameters is reduced.

%--------------------------------------------------------------------------------------
% REFERENCES
%--------------------------------------------------------------------------------------


\begin{thebibliography}{10}

%\cite{Baumann:2010tm}
\bibitem{Baumann:2010tm} 
  D.~Baumann, A.~Nicolis, L.~Senatore and M.~Zaldarriaga,
  ``Cosmological Non-Linearities as an Effective Fluid,''
  JCAP {\bf 1207}, 051 (2012)
  [\href{http://arxiv.org/abs/1004.2488}{{\tt arXiv:1004.2488}}] [astro-ph.CO].
  %[arXiv:1004.2488 [astro-ph.CO]].
  %%CITATION = ARXIV:1004.2488;%%
  %47 citations counted in INSPIRE as of 30 Jul 2013
  
%\cite{Carrasco:2012cv}
\bibitem{Carrasco:2012cv} 
  J.~J.~M.~Carrasco, M.~P.~Hertzberg and L.~Senatore,
  ``The Effective Field Theory of Cosmological Large Scale Structures,''
  JHEP {\bf 1209}, 082 (2012)
 [\href{http://arxiv.org/abs/1206.2926}{{\tt arXiv:1206.2926}}] [astro-ph.CO].
  %%CITATION = ARXIV:1206.2926;%%
  
   
  %\cite{Carrasco:2013mua}
\bibitem{Carrasco:2013mua} 
  J.~J.~M.~Carrasco, S.~Foreman, D.~Green and L.~Senatore,
  ``The Effective Field Theory of Large Scale Structures at Two Loops,''
  JCAP {\bf 1407}, 057 (2014)
  %[arXiv:1310.0464 [astro-ph.CO]].
    [\href{http://arxiv.org/abs/1310.0464}{{\tt arXiv:1310.0464}}]   [astro-ph.CO].
  %%CITATION = ARXIV:1310.0464;%%
  %16 citations counted in INSPIRE as of 08 Aug 2014
  
  %\cite{Pajer:2013jj}
\bibitem{Pajer:2013jj} 
  E.~Pajer and M.~Zaldarriaga,
  ``On the Renormalization of the Effective Field Theory of Large Scale Structures,''
  JCAP {\bf 1308}, 037 (2013)
  [\href{http://arxiv.org/abs/1301.7182}{{\tt arXiv:1301.7182}}] [astro-ph.CO].
  %[arXiv:1301.7182 [astro-ph.CO]].
  %%CITATION = ARXIV:1301.7182;%%
  %11 citations counted in INSPIRE as of 07 Mar 2014
  
     %\cite{Carrasco:2013sva}
\bibitem{Carrasco:2013sva} 
  J.~J.~M.~Carrasco, S.~Foreman, D.~Green and L.~Senatore,
  ``The 2-loop matter power spectrum and the IR-safe integrand,''
  JCAP {\bf 1407}, 056 (2014)
  [\href{http://arxiv.org/abs/1304.4946}{{\tt arXiv:1304.4946}}] [astro-ph.CO].
  %arXiv:1304.4946 [astro-ph.CO].
  %%CITATION = ARXIV:1304.4946;%%
  %13 citations counted in INSPIRE as of 08 Aug 2014

   %\cite{Mercolli:2013bsa}
\bibitem{Mercolli:2013bsa} 
  L.~Mercolli and E.~Pajer,
  ``On the Velocity in the Effective Field Theory of Large Scale Structures,''
  JCAP {\bf 1403}, 006 (2014)
  [\href{http://arxiv.org/abs/1307.3220}{{\tt arXiv:1307.3220}}]  [astro-ph.CO].
  %arXiv:1307.3220 [astro-ph.CO].
  %%CITATION = ARXIV:1307.3220;%%
  %4 citations counted in INSPIRE as of 07 Mar 2014
  
  
  %\cite{Carroll:2013oxa}
\bibitem{Carroll:2013oxa} 
  S.~M.~Carroll, S.~Leichenauer and J.~Pollack,
  ``A Consistent Effective Theory of Long-Wavelength Cosmological Perturbations,''
  Phys.\ Rev.\ D {\bf 90}, 023518 (2014)
  %arXiv:1310.2920 [hep-th].
   [\href{http://arxiv.org/abs/1310.2920}{{\tt arXiv:1310.2920}}] [hep-th].
  %%CITATION = ARXIV:1310.2920;%%
  %15 citations counted in INSPIRE as of 29 Jul 2014

  
%\cite{Porto:2013qua}
\bibitem{Porto:2013qua}
  R.~A.~Porto, L.~Senatore and M.~Zaldarriaga,
  ``The Lagrangian-Space Effective Field Theory of Large Scale Structures,''
  JCAP {\bf 1405}, 022 (2014)
 % [arXiv:1311.2168 [astro-ph.CO]].
    [\href{http://arxiv.org/abs/1311.2168}{{\tt arXiv:1311.2168}}]   [astro-ph.CO].
  %%CITATION = ARXIV:1311.2168;%%
  %27 citations counted in INSPIRE as of 20 Dec 2014
  
 
  
  %\cite{Senatore:2014via}
\bibitem{Senatore:2014via} 
  L.~Senatore and M.~Zaldarriaga,
  ``The IR-resummed Effective Field Theory of Large Scale Structures,''
  JCAP {\bf 1502}, 013 (2015)
  % [arXiv:1404.5954 [astro-ph.CO]].
   [\href{http://arxiv.org/abs/1404.5954}{{\tt arXiv:1404.5954}}]   [astro-ph.CO].
  %%CITATION = ARXIV:1404.5954;%%
  %16 citations counted in INSPIRE as of 25 Feb 2015
  
     %\cite{Angulo:2014tfa}
\bibitem{Angulo:2014tfa} 
  R.~E.~Angulo, S.~Foreman, M.~Schmittfull and L.~Senatore,
  ``The One-Loop Matter Bispectrum in the Effective Field Theory of Large Scale Structures,''
  %arXiv:1406.4143 [astro-ph.CO].
    [\href{http://arxiv.org/abs/1406.4143}{{\tt arXiv:1406.4143}}]   [astro-ph.CO].
  %%CITATION = ARXIV:1406.4143;%%
  
  %\cite{Baldauf:2014qfa}
\bibitem{Baldauf:2014qfa} 
  T.~Baldauf, L.~Mercolli, M.~Mirbabayi and E.~Pajer,
  ``The Bispectrum in the Effective Field Theory of Large Scale Structure,''
  %arXiv:1406.4135 [astro-ph.CO].
    [\href{http://arxiv.org/abs/1406.4135}{{\tt arXiv:1406.4135}}]   [astro-ph.CO].
  %%CITATION = ARXIV:1406.4135;%%
  
  %\cite{Senatore:2014eva}
\bibitem{Senatore:2014eva} 
  L.~Senatore,
  ``Bias in the Effective Field Theory of Large Scale Structures,''
    [\href{http://arxiv.org/abs/1406.7843}{{\tt arXiv:1406.7843}}]  [astro-ph.CO].
  %arXiv:1406.7843 [astro-ph.CO].
  %%CITATION = ARXIV:1406.7843;%%
  
  %\cite{Senatore:2014vja}
\bibitem{Senatore:2014vja}
  L.~Senatore and M.~Zaldarriaga,
  ``Redshift Space Distortions in the Effective Field Theory of Large Scale Structures,''
  %arXiv:1409.1225 [astro-ph.CO].
      [\href{http://arxiv.org/abs/1409.1225}{{\tt arXiv:1409.1225}}] [astro-ph.CO].  
  %%CITATION = ARXIV:1409.1225;%%
  %2 citations counted in INSPIRE as of 21 Dec 2014

  %\cite{Lewandowski:2014rca}
\bibitem{Lewandowski:2014rca}
  M.~Lewandowski, A.~Perko and L.~Senatore,
  ``Analytic Prediction of Baryonic Effects from the EFT of Large Scale Structures,''
 % arXiv:1412.5049 [astro-ph.CO].
  [\href{http://arxiv.org/abs/1412.5049}{{\tt arXiv:1412.5049}}] [astro-ph.CO].
  %%CITATION = ARXIV:1412.5049;%%

  %\cite{Mirbabayi:2014zca}
\bibitem{Mirbabayi:2014zca}
  M.~Mirbabayi, F.~Schmidt and M.~Zaldarriaga,
  ``Biased Tracers and Time Evolution,''
 % arXiv:1412.5169 [astro-ph.CO].
  [\href{http://arxiv.org/abs/1412.5169}{{\tt arXiv:1412.5169}}] [astro-ph.CO].
  %%CITATION = ARXIV:1412.5169;%%



%\cite{formalism}
\bibitem{formalism} 
  J.~J.~M.~Carrasco, S.~Foreman, and L.~Senatore,
  %``\sfnote{Formalism/exact time-dependence paper},''
    to appear.



  %\cite{Lewis:1999bs}
\bibitem{Lewis:1999bs} 
  A.~Lewis, A.~Challinor and A.~Lasenby,
  ``Efficient computation of CMB anisotropies in closed FRW models,''
  Astrophys.\ J.\  {\bf 538}, 473 (2000)
 [\href{http://arxiv.org/abs/astro-ph/9911177}{{\tt astro-ph/9911177}}].
  %%CITATION = ASTRO-PH/9911177;%% 
   

 %\cite{Heitmann:2008eq,Heitmann:2009cu,Lawrence:2009uk,Heitmann:2013bra}  
%\cite{Heitmann:2008eq}
\bibitem{Heitmann:2008eq} 
  K.~Heitmann, M.~White, C.~Wagner, S.~Habib and D.~Higdon,
  ``The Coyote Universe I: Precision Determination of the Nonlinear Matter Power Spectrum,''
  Astrophys.\ J.\  {\bf 715}, 104 (2010)
  [\href{http://arxiv.org/abs/0812.1052}{{\tt arXiv:0812.1052}}]  [astro-ph]. \\
  %[arXiv:0812.1052 [astro-ph]].
  %%CITATION = ARXIV:0812.1052;%%
  %85 citations counted in INSPIRE as of 01 Jul 2013
%%\cite{Heitmann:2009cu}
%\bibitem{Heitmann:2009cu} 
  K.~Heitmann, D.~Higdon, M.~White, S.~Habib, B.~J.~Williams and C.~Wagner,
  ``The Coyote Universe II: Cosmological Models and Precision Emulation of the Nonlinear Matter Power Spectrum,''
  Astrophys.\ J.\  {\bf 705}, 156 (2009)
  [\href{http://arxiv.org/abs/0902.0429}{{\tt arXiv:0902.0429}}]  [astro-ph.CO]. \\
 % [arXiv:0902.0429 [astro-ph.CO]].
  %%CITATION = ARXIV:0902.0429;%%
  %46 citations counted in INSPIRE as of 01 Jul 2013
%   %\cite{Lawrence:2009uk}
%\bibitem{Lawrence:2009uk} 
  E.~Lawrence, K.~Heitmann, M.~White, D.~Higdon, C.~Wagner, S.~Habib and B.~Williams,
  ``The Coyote Universe III: Simulation Suite and Precision Emulator for the Nonlinear Matter Power Spectrum,''
  Astrophys.\ J.\  {\bf 713}, 1322 (2010)
    [\href{http://arxiv.org/abs/0912.4490}{{\tt arXiv:0912.4490}}]  [astro-ph.CO]. \\
  %[arXiv:0912.4490 [astro-ph.CO]].
  %%CITATION = ARXIV:0912.4490;%%
  %51 citations counted in INSPIRE as of 01 Jul 2013
%%\cite{Heitmann:2013bra}
%\bibitem{Heitmann:2013bra} 
  K.~Heitmann, E.~Lawrence, J.~Kwan, S.~Habib and D.~Higdon,
  ``The Coyote Universe Extended: Precision Emulation of the Matter Power Spectrum,''
  Astrophys.\ J.\  {\bf 780}, 111 (2014)
  [\href{http://arxiv.org/abs/1304.7849}{{\tt arXiv:1304.7849}}]  [astro-ph.CO].
  %[arXiv:1304.7849 [astro-ph.CO]].
  %%CITATION = ARXIV:1304.7849;%%
  %8 citations counted in INSPIRE as of 27 Mar 2014
  
  
  
      %\cite{Carlson:2009it}
\bibitem{Carlson:2009it} 
  J.~Carlson, M.~White and N.~Padmanabhan,
  ``A critical look at cosmological perturbation theory techniques,''
  Phys.\ Rev.\ D {\bf 80}, 043531 (2009)
  [\href{http://arxiv.org/abs/0905.0479}{{\tt arXiv:0905.0479}}] [astro-ph.CO].
  %%CITATION = ARXIV:0905.0479;%%

  %\cite{Hahn:2004fe}
\bibitem{CUBA} 
  T.~Hahn,
  ``CUBA: A Library for multidimensional numerical integration,''
  Comput.\ Phys.\ Commun.\  {\bf 168}, 78 (2005)
 % [hep-ph/0404043v2].
  %%CITATION = HEP-PH/0404043;%%
  %191 citations counted in INSPIRE as of 12 Apr 2013
  [\href{http://arxiv.org/abs/hep-ph/0404043v2}{{\tt hep-ph/0404043v2}}].

   %\cite{Eisenstein:1997ik}
\bibitem{Eisenstein:1997ik} 
  D.~J.~Eisenstein and W.~Hu,
  ``Baryonic features in the matter transfer function,''
  Astrophys.\ J.\  {\bf 496}, 605 (1998)
  [\href{http://arxiv.org/abs/astro-ph/9709112}{{\tt astro-ph/9709112}}].
  %%CITATION = ASTRO-PH/9709112;%%

%\cite{Munshi:2006fn}
\bibitem{Munshi:2006fn} 
  D.~Munshi, P.~Valageas, L.~Van Waerbeke and A.~Heavens,
  ``Cosmology with Weak Lensing Surveys,''
  Phys.\ Rept.\  {\bf 462}, 67 (2008)
 % [astro-ph/0612667].
    [\href{http://arxiv.org/abs/astro-ph/0612667}{{\tt astro-ph/0612667}}].
  %%CITATION = ASTRO-PH/0612667;%%
  %150 citations counted in INSPIRE as of 01 Feb 2015

%\cite{Takahashi:2012em}
\bibitem{Takahashi:2012em} 
  R.~Takahashi, M.~Sato, T.~Nishimichi, A.~Taruya and M.~Oguri,
  ``Revising the Halofit Model for the Nonlinear Matter Power Spectrum,''
  Astrophys.\ J.\  {\bf 761}, 152 (2012)
 % [arXiv:1208.2701 [astro-ph.CO]].
    [\href{http://arxiv.org/abs/1208.2701}{{\tt arXiv:1208.2701}}] [astro-ph.CO].
  %%CITATION = ARXIV:1208.2701;%%
  %77 citations counted in INSPIRE as of 01 Feb 2015



  

  
  %\cite{Lewis:2006fu}
\bibitem{Lewis:2006fu} 
  A.~Lewis and A.~Challinor,
  ``Weak gravitational lensing of the CMB,''
  Phys.\ Rept.\  {\bf 429}, 1 (2006)
   [\href{http://arxiv.org/abs/astro-ph/0601594}{{\tt astro-ph/0601594}}].
  %[astro-ph/0601594].
  %%CITATION = ASTRO-PH/0601594;%%
  %281 citations counted in INSPIRE as of 15 Dec 2014
  
  %\cite{vanEngelen:2012va}
\bibitem{vanEngelen:2012va} 
  A.~van Engelen, R.~Keisler, O.~Zahn, K.~A.~Aird, B.~A.~Benson, L.~E.~Bleem, J.~E.~Carlstrom and C.~L.~Chang {\it et al.},
  ``A measurement of gravitational lensing of the microwave background using South Pole Telescope data,''
  Astrophys.\ J.\  {\bf 756}, 142 (2012)
  %[arXiv:1202.0546 [astro-ph.CO]].
      [\href{http://arxiv.org/abs/1202.0546}{{\tt arXiv:1202.0546}}] [astro-ph.CO].
  %%CITATION = ARXIV:1202.0546;%%
  %102 citations counted in INSPIRE as of 15 Dec 2014
  
  %\cite{Das:2013zf}
\bibitem{Das:2013zf} 
  S.~Das, T.~Louis, M.~R.~Nolta, G.~E.~Addison, E.~S.~Battistelli, J.~R.~Bond, E.~Calabrese and D.~C.~M.~J.~Devlin {\it et al.},
  ``The Atacama Cosmology Telescope: temperature and gravitational lensing power spectrum measurements from three seasons of  data,''
  JCAP {\bf 1404}, 014 (2014)
  %[arXiv:1301.1037 [astro-ph.CO]].
      [\href{http://arxiv.org/abs/1301.1037}{{\tt arXiv:1301.1037}}] [astro-ph.CO].
  %%CITATION = ARXIV:1301.1037;%%
  %126 citations counted in INSPIRE as of 15 Dec 2014
  

%\cite{Ade:2015zua}
\bibitem{Ade:2015zua} 
  P.~A.~R.~Ade {\it et al.}  [Planck Collaboration],
  ``Planck 2015 results. XV. Gravitational lensing,''
  %arXiv:1502.01591 [astro-ph.CO].
      [\href{http://arxiv.org/abs/1502.01591}{{\tt arXiv:1502.01591}}] [astro-ph.CO].
  %%CITATION = ARXIV:1502.01591;%%
  
  %\cite{Story:2014dwa}
\bibitem{Story:2014dwa} 
  K.~T.~Story, D.~Hanson, P.~A.~R.~Ade, K.~A.~Aird, J.~E.~Austermann, J.~A.~Beall, A.~N.~Bender and B.~A.~Benson {\it et al.},
  ``A Measurement of the Cosmic Microwave Background Gravitational Lensing Potential from 100 Square Degrees of SPTpol Data,''
 % arXiv:1412.4760 [astro-ph.CO].
      [\href{http://arxiv.org/abs/1412.4760}{{\tt arXiv:1412.4760}}] [astro-ph.CO].
  %%CITATION = ARXIV:1412.4760;%%

%\cite{Baldauf:2011bh}
\bibitem{Baldauf:2011bh}
  T.~Baldauf, U.~Seljak, L.~Senatore and M.~Zaldarriaga,
  ``Galaxy Bias and non-Linear Structure Formation in General Relativity,''
  JCAP {\bf 1110} (2011) 031
 % [arXiv:1106.5507 [astro-ph.CO]].
  [\href{http://arxiv.org/abs/1106.5507 }{{\tt arXiv:1106.5507}}] [astro-ph.CO].
  %%CITATION = ARXIV:1106.5507;%%
  %58 citations counted in INSPIRE as of 02 mar 2015

%\cite{Wagner:2014aka}
\bibitem{Wagner:2014aka}
  C.~Wagner, F.~Schmidt, C.~T.~Chiang and E.~Komatsu,
  ``Separate Universe Simulations,''
  Mon.\ Not.\ Roy.\ Astron.\ Soc.\  {\bf 448} (2015) 11
%  [arXiv:1409.6294 [astro-ph.CO]].
  [\href{http://arxiv.org/abs/1409.6294 }{{\tt arXiv:1409.6294}}] [astro-ph.CO].
  %%CITATION = ARXIV:1409.6294;%%
  %2 citations counted in INSPIRE as of 02 Mar 2015


   
\end{thebibliography}
\end{document}